\documentclass[bibyear]{aa} 

\usepackage{graphicx}
\usepackage{txfonts}
\begin{document} 

   \title{Extreme submillimetre starburst galaxies}

   \subtitle{ }

   \author{M. Rowan-Robinson
             \inst{1}
          \and Lingyu Wang
          \inst{2,3}
           \and Duncan Farrah
            \inst{4,5}
           \and Dimitra Rigopoulou
            \inst{6}
            \and Carlotta Gruppioni
            \inst{7}
             \and Mattia Vaccari \inst{8,10}
             \and Lucia Marchetti \inst{8,9}
            \and David L. Clements \inst{1}
            \and William J. Pearson \inst{2,3}
          }

   \institute{Astrophysics Group, Imperial College London, Blackett Laboratory, Prince Consort Road, London SW7 2AZ, UK\\
              \email{mrr@imperial.ac.uk}
         \and
             SRON, Groningen 9747 AD, Netherlands\\
         \and
             Kapteyn Astronomical Institute, University of Groningen, Groningen 9747 AD, Netherlands\\
         \and
              Department of Physics and Astronomy, University of Hawaii, 2505 Correa Road, Honolulu, HI 96822, USA\\
         \and
            Institute for Astronomy, 2680 Woodlawn Drive, University of Hawaii, Honolulu, HI 96822, USA\\
          \and
          Department of Astrophysics, University of Oxford, Keble Rd, Oxford OX1 3RH\\ 
          \and
           INAF, Osservatorio Astronomico di Bologna, via Ranzani
           1, I-40127 Bologna, Italy\\
           \and
           Department of Physics and Astronomy, University of the Western Cape, Robert Sobukwe Road, 
           7535, Bellville, Cape Town, South Africa\\
           \and
            Department of Astronomy, University of Cape Town, Rondebosch 7701, South Africa\\
           \and
            INAF, Istituto di Radioastronomia, via Gobetti 101, I-40129 Bologna, Italy\\
             }

   \date{Received  xxx; accepted xxxx}

 
  \abstract
  { We use two catalogues, a {\it Herschel} catalogue selected at 500
    $\mu$m ({\it HerMES}) and an {\it IRAS} catalogue selected at 60 $\mu$m
    (RIFSCz), to contrast the sky at these two wavelengths. Both
    surveys demonstrate the existence of ``extreme'' starbursts, with
    star-formation rates (SFRs) $> 5000 M_{\odot} yr^{-1}$.  The
    maximum intrinsic star-formation rate appears to be
    $\sim30,000 M_{\odot} yr^{-1}$.  The sources with apparent SFR
    estimates higher than this are in all cases either lensed systems,
    blazars, or erroneous photometric redshifts.

    At redshifts of 3 to 5, the time-scale for the {\it Herschel}
    galaxies to make their current mass of stars at their present rate
    of star formation $\sim 10^8$ yrs, so these galaxies are making a
    significant fraction of their stars in the current star-formation
    episode.  Using dust mass as a proxy for gas mass, the {\it
      Herschel} galaxies at redshift 3 to 5 have gas masses comparable
    to their mass in stars.  Of the 38 extreme starbursts in our {\it
      Herschel} survey for which we have more complete spectral energy
    distribution (SED) information, 50$\%$ show evidence for
    QSO-like optical emission, or exhibit AGN dust tori in the
    mid-infrared SEDs. In all cases however the infrared luminosity is
    dominated by a starburst component. We derive a mean covering
    factor for AGN dust as a function of redshift and derive black
    hole masses and black hole accretion rates. There is a universal
    ratio of black-hole mass to stellar mass in these high redshift
    systems of $\sim 10^{-3}$, driven by the strong period of
    star-formation and black-hole growth at $z = 1-5$.  }


   \maketitle
%

\section{Introduction}

A key discovery from the Infrared Astronomical Satellite ({\it IRAS})
surveys was the existence of galaxies with remarkably high infrared
(rest-frame $1-1000\,\mu$m) luminosities. {\it IRAS} found thousands of
galaxies with infrared luminosities $> 10^{12} L_{\odot}$, termed
``ultraluminous'' infrared galaxies or ULIRGs (Soifer et al 1984,
Aaronson and Olszewski 1984, Houck et al 1985, Joseph and Wright 1985,
Allen et al 1985, Lawrence et al 1986), and over a hundred with
luminosities $> 10^{13} L_{\odot}$, termed ``hyperluminous'' infrared
galaxies or HyLIRGs (Rowan-Robinson et al 2000, Rowan-Robinson and
Wang 2010).  All-sky surveys in the mid-infrared with WISE also
uncovered comparably luminous systems (Eisenhardt et al 2012). While
rare locally, infrared-luminous systems rise dramatically in number
with increasing redshift, until at $z>1$ they host a substantial,
possibly dominant fraction of the comoving infrared luminosity density
(Le Floc’h et al 2005, Perez-Gonzalez et al 2005).  Infrared template
modelling, and other follow-up, has shown that both starbursts and AGN
dust tori can contribute to these very high infrared luminosities,
with implied star formation rates (SFRs) exceeding $1000\,M_{\odot}$
yr$^{-1}$.  Selection at 22 (WISE) or 25 ({\it IRAS}) $\mu$m favours
dominance by AGN dust tori, while selection at 60 $\mu$m or longer
wavelengths favours dominance by star formation. In order to
obtain the most robust selection possible we here use infrared
template modelling to select sources on the basis of their
star-formation rate, rather than their infrared luminosity.

More recent surveys at longer wavelengths by {\it Herschel} have drawn
attention to even more extreme objects, with star-formation rates in
excess of $10,000\,M_{\odot} yr^{-1}$ in some cases (Rowan-Robinson et
al 2016).  
 The existence of these ``extreme'' starbursts poses a
fundamental problem for semi-analytic models of galaxy formation. The
observed number density of extreme starbursts with
SFRs$>$1000$\,$M$_{\odot}$yr$^{-1}$ (Dowell et al 2014, Asboth et al
2016) is factors of several above model predictions, while the extreme
starbursts with SFRs $>$ 3000$\,$M$_{\odot}$yr$^{-1}$ do not exist
{\itshape at all} in models (Lacey et al 2010, 2016, Gruppioni et al
2011, 2015, Hayward 2013, Henriques et al 2015). The issue for the
models is that neither mergers or cold accretion should produce such
high SFRs; mergers because they cannot channel enough gas to the
centers of haloes (e.g. fig. 1 of Narayanan et al 2010, Dave et al
2010), and cold accretion because massive haloes inhibit the gas flow
on to central galaxies via shock heating (Birnbolm and Dekel 2003,
Keres et al 2005, Narayanan et al 2015). The models could potentially
reproduce SFRs of $>$3000$\,$M$_{\odot}$yr$^{-1}$ at $z>1$ if feedback
is turned off completely, but would then strongly overpredict the
$z=0$ galaxy mass function.
 
 There is thus a pressing need to
confirm the existence of systems with such high star formation rates,
especially at high redshifts, understand how efficient surveys at
different wavelengths are at uncovering them, and to understand the
relation between their stellar mass and black hole mass assembly
events. In this paper we undertake such a study, by examining and
contrasting the selection of extreme starburst galaxies from two
surveys, one at 60\,$\mu$m and one at 500\,$\mu$m. There are four
reasons for using a 60 $\mu$m ({\it IRAS}) sample as well as a 500
$\mu$m one.  Firstly the contrast between 500 and 60 $\mu$m surveys
brings out what is distinct about the 500 $\mu$m sky.  Secondly we find that the 60
$\mu$m sample helps us delineate the maximum possible rate of
star-formation in galaxies (section 4).  Thirdly our 60
$\mu$m survey is free of the problems of confusion and blending which are
issues at submillimetre wavelengths, because of the small numbers of sources per beam.
Confusion only became an issue for IRAS in the Galactic plane and in the very deepest {\it IRAS}
surveys at the North Ecliptic Pole (Hacking and Houck 1987).  Finally in our analysis of AGN
(section 7) the 60 $\mu$m sample provides us with a useful low
redshift benchmark.
 
 This paper is structured as follows. Section
2 describes our sample selection strategy from the {\it IRAS} and
{\itshape Herschel} surveys. In section 4 we outline how candidate
lenses are removed from the samples. We then describe how stellar
masses, gas masses, and star formation rates are computed for each
source in section 5. Using these derived quantities, we then examine
the properties of the extreme starbursts in the sample in section 6,
and the role of AGN in section 7.

A cosmological model with $\Lambda=0.7$, $h_0=0.72$ has been used
throughout.  If we were to use $H_0 = 67 km/s/Mpc$ (Planck
collaboration 2014) then luminosities and star-formation rates would
increase by 15.5$\%$.
 

\section{Sample Selection}

We select sources from two catalogs; the Revised {\it IRAS} Faint Source
Survey Redshift Catalogue (RIFSCz, Wang et al 2014), and the {\it
  Herschel} Multi-tiered Extragalactic Survey ({\it HerMES}, Oliver et al
2012). Below, we describe each catalogue in turn.  
 
\subsection {IRAS}
The Revised {\it IRAS}
Faint Source Survey Redshift (RIFSCz) Catalogue (Wang et al 2014a) is
a 60 $\mu$m survey for galaxies over the whole sky at $|b|>20^o$,
which incorporates data from the SDSS, 2MASS, WISE, and {\it Planck}
all-sky surveys to give wavelength coverage from 0.36-1380 $\mu$m.
Since publication of Wang et al (2014) {\it Akari} fluxes have been
added to the catalogue, using a search radius of 1 arc min. An
aperture correction needs to be applied to {\it Akari} 65 and 90
$\mu$m fluxes to give consistency with {\it IRAS} photometry
(Rowan-Robinson and Wang 2015).  Furthermore the optical and
near-infrared photometry of 1271 catalogued nearby galaxies has been
improved, following a systematic trawl through the NASA$/$IPAC
Extragalactic Database.  Wang et al (2014) found that 93$\%$ of RIFSCz
sources had optical or near infrared counterparts with spectroscopic
or photometric redshifts.  The photometric redshifts primarily make
use of 2MASS and SDSS photometric data.  Thus for 93$\%$ of the
catalogue the prime selection effect is the 60 $\mu$m sensitivity
limit of the {\it IRAS} Faint Source Survey ($\sim0.36mJy$).  Table 1 summarises the number
of RIFSCz galaxies by waveband.

\begin{table}
\caption{RIFSCz catalogue by band}             
\label{table:1}     
\centering                          
\begin{tabular}{lll }        
\hline\hline
            \noalign{\smallskip}            
  Wavelength & Survey& Number of \\
 ($\mu$m)& &sources\\
\hline
3.4           &  WISE  &  48603      \\
4.6           &  WISE  &  48603      \\
12            &  WISE  &  48591      \\
12            &  IRAS  &  4476      \\
22            &  WISE  &  48588      \\
25            &  IRAS  &  9608      \\
60            &  IRAS  &  60303      \\
65            &  AKARI  &  857      \\
90            &  AKARI  & 18153      \\
100            &  IRAS  & 30942      \\
140            &  AKARI  & 3601      \\
160           &  AKARI  &  739      \\
350            &  PLANCK  & 2275      \\
550           &  PLANCK  & 1152      \\
850           &  PLANCK  & 616     \\
1380           &  PLANCK  & 150     \\           
\hline                                   
\end{tabular}
\end{table}

\subsection {{\it Herschel}}
The {\it HerMES} survey allows us to construct a 500 $\mu$m sample of
galaxies in areas in which we have deep optical and infrared data from the
{\it Spitzer}-SWIRE survey (Lonsdale et al 2003, Rowan-Robinson et al 2008,
2014, 2016) over a total area of 26.3 sq deg in five fields (see Table
1 of Rowan-Robinson et al 2016). Aperture corrections are applied at optical, near and mid-infrared
wavelengths to ensure that all SEDs are based on integrated flux-densities (Rowan-Robinson et al 2013).
Selection at 500 $\mu$m, rather than
say 250 $\mu$m, gives us greater visibility of the high redshift
(z$>$3) universe due to the intrinsic shape of starburst SEDs at
far-infrared wavelengths (Franceschini et al 1991) and has the benefit
of ensuring detection also at 350, and in most cases 250 $\mu$m, to
give valuable SED information.  
The association of {\it Herschel} sources with SWIRE 24$\mu$m sources uses a likelihood which combines
the positional disagreement between 250$/$350$\mu$m and 24$\mu$m positions and the agreement of the observed
500$\mu$m flux with that predicted from automatic template fits to the SWIRE 4.5-170$\mu$m data. We
have argued previously (Rowan-Robinson et al (2014) that the use of SED information is essential in the
association process.  Assignment of submillimetre flux to counterparts based purely on positional
agreement can lead to physically unrealistic SEDs.
The complete {\it HerMES}-SWIRE 500 $\mu$m
catalogue comprises sources in the Lockman, XMM, ELAIS-S1, ELAIS-N1
and CDF-S fields, and consists of 2181 galaxies.  In the
Lockman+XMM+ELAIS-S1 areas there are a further 833 good quality
500+350 $\mu$m sources which are not associated with {\it Spitzer}-SWIRE
galaxies, for which Rowan-Robinson et al (2016) have estimated
redshifts from their submillimetre colours.  Thus for all {\it HerMES}
500$\mu$m sources we have an estimate of redshift and hence of
infrared luminosity (and star-formation rate).  The prime selection
effect on this sample is therefore the 500 $\mu$m flux-density limit of the
survey.

We performed a check of the surface density of 500 $\mu$m sources in
the {\it HerMES} survey using data from the COSMOS area (Scoville et al
2007), which was surveyed as part of the {\it HerMES} project.  Photometric
redshifts for COSMOS have been discussed by Ilbert et al (2013) and
Laigle et al (2016).
There are 181 500 $\mu$m sources with flux greater than 25 mJy, the
flux limit we used in Rowan-Robinson et al (2016), and which also have
350 $\mu$m detections, in the 2.0 sq deg of the COSMOS survey.  All
have 24 $\mu$m associations. This yields a 500 $\mu$m source-density
of 90 per sq deg, similar to that found in the 26.3 sq deg of our
sample.

\subsection {Comparison with other studies}
Schulz et al (2017) have published a new IPAC SPIRE catalogue (HPSC)
which analyses data taken in all {\it Herschel} SPIRE programmes in a
homogeneous way, using a blind source detection approach.  This would
appear to offer the opportunity of a much larger sample of SPIRE
galaxies.  We used the HPSC catalogue to create a 250-350-500 $\mu$m
list as in Rowan-Robinson et al (2014).  When we associated this list
with the SWIRE photometric redshift catalogue (Rowan-Robinson et al
2013), we found only about half of the 2181 sources. This is an issue
acknowledged in the HPSC explanatory supplement, which they attribute
to blending of SPIRE sources in their detection procedure.

We also associated this HPSC 500 $\mu$m catalogue with RIFSCz, finding
1640 associations.  Many of these were also detected by {\it Planck}
and so we can make a direct comparison of 350 and 500 $\mu$m fluxes in
the two surveys.  The sources in common to HPSC, RIFSCz and {\it Planck}
tend to be low redshift galaxies.
We find that these galaxies need an aperture correction of
k*delmag to the SWIRE fluxes, where delmag = $J_{ext} - J_{ps}$ is the
J-band aperture correction and k=0.15 at 350, and 0.10 at 500 $\mu$m,
to get agreement of SPIRE and {\it Planck} fluxes.  Previously Wang et
al (2014) reported the need for aperture corrections to be applied to
WISE fluxes at 12 and 22 $\mu$m.  The latest version of RIFSCz
(http:$//$mattiavaccari.net:$/$df$/$mrr$/$readmeRIFSCz) thus provides
a comprehensive collection of fluxes, with aperture corrections where
necessary, from optical (SDSS), near infrared (2MASS), mid and far
infrared (WISE, {\it IRAS}, {\it Akari}), through to submillimetre and
millimetre ({\it Herschel} and {\it Planck}).

Koprowski et al (2017) have used a SCUBA2 survey at 850 $\mu$m to
estimate the rest-frame 250 $\mu$m luminosity-density and then
translated this to a star-formation-rate-density assuming a universal
submillimetre SED.  They cast doubt on the reality of the high
star-formation rates found by Rowan-Robinson et al (2016) at z = 4-6.
There are some flaws in the Koprowski et al analysis.  Firstly their
850 $\mu$m detection threshold is set at 3.5 $\sigma$, which means
they are heavily into the confusion regime.  Our strategy of
thresholding at 5 $\sigma$, made possible by the excellent
submillimetre sensitivity of {\it Herschel-SPIRE}, ensures that
problems of confusion and source blending are greatly reduced (see
section 3).  Secondly, they associate their submillimetre sources with
other multi wavelength data using the nearest bright 8 or 24 $\mu$m,
or 1.4 GHz, source, thus potentially biassing their associations against more
probable (in terms of their SED) higher redshift galaxies.  Finally
because their survey is at a single submillimetre wavelength they have
no reliable way of estimating the star-formation rate.  It is simply
not true that all submillimetre galaxies have a common submillimetre
SED (e.g. Rowan-Robinson et al 2014).  The high star-formation
rates we find are supported by the {\it IRAS} RIFSCz sample (see section 5
below) which is not subject to any of the submillimetre confusion or
blending issues.  
Novak et al (2017) have nicely confirmed
Rowan-Robinson et al (2016)’s star-formation-rate-density from z = 0
to 5 with radio estimates from a VLA survey.

 \begin{table*}
\caption{Contrast between the 60 $\mu$m selected RIFSCz catalogue and
  the 500 $\mu$m selected catalogue from the {\it HerMES} survey}             
\label{table:2}      
\centering                          
\begin{tabular}{lll }        
\hline\hline    
 \noalign{\smallskip}     
 & IRAS-FSS & HerMES-SWIRE\\
 & 60 $\mu$m & 500 $\mu$m\\
\hline
number of sources & 60303 & 2181\\
effective area (sq deg) & 27143 & 26.3\\
surface-density of lensed galaxies & 0.001 per deg$^{2}$ & 10 per deg$^{2}$\\
fraction of Ultraluminous galaxies & 8 $\%$ & 70$\%$ \\
fraction of Hyperluminous galaxies & 0.7$\%$ & 25$\%$\\
fraction of galaxies with standard cirrus & 42$\%$ & 34$\%$\\
fraction of galaxies with cool or cold cirrus & 2.5$\%$ & 29$\%$\\
redshift $>$ 0.3 & 4$\%$ & 88$\%$\\
\hline
\end{tabular}
\end{table*}

Table 2 shows a comparison of the sky seen at 60 and at 500 $\mu$m, as
seen in the RIFSCz and {\it HerMES}-SWIRE catalogues (Rowan-Robinson et al
2014). The most striking contrasts of 500$\mu$m selection, compared to
60 $\mu$m selection, are (i) a much higher fraction of high redshift
galaxies (as predicted by Franceschini et al 1991), (ii) a much higher
fraction of lensed objects (as predicted by Blain et al 2002), (iii) a
much higher fraction of galaxies with cool or cold dust (RR et al
2010, 2016, Rowan-Robinson and Clements 2015).

\begin{figure}
\includegraphics[width=9cm]{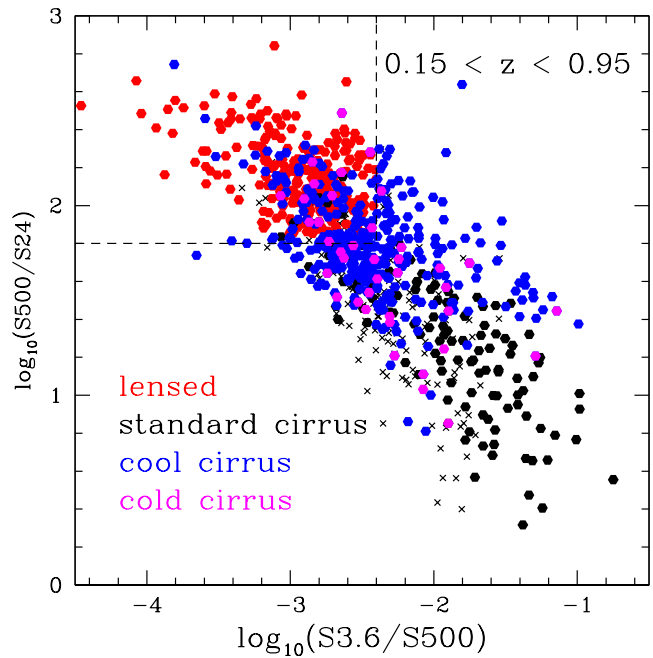}
\caption{ $S_{24}/S_{500}$ vs. $S_{3.6}/S_{500}$, illustrating the
  diagnostic ratios used by Rowan-Robinson et al (2014) to select
 lensed objects, here plotted for the sources in {\it HerMES}
  (Lockman+XMM+ELAIS S1+CDF-S+ELAIS N1) with $0.15<z<0.95$. Red filled
  circles: lensed galaxy candidates,large black filled circles:
  galaxies with standard cirrus components, blue filled circles:
  galaxies with cool cirrus components, magenta filled circles:
  galaxies in Lockman with cold cirrus components, small black dots:
  non-cirrus galaxies.}
\end{figure}

\section{Confusion and source blending}

An important issues for ground- and space-based submillimetre surveys
is confusion and source-blending.  For a random distribution of point-sources
characterised by differential source-counts $dN/dS = n(S)$, where S is the
flux-density, observed
with a telescope of specified beam, the measured responses are
characterised by the probability of an observed deflection D, P(D).
Scheuer (1957) gave the formalism for calculating P(D) and Condon
(1974) used this to calculate the confusion noise for a telescope with
a Gaussian beam of dispersion $\Theta$ and a power-law source-count
distribution $n(S) = k S^{-\gamma}$.  Integrating $S^2 P(D)$ from
$S = 0$ to $D_c$ to evaluate the rms dispersion $\sigma^2$ gives
\begin{equation}
	\sigma = (k \Omega_e/(3-\gamma)^{1/2} D_c^{(3-\gamma)/2}  
\end{equation}
\noindent
where the effective telescope beam $\Omega_e = \Omega/(\gamma-1)$ and the Gaussian beam area 
\begin{equation}
\Omega= 2 \pi \Theta^2 = 1.14 \Theta_{FWHP}^2.
\end{equation}
Thresholding at a multiple q of $\sigma$, $D_c = q \sigma$, yields 
\begin{equation}
\sigma = (q^{(3-\gamma)}/(3-\gamma))^{1/(\gamma-1)} (k \Omega_e)^{1/(\gamma-1)}
\end{equation}
\noindent
(Condon’s eqn (14)).
\noindent
We can use this to calculate the number of sources per Gaussian beam at the $q \sigma$ limit (after some cancellation):
\begin{equation}
	N(q\sigma) \Omega = (3-\gamma)/q^2
\end{equation}

\noindent
Results for q = 5 and different values of $\gamma$ are given in Table 3.
Franceschini (1982) has an expression equivalent to eqn (4) in his equation (14).
\begin{table}
\caption{Number of beams per source for 5-$\sigma$ survey, as function of $\gamma$}
\begin{tabular}{llllll}
\hline\hline
$\gamma$ = & 1.5 & 2.5 & 2.6 & 2.7 & 2.8\\
no. of beams per source = & 17 & 50 & 61 & 83 & 125\\
\hline
\end{tabular}
\end{table}

Hacking and Houck (1987) repeated the Condon calculation and
gave a table of results on $\alpha=\gamma-1$, the beam size and
$\sigma$ for their deep 12 and 60 $\mu$m survey. They confirm that at
60 $\mu$m, where $\gamma$=2.5, their survey is confusion limited at
their $5-\sigma$ limit of 50 mJy, at a source-density of 1 source per
50 beams.  Hogg (2001) carried out simulations of position error, flux
error and completeness for $\alpha = 1-1.5$ (relevant to optical
galaxy counts.  He gives a rule of thumb of 30 beams per source for
avoiding effects of confusion, but finds that for $\alpha \ge 1.5$,
the requirement should be 50 beams per source.

For applications to {\it Herschel} 500 $\mu$m surveys we note that
$\Theta_{FWHM} = 36.6$ arcsec, so $\Omega = 1/8486$ sq deg.  From
Bethermin et al (2012)’s 500 $\mu$m differential source-counts we find
the source count slope at 6-24 mJy ($\sim 1-5 \sigma$) is
$\gamma \sim$ 2.65, so eqn (1) predicts the $5-\sigma$ limit as 1
source per 71 beams.  In Rowan-Robinson et al (2014) we found 1335
$5-\sigma$ 500 $\mu$m sources in the 7.53 sq deg of the Lockman-SWIRE
area, which would correspond to 1 source per 48 beams.  Of course the true
source-counts can not be a power-law at all fluxes and this would modify 
the calculation slightly.

We also calculate the probability of a blend of a source fS with a
second source (1-f)S, where S=5$\sigma$, within the telescope beam,
for an assumed source-count slope $\gamma = 2.65$ and find
\begin{equation}
	p(fS,(1-f)S)= [(3-\gamma)/q^2]^2 f^{-\gamma+1} (1-f)^{-\gamma+1}.
\end{equation}
\noindent
Values of the relative probability of different blending cases, p(fS,(1-f)S)$/$p(S), 
for different values of f are given in Table 4 for 500 and 350 $\mu$m.  A similar 
expression can be derived for blends of three sources, p(fS,gS,(1-f-g)S),
and values for the relative probability of three-source blends are also given in Table 4.

The probability of a roughly equal-flux blend is extremely
low at 500 $\mu$m and even lower at 350 $\mu$m, where all the sources have to be
detected to be in our sample.  Most of the sources are also detected at 250 $\mu$m, 
where the probabilities are lower still, by a further factor of 2.

These probabilities apply to an unclustered distribution of sources.
Source confusion will be enhanced by intrinsic
clustering of galaxies (Barcons 1992, Scott et al 2002, Bethermin et
al 2017). 

\begin{table}
\caption{Relative probability of two-source blends,  p(fS,(1-f)S$/$p(S), and three-source
blends, p(fS,gS,(1-f-g)S$/$p(S), where S = 5$\sigma$ and $\gamma = 2.65$}
\begin{tabular}{llll}
\hline\hline
 & 500$\mu$m & 350 $\mu$m & 250 $\mu$m\\
\hline
two-source blends & & &\\
0.8 S, 0.2 S & 0.29 & 0.145 & 0.07\\
0.7 S, 0.3 S & 0.18 & 0.09 & 0.045\\
0.6 S, 0.4 S & 0.15 & 0.075 & 0.04\\
0.5 S, 0.5 S & 0.14 & 0.07 & 0.035\\
three-source blends & & &\\
0.6 S, 0.2 S, 0.2 S & 0.09 & 0.045 & 0.02\\
0.4 S, 0.3 S, 0.3 S & 0.05 & 0.025 & 0.012\\
\hline
\end{tabular}
\end{table}
\noindent

Scott et al (2002), carried out simulations of the SCUBA 850 $\mu$m
8-mJy survey.  From their tables we see that to achieve better than
90$\%$ completeness, positional error $< 20\%$ of the beam width, and
flux-boost $<5\%$, we need to threshold at $5-\sigma$.  
Michalowski et al (2017) found through ALMA follow-up that the fraction of
bright SCUBA 850$\mu$m sources ($S_{850}>4 mJy$) significantly affected by blending is small (15-20$\%$).
Hill et al (2018) have observed 103 bright SCUBA 850$\mu$m sources ($S_{850}>8 mJy$) with the SMA interferometer
and found that the probability of a source being resolved into two or more sources of comparable flux-density is 15$\%$.
Simulations of {\it Herschel} 500 $\mu$m surveys have been carried out by Nguyen et al
(2010), Roseboom et al (2010), Wang et al (2014b) Valiante et al
(2016) and Bethermin et al (2017). The Valiante et al study finds that
with a $5-\sigma$ threshold completeness is 97$\%$ and flux-boost is
2$\%$.  The Bethermin et al (2017) simulation suggests that even
allowing for clustering of sources, selection at $5-\sigma$ ensures that
the average flux-boosting at 250, 350 and 500 $\mu$m is 13, 21 and 34$\%$ respectively. 
We have tested the effect of deboosting by these quantities on our extreme starburst sample (section 6 below)
and find that the
resulting infrared luminosities and star-formation rates are reduced by a median value of
0.08 dex, an amount that would be almost exactly compensated by changing the Hubble
constant from 72 to 67.  A few examples have been found of Herschel sources which are identified
as distant clusters (e.g. Clements et al 2014, 2016, Wang et al 2016) but these tend to be
extended or multiple submillimetre sources.

It is worth noting that thresholding at $3.5\sigma$, as has been
rather widespread in submillimetre surveys, entails a
probability of blended sources four times higher than thresholding at $5\sigma$.

In conclusion the problems of observing in a confused region of sky
(flux-blending, increased positional error, flux-boosting) can be
greatly reduced by thresholding at $5-\sigma$.  We discuss
the issue of blended sources further in section 6.

\section{Lensed galaxy diagnostics}
One of the most serious issues for cosmological analysis of a
submillimetre-selected sample is the high incidence of lensed objects.
Negrello et al (2010) argued that a high proportion of 500 $\mu$m sources
with S500 $>$ 100 mJy are likely to be lensed.  Wardlow et al (2013)
showed that, after exclusion of blazers and local spirals, more than 78 $\%$ of
such sources are confirmed lensed sources.  Negrello et al (2010) plotted SEDs of
confirmed lensed sources and showed that at optical and near infrared
wavelengths we see the lensing galaxy while at submillimetre
wavelengths we see emission from the lensed galaxy. Rowan-Robinson et
al (2014) modelled SEDs of 300 {\it Herschel} sources in the Lockman-SWIRE
area and identified 36 candidate lensed galaxies in this way.  They
showed how lensing candidates can be extracted by a set of
colour-colour constraints (including submillimetre colour constraints
suggested by Wardlow et al (2013)).

Figure 1 illustrates the 3.6-24-500 $\mu$m diagnostic ratios used by
Rowan-Robinson et al (2014) to select lensed objects.  It is a plot of
S500$/$S24 versus S3.6$/$S500, with candidate lensed objects shown in
red, normal cirrus galaxies shown in black, galaxies with cool dust
($T_{dust}\sim$ 14-19K) shown in blue and galaxies with cold dust
($T_{dust}\sim$ 9-13K) shown in magenta.  The colour selection shown,
with others, is remarkably effective at identifying lensed galaxy
candidates. In particular the two confirmed lenses in the
SWIRE-Lockman area studied by Wardlow et al (2013) satisfy these
colour contsraints.  Details of the table of the 275 {it HerMES}-SWIRE
(Lock+XMM+ES1+CDFS+EN1) lensed galaxy candidates are given at

http:$//$mattiacaccari.net$/$df$/$mrr$/$readmespirerev.
These sources are not used in the subsequent analysis.  ALMA or HST
imaging would be highly desirable to confirm the reality of these
lensed galaxy candidates.

For {\it IRAS} FSS (RIFSCz) sources we can not use this colour-colour
diagnostic. Instead the infrared luminosity, or inferred
star-formation rate, is a good indicator of lensing.  There do not
seem to be any cases where the true, unlensed star-formation rate is
$> 10^{4.5} M_{\odot}/yr$ (see Fig 2R).  Table 5 lists 22 RIFSCz
objects with star-formation rate, calculated by the automated
template-fitting code, $> 10^{4.5} M_{\odot} yr^{-1}$.  Four are known
lenses. One (F14218+3845) has been imaged with HST and shows no
evidence of lensing (Farrah et al 2002): Rowan-Robinson and Wang
(2010) point out that there is a discrepancy between the ISO 90 $\mu$m
flux and the {\it IRAS} 60 and 100 $\mu$m fluxes and if the former is
adopted a much lower SFR (4,400 $M_{\odot} /yr$) is obtained.  Three are
blazars, for which the submillimetre emission is non-thermal, one
object is more probably associated with a z=0.032 Zwicky galaxy, and
three have photometric redshifts $>$4 which their SEDs show are
implausible: these 7 have been removed from Fig 2R).  We are left with
10 new candidate lenses, of which 5 have spectroscopic
redshifts. These 22 sources have been removed from the subsequent
analysis.

\begin{table*}
\caption{RIFSCz objects with apparent SFRs $>10^{4.5} M_{\odot} yr^{-1}$}
\begin{tabular}{llllll}
\hline
\hline
\smallskip
 IRAS name & RA(J2000) & Dec(J2000) & Redshift  & $log_{10}$ SFR & Notes\\
& &&& ($M_{\odot}/yr$) &\\
\hline
candidate lensed objects\\
\hline
F02416-2833  &  40.953415 &  -28.343891 &  1.514000  &  4.54 & \\
F03445-1359  &  56.718334 &  -13.844521 &  (1.14)      &  4.54 & \\
F08105+2554  & 123.380363 &   25.750853 &  1.512380  &  5.08 &
                                                               Lensed\\
F08177+4429  & 125.316353 &   44.333546 &  (2.65)      &  5.78 & \\
F08279+5255  & 127.923744 &   52.754921 &  3.912200  &  6.89 &
                                                               Lensed\\
F10018+3736  & 151.207672 &   37.362133 &  1.684160  &  5.30 & \\
F10026+4949  & 151.469330 &   49.579998 &  1.120000  &  4.54 & Unlensed\\
F10119+1429  & 153.657822 &   14.251303 &  1.550000  &  5.01 & \\
F10214+4724  & 156.144012 &   47.152695 &  2.285600  &  5.05 & Lensed
  \\
F10534+3355  & 164.055649 &   33.661686 &  (1.17)      &  4.50 & \\
F13445+4128  & 206.656906 &   41.225357 &  (1.33)      &  4.75 & \\
F13510+3712  & 208.286133 &   36.964321 &  1.311000  &  4.70 & \\
F14132+1144  & 213.942673 &   11.495399 &  2.550000  &  5.80 & Lensed
  \\
F14218+3845  & 215.981201 &   38.530708 &  1.209510 &  4.98 &
                                                              Unlensed, see text \\
F23265+2802  & 352.262146 &   28.312298 &  (1.90)      &  5.50 & \\
\hline
wrong ID, wrong redshift or blazars\\
\hline
F02263-0351 & 37.221718  & -3.626988 &  2.055000  &  5.51? & blazar\\
F00392+0853  & 10.453402  &   9.173513 &  (4.62?)    &  7.19? & alias at  z=1.4\\
F06389+8355  & 102.896248 &   83.865295 &  (4.50?)   &  6.83? & alias at z=1.4\\
F13080+3237 & 197.619431 & 32.345490 &  0.998010  &  4.52? & blazar\\
F15419+2751 & 236.008347 & 27.697693 &  (2.02)  &  5.55? & Zwicky gal z=0.032\\
F16360+2647  & 249.522308 &   26.694941 &  (4.55?)   &  7.23? & z=0.066 2MASS gal at 0.27’\\
F22231-0512 & 336.446899 &  -4.950383	&  1.404000  &  5.25? & blazar 3C446\\
\hline
\end{tabular}
\end{table*}

\begin{table*}
\caption{RIFSCz objects with extreme SFRs ($>5000 M_{\odot} yr^{-1}$)}
\begin{tabular}{llllll}
\hline
\hline
\smallskip
 IRAS name & RA(J2000) & Dec(J2000) & Redshift  & opt type & $log_{10}$ SFR\\
& &&& & ($M_{\odot}/yr$)\\
\hline
F00167-1925 & 4.824683 &  -19.138355 & (0.82) & Scd & 4.01\\
F01175-2025 & 19.983685 & -20.172934 & 0.8137 & QSO & 3.92\\
F02314-0832 & 38.473282 &  -8.319294 & 1.1537 & QSO & 4.40\\
F04099-7514 & 62.201244 & -75.105988 & 0.6940 & E & 3.90\\
F07523+6348 & 119.230530 &  63.678543 & (0.77) & QSO & 3.79\\
F08010+1356 & 120.967873 &  13.795245 & (1.34) & Sab & 4.26\\
F10328+4152 & 158.926239 & 41.615841 & (0.90) & Sab & 3.73\\
F12431+0848 & 191.435791 &  8.524883 & 0.9380 & Sbc & 4.15\\
F13073+6057 & 197.320648 & 60.702477 & (1.01) & QSO & 4.22\\
F13408+4047 & 205.720627 & 40.533772 & 0.9058 & QSO & 3.78\\
F13489+0524 & 207.858673 & 5.158453 & 0.6202 & E & 3.78\\
F14165+0642 & 214.784088 & 6.476324 & 1.4381 & QSO & 3.70\\
F15104+3431 & 228.108719 & 34.336456 & 0.8554 & QSO & 3.95\\
F15307+3252 & 233.183395 & 32.71295 & 0.9227 & sb & 3.91\\
F15415+1633 & 235.966370 & 16.406157 & 0.8500 & QSO & 4.01\\
F16042+6202 & 241.252289 & 61.907372 & (0.99) & Sab & 3.70\\
F16501+2109 & 253.077240 & 21.078678 & (1.17) & Sab & 3.86\\
F17135+4153 & 258.781433 & 41.831528 & (0.90) & Sbc & 3.88\\
F21266+1741 & 322.241943 & 17.914932 & 0.8340 & Sab & 3.74\\
\hline
\end{tabular}
\end{table*}

\begin{figure*}
\includegraphics[width=8cm]{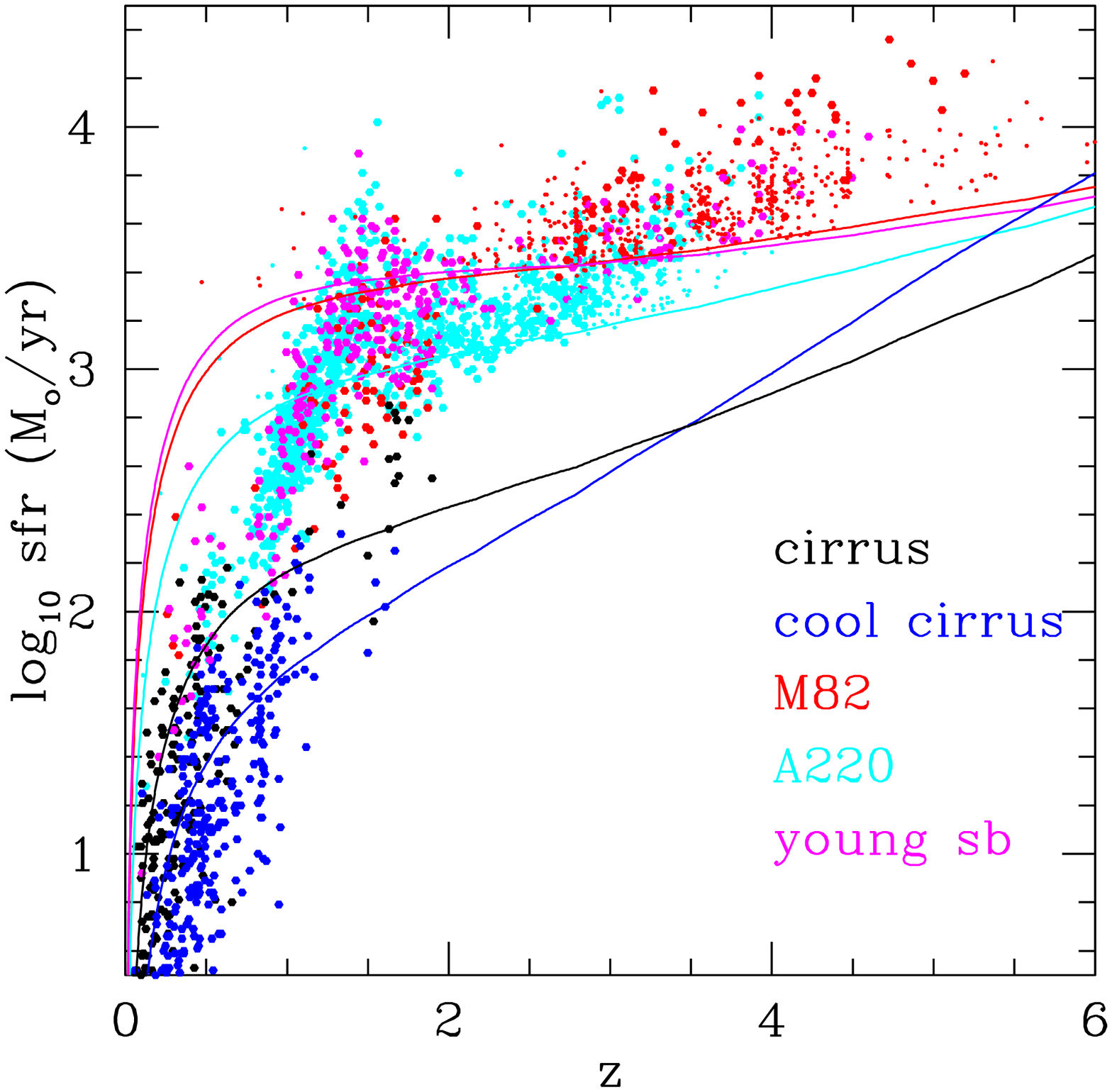}
\includegraphics[width=8cm]{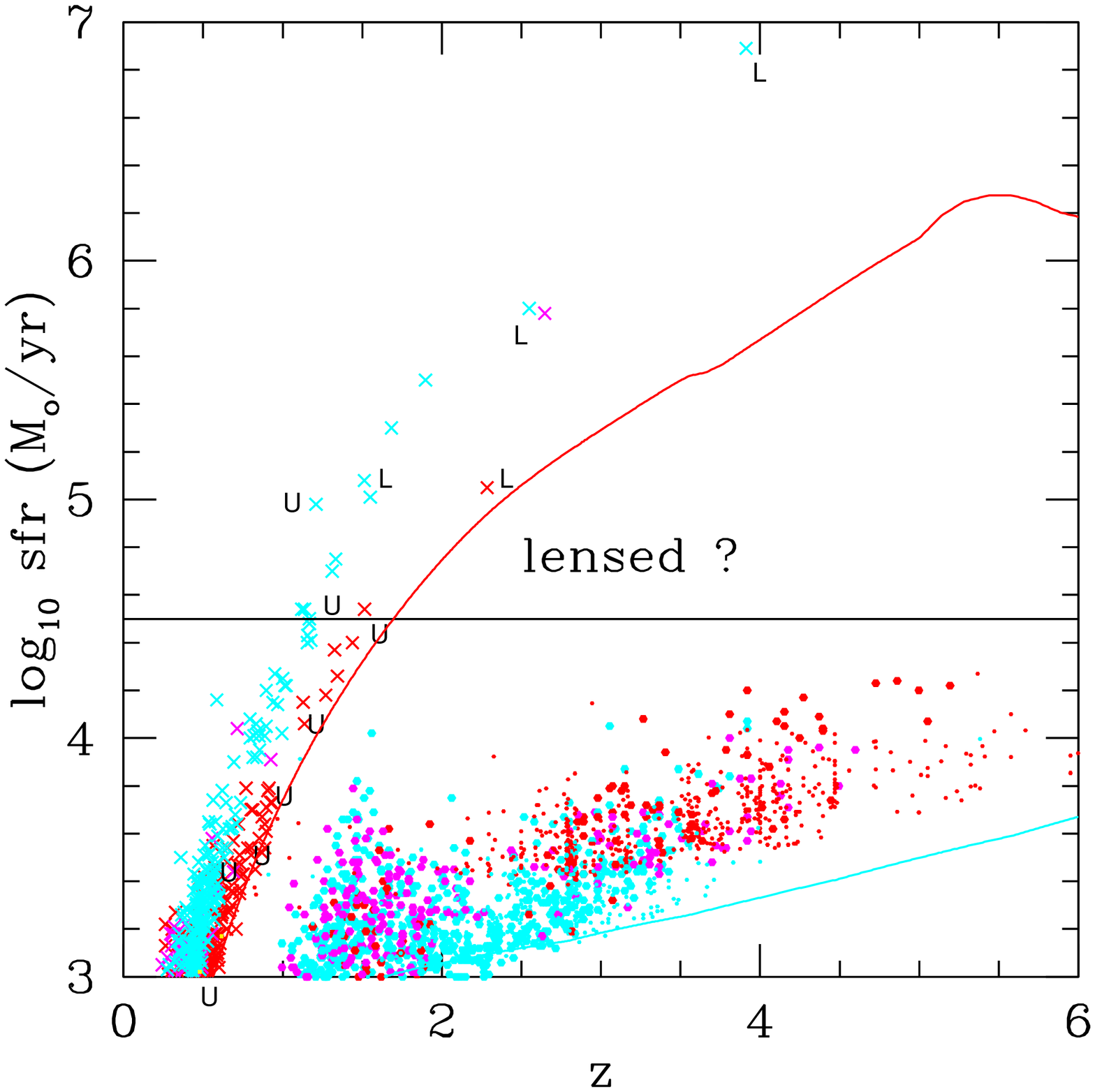}
\caption{L: Star formation rate versus redshift for {\it HerMES} Lockman+XMM+ES1 galaxies, with loci showing 500 $\mu$m selection limits for each template type. Small dots are unidentified sources.
R: Star-formation rate versus redshift for extreme starbursts from both {\it HerMES} (Lockman+XMM+ES1, filled circles) and from the {\it IRAS} RIFSCz catalogue (crosses).
Known lenses are indicated by L and cases known to be unlensed indicated by U.   The unlensed object apparently above the 30,000 $M_{\odot}/yr$ line (F14218+8345) is discussed in section 4.
Typical 60 and 500 $\mu$m selection limits are indicated by the red and cyan loci. 
}
\end{figure*}

\section{Stellar mass, dust (and gas) mass, star formation rate}

We derive stellar masses, dust \& gas masses, and star formation rates
for both the RIFSCz and {\it HerMES} sources by fitting model Spectral
Energy Distributions (SEDs) to the catalogue data.  Our approach of
fitting optical and near infrared SEDs with templates based on stellar
synthesis codes (Babbedge et al 2006, Rowan-Robinson et al 2008)
allows us to estimate stellar masses.  The templates are derived using
simple stellar populations, each weighted by a different star formation
rate and specified extinction (Berta et al 2004). An empirical correction is applied 
to allow for the variation of mass-to-light ratio with age (Rowan-Robinson et al 2008). 
A Salpeter mass-function is assumed.

Similarly, fitting mid infrared, far
infrared and submillimetre data with templates based on radiative
transfer models (Efstathiou et al 2000, 2003, Rowan-Robinson et al
2010, 2013, 2016), allows us to estimate star formation rates and dust
masses. 

\begin{table*}
\caption{A comparison between our new star-formation rate density, and that previously published in Rowan-Robinson et al (2016).}
\label{tab:sfrdcomp}
\begin{tabular}{lllllllll}
\hline\hline
\smallskip
Mean Redshift 	& 0.5-1.0  & 1.0-1.5   & 1.5-2.0   & 2.0-2.5   & 2.5-3.0   & 3.0-3.5  & 3.5-4.0 & 4.0-4.5      \\
\hline
Old SFRD ($\log_{10}(\phi)$)   & -1.28$\pm$0.21 & -0.95$\pm$0.11 & -1.06$\pm$0.13 & -1.05$^{+0.27}_{-0.09}$ & -0.82$^{+0.18}_{-0.36}$ & -0.99$^{+0.29}_{-0.46}$ & -0.82$^{+0.18}_{-0.36}$ & -0.79$^{+0.14}_{-0.41}$\\
($M_{\odot} yr^{-1} Mpc^{-3}$) &&&&&& \\
New SFRD              & -1.27$\pm$0.10 & -0.93$\pm$0.11 & -1.03$\pm$0.18 & -0.90$\pm$0.08 & -0.99$^{+0.25}_{-0.07}$ & -1.06$\pm$0.18 & -0.89$^{+0.21}_{-0.46}$ & -0.85$^{0.09}_{-0.58}$  \\
\hline
\end{tabular}
\end{table*}

\begin{figure*}
\includegraphics[width=6.0cm]{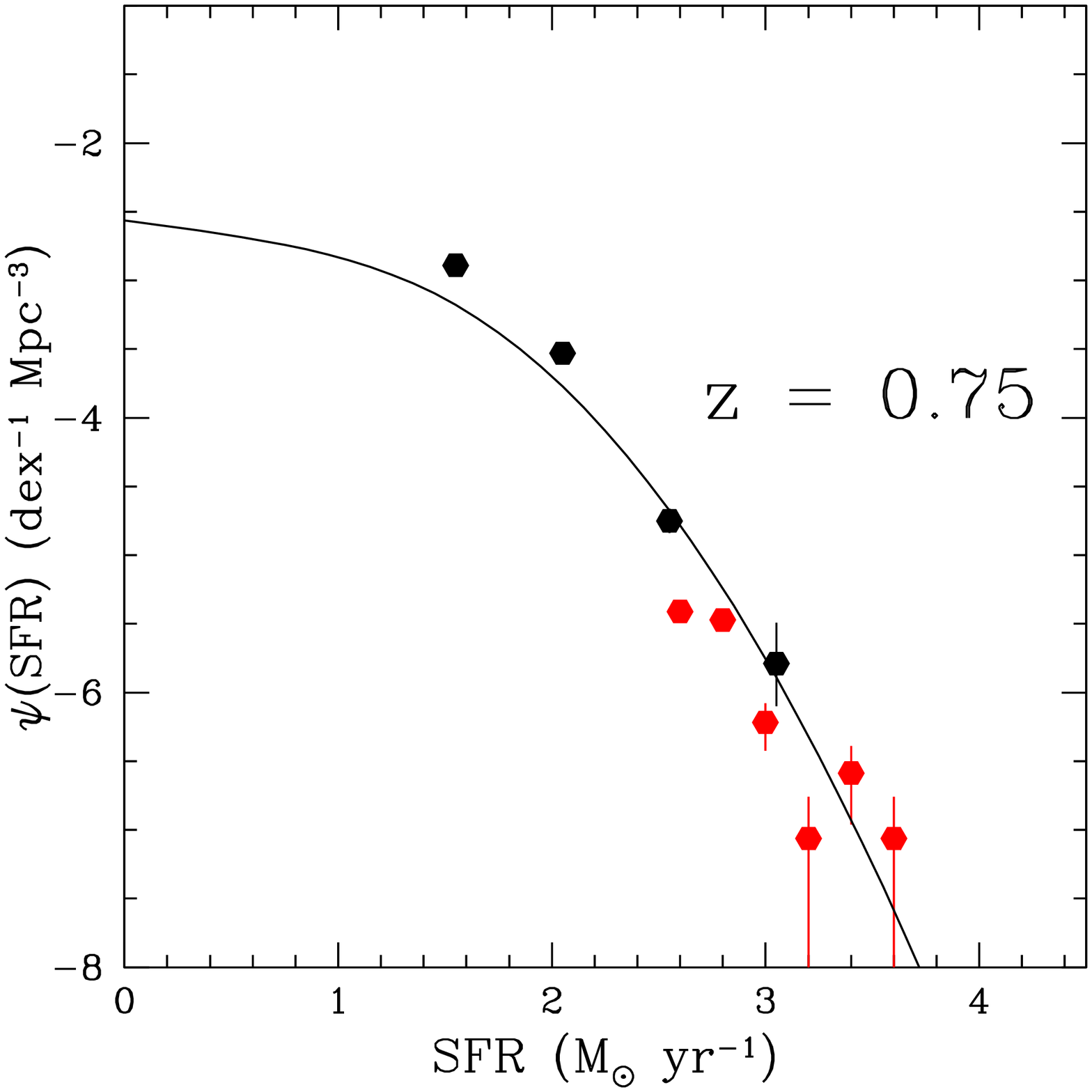}
\includegraphics[width=6.0cm]{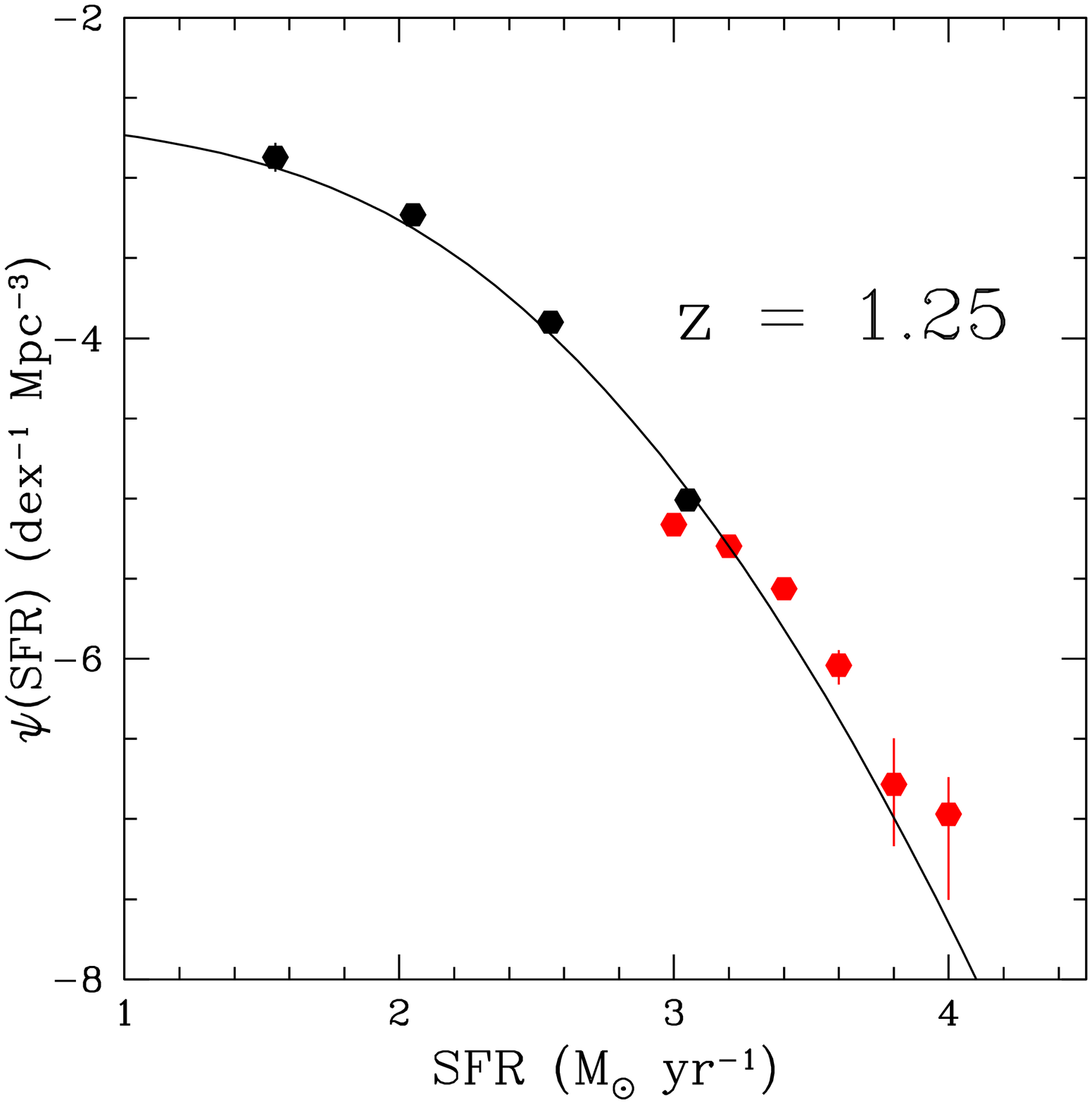}
\includegraphics[width=6.0cm]{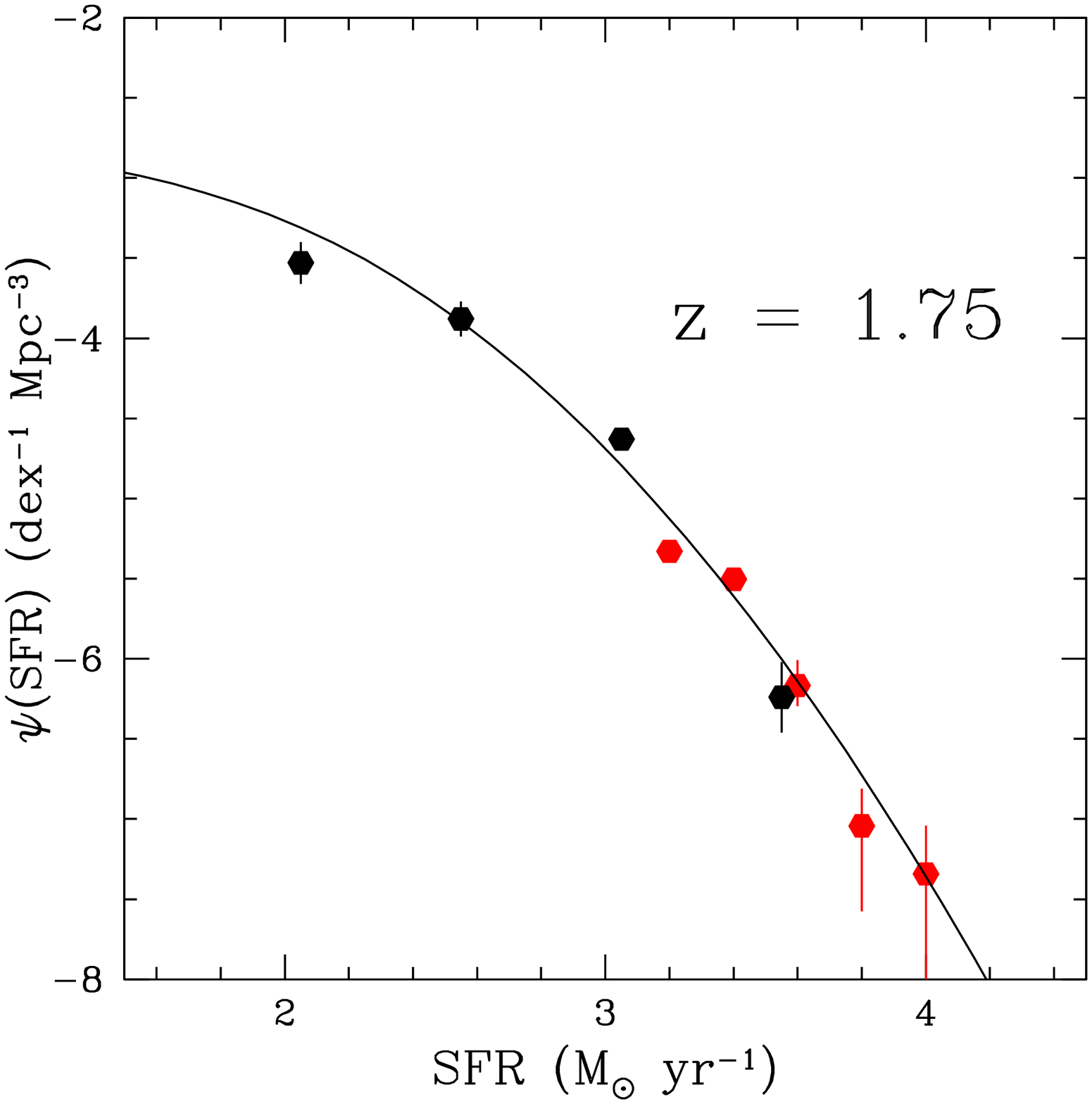}
\includegraphics[width=6.0cm]{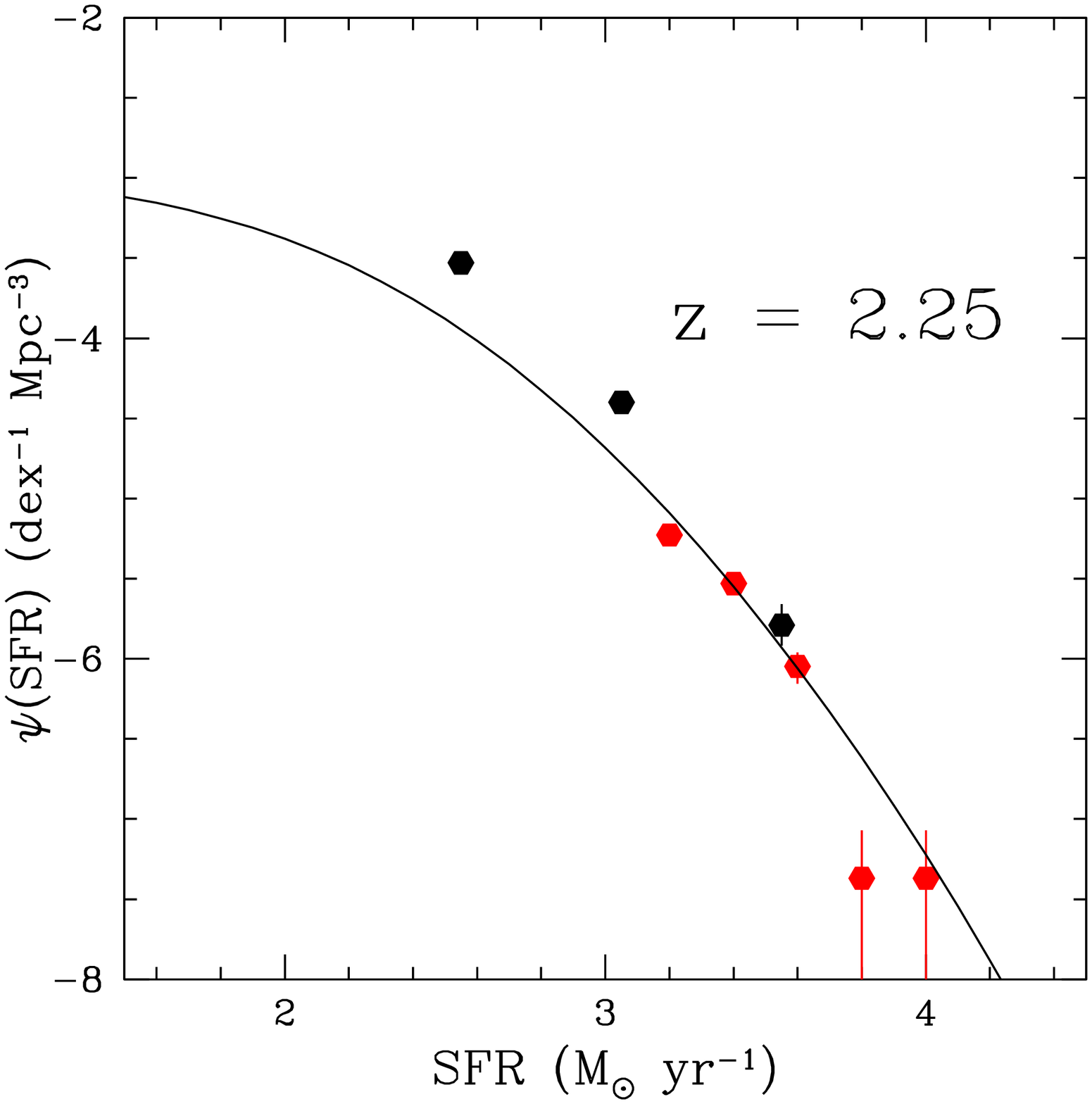}
\includegraphics[width=6.0cm]{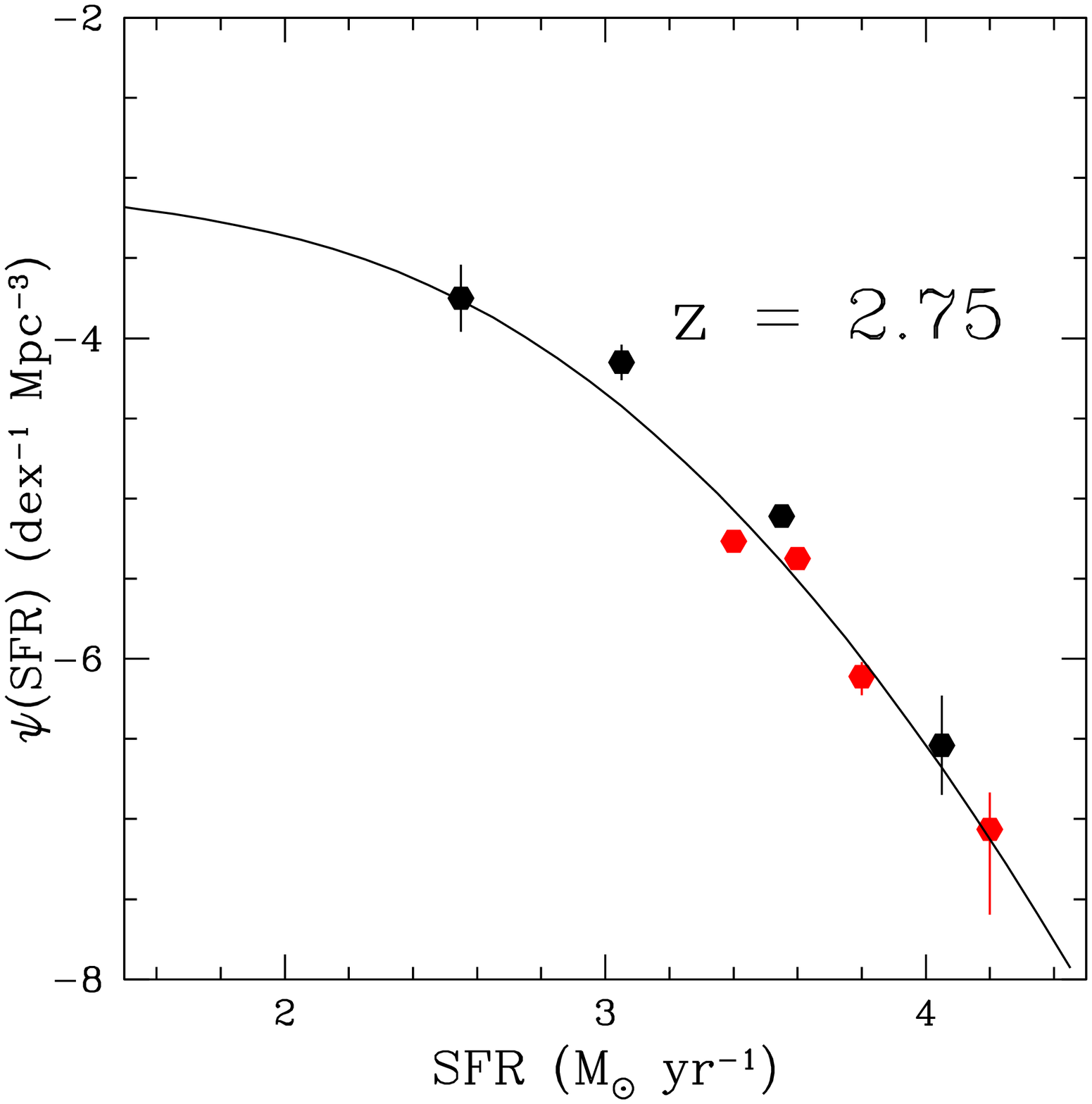}
\includegraphics[width=6.0cm]{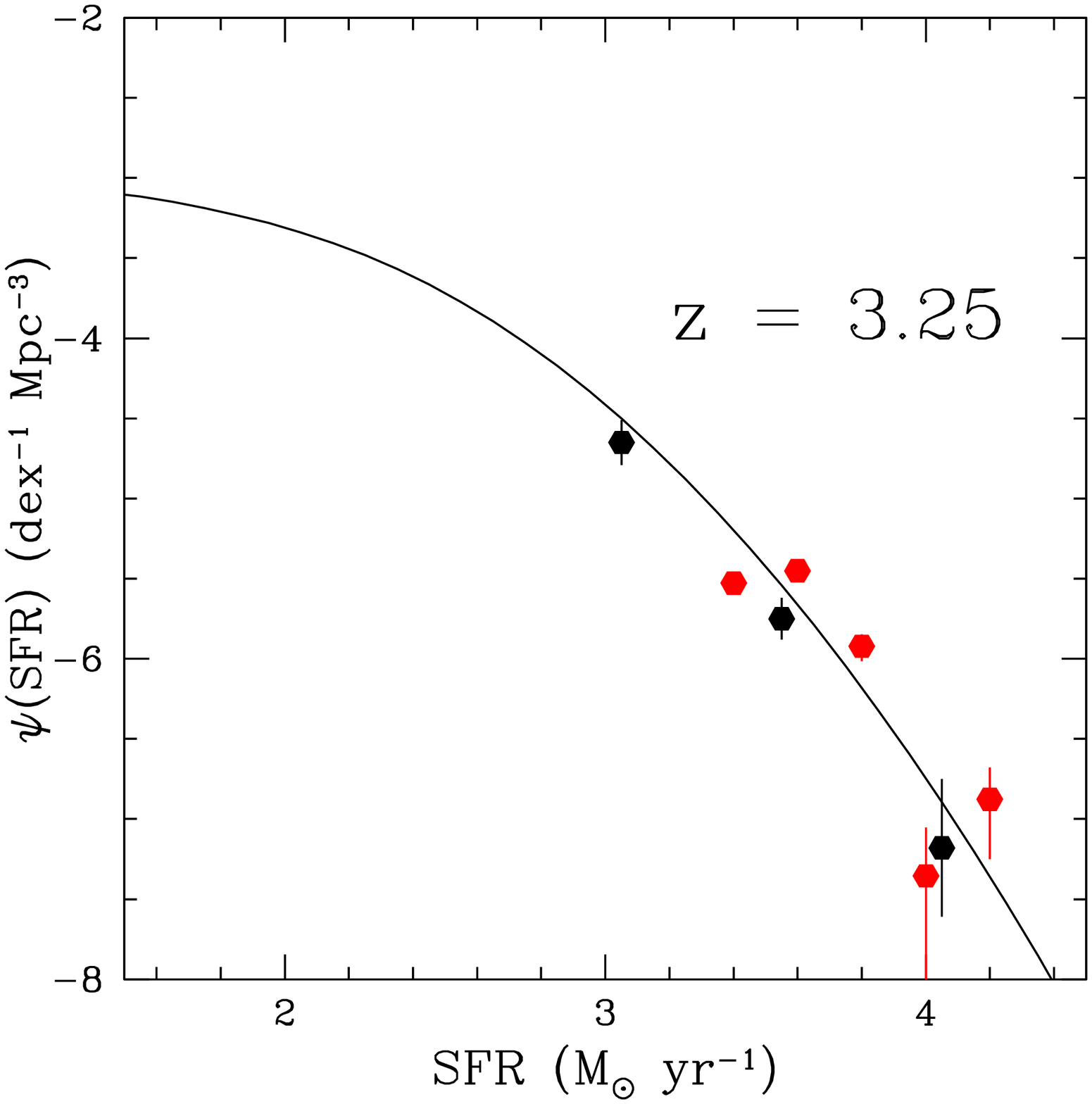}
\caption{The revised star-formation rate functions for $z = 0.25-3.25$, using a least-squares
normalization at $24-500\,\mu$m.  
Black dots: data from Gruppioni et al (2013), red dots: present work.}
\label{fig:newsfr}
\end{figure*}

In the automated fitting of infrared SED templates and calculation of
infrared luminosities and other derived quantities, we previously
normalised the SEDs at 8 $\mu$m, if the source was detected there, or
at 24 $\mu$m otherwise.  In studying the SEDs of galaxies with very
high star-formation rates, we have found that normalisation at 8$\mu$m
for sources at z = 1.5-3.5 can result in poor estimates of the
infrared luminosity, because for many sources with z $>$ 1.5 the 8
$\mu$m emission is dominated by starlight. For z $>$ 3.5 we already
required normalisation to be at 24 $\mu$m (for this reason). 

We have therefore switched to normalisation (and luminosity estimation) based on a least-squares
fit at 24-500$\mu$m for all sources.  We have required a 24 $\mu$m detection
in order to associate a {\it Herschel} source with a SWIRE photometric
redshift catalogue source, so all sources have 24, 350 and 500 $\mu$m detections. 
This change significantly reduces the
number of very high luminosity (and high star-formation rate)
galaxies. From detailed SED modelling, we estimate the uncertainty in our corrected luminosities and
star-formation rates as $\pm$0.1 dex.  The star-formation rates are calculated for
a 0.1-100 $M_{\odot}$ Salpeter IMF.  Changing to a Miller-Scalo IMF would increase the
star-formation rates by a factor 3.3, while changing the mass range to 1.6-100 $M_{\odot}$,
ie forming A, B, O stars only, would reduce them by a factor 3.1 (Rowan-Robinson et al 1997).

Figure 2L shows our revised plot of star-formation rate (SFR) against
redshift for {\it HerMES}-SWIRE galaxies, which can be compared with Fig 2L
of Rowan-Robinson et al (2016). Details of the revised {\it HerMES}-SWIRE
catalogue are given at
http:$//$mattiavaccari.net$/$df$/$mrr$/$readmespirerev.
The revised luminosities have some effect on the bright end of the star-formation
rate functions.  In figure 3 we show the star-formation rate functions
for $z = 0.75-3.25$, derived using the new least-squares normalisation.
The tendency of the bright end of the function to be overestimated
relative to the model fits (figure 9 of Rowan-Robinson et al 2016)
has disappeared. The new parametric fits give star-formation rate
densities that differ from the values of Rowan-Robinson et al (2016)
by $< 1 \sigma$. A comparison between our SFRD and those previously
reported is also given in Table 7. The effect on the derived
star-formation-rate density from $z = 0-6$ is negligible. For $z>4.5$
there is no change, but these SFRDs are based almost entirely on sources with no 
association with SWIRE galaxies and so are very uncertain.

Figure 2R shows the SFR against redshift for {\it HerMES}-SWIRE and RIFSCz
galaxies with SFR $> 1000 M_{\odot}/yr$.  Typical {\it IRAS} 60 $\mu$m and
{\it Herschel} 500 $\mu$m detection limits are indicated.  The highest
star-formation rates significantly exceed the highest rates found by
Weedman and Houck (2008) at 0 $<$ z $<$ 2.5.  There appears to be a
natural upper limit to the SFR of 30,000 $M_{\odot}/yr$.  No
{\it HerMES}-SWIRE galaxies are found above this value and the {\it IRAS}-FSS
galaxies above this limit are probably gravitational lenses (see
previous section and Table 5).  This limit could represent an
Eddington-type radiation pressure limit on the star-formation rate of
the kind postulated by Elmegreen (1983), Scoville et al (2001), and
Murray et al (2005).  Scoville et al (2001) give a limit for $L/M_*$
of 500 $L_{\odot}/M_{\odot}$, which would translate to SFR
$< 10^{4.5} M_{\odot}/yr$ for $M_* < 10^{11.5} M_{\odot}$.

We can use the dust mass as a proxy for gas mass, assuming a
representative value for $M_{gas}/M_{dust}$.  Magdis et al (2011) have
summarised values of $M_{gas}/M_{dust}$ as a function of metallicity
for local galaxies, and shown that a redshift 4 galaxy lies on the
same relation, with $M_{gas} \sim 100 M_{dust}$ (cf also Chen et al
2013). We use this ratio to estimate $M_{gas}$ and then compare this
with our stellar mass estimates.  Figure 4L illustrates the behaviour
of the ($M_{gas})/M_*$ ratio as a function of redshift in the {\it HerMES}
galaxy sample.  For {\it HerMES} galaxies with z$>$1, $M_{gas}$ is
comparable with $M_{stars}$, so these are very gas-rich galaxies (as
noted by Rowan-Robinson et al 2010).  Very high gas fractions have
been found in galaxies with z $>$ 1 by Daddi et al (2010), Tacconi et
al (2010, 2013), and Carrilli and Walter (2013).  At low z,
$100 M_{dust} \sim 0.01-0.1 M_*$ so these galaxies have already
consumed most of their gas in star-formation.

Figure 4R shows $M_*/SFR$ as a function of redshift.  It is apparent
that the time to double the stellar mass at z = 3-5 is $\sim 10^8$
yrs.  In some objects the gas-depletion time is as low as 1-3x$10^7$
yrs (cf Rowan-Robinson 2000, Carilli and Walter 2013).  The Scoville
et al (2001) Eddington limit quoted above translates to
$M_*/SFR \sim 10^7$ yrs.

The picture that emerges is that the {\it Herschel} galaxies at z$>$3
are in the process of making most of the stars in the galaxy.
Essentially these are metal factories.  However we are not seeing
monolithic galaxy formation of the kind postulated by Partridge and
Peebles (1967), even though the star-formation rates and time-scales
are similar to those they suggested, because we can see from the
optical and near infrared SEDs that there has been an earlier
generation of star-formation at least 1 Gyr prior to the
star-formation we are witnessing.  This is evidenced by the classic
0.4-2$\mu$m SED profile of evolved red giant stars seen in the SEDs of
many of these galaxies (cf Fig 9 of Bruzual and Charlot 2003). Between
z = 1 and the present epoch we see a dramatic decline in the gas
content and star-formation rate.  For z $<$0.5 the gas depletion
time-scale is longer than the age of the universe so these are
galaxies that must have had a much higher rate of star-formation in
the past.

\begin{figure*}
\includegraphics[width=8cm]{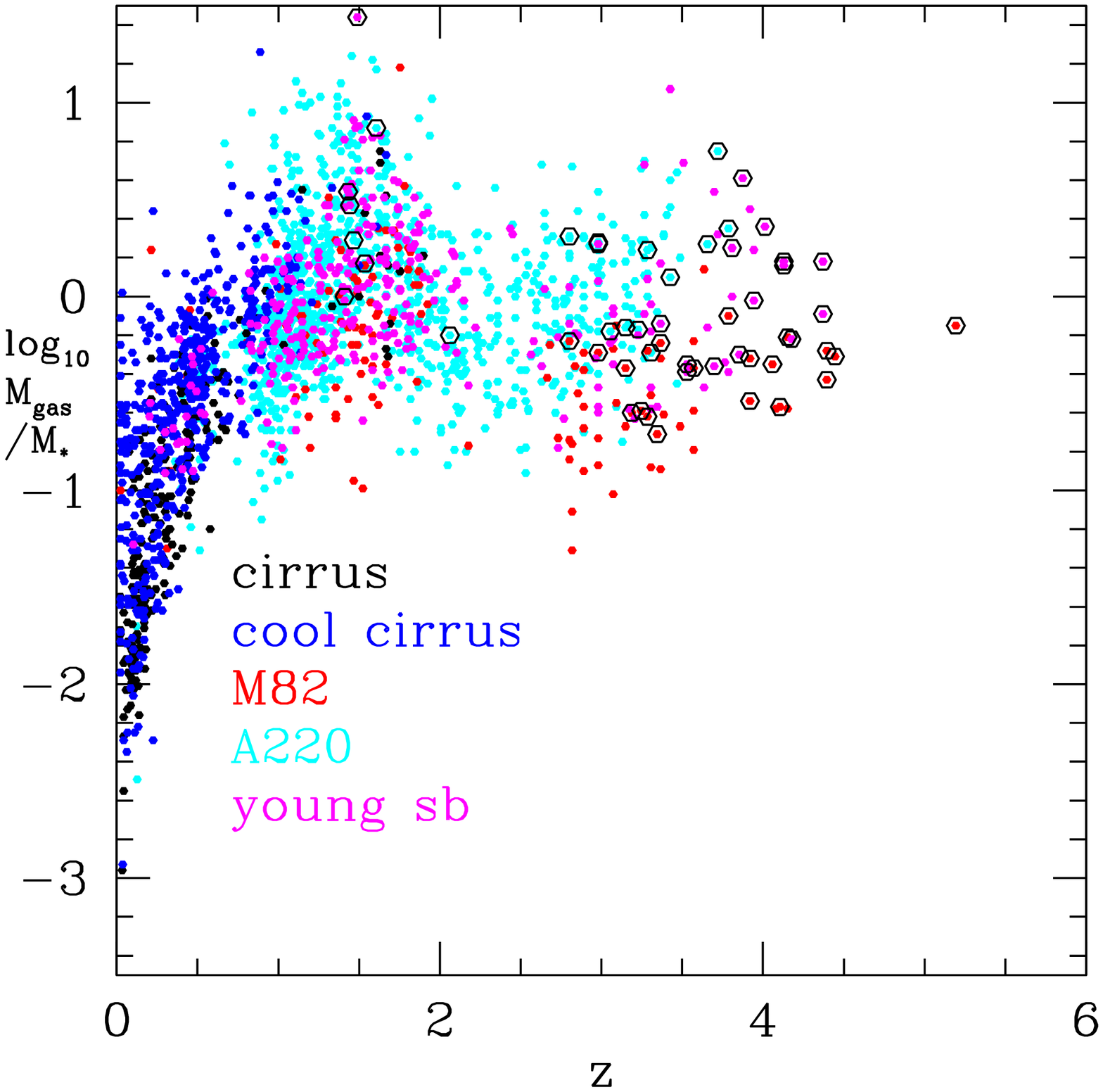}
\includegraphics[width=8cm]{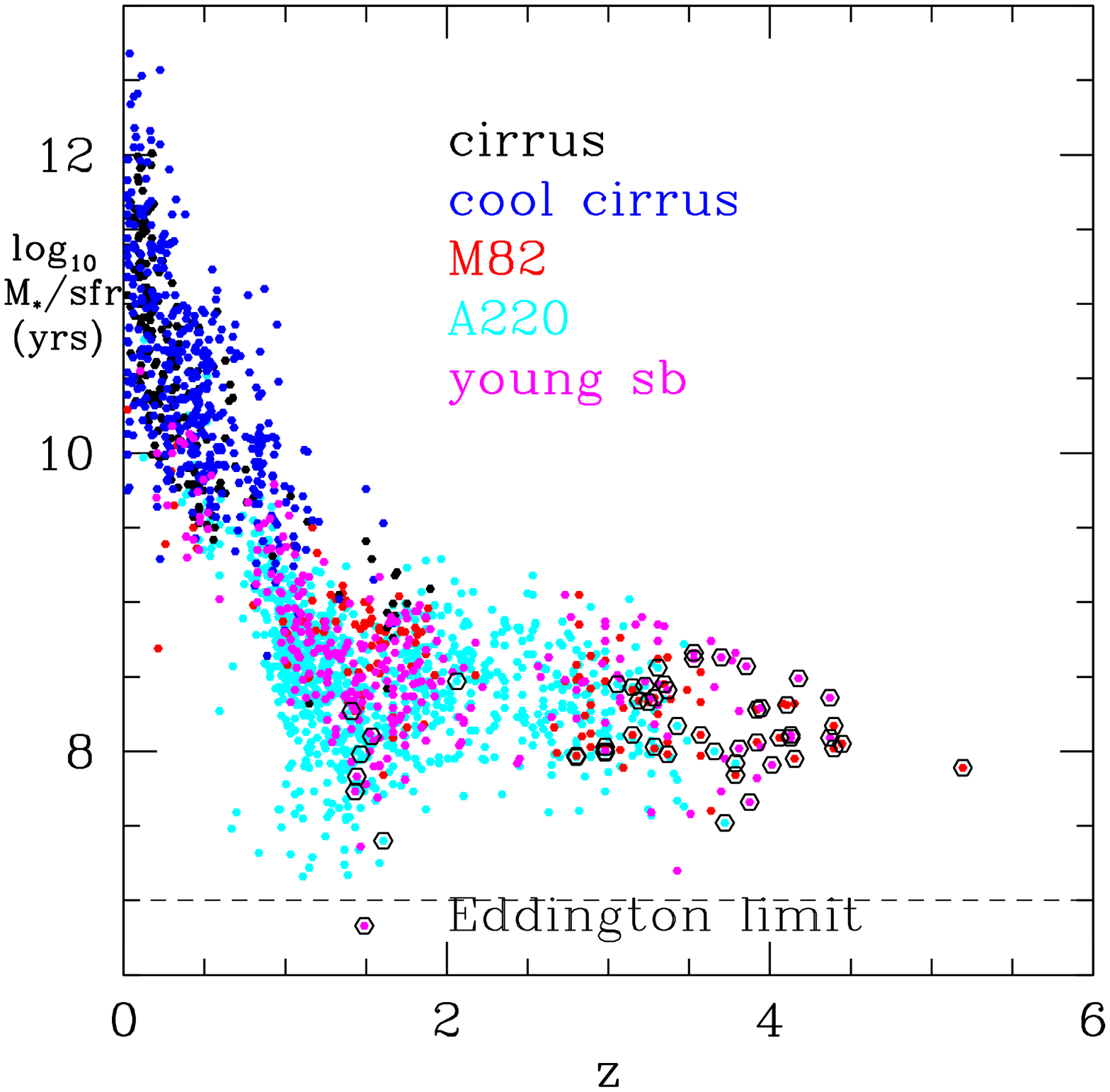}
\caption{L:  $M_{gas}/M_*$ vs. redshift (where $M_{gas}=100
M_{dust}$). Circled points are the `extreme' starbursts, those with SFR$> 5000\,M_{\odot} yr^{-1}$.
R:  $M_*/SFR$ (or, the time-scale needed to make the observed mass of
stars at the present star-formation rate) vs. redshift for {\it HerMES}
galaxies. The source are labelled by their dominant infrared template type. The candidate extreme starbursts are shown circled.}
\end{figure*}

\section{ Extreme starbursts}

Here we look in more detail at the galaxies in the {\it HerMES}-SWIRE survey
with implied star-formation rates greater
than 5000 $M_{\odot} / yr$.  Previously, detailed studies have been
presented of just two objects in this class: Rowan-Robinson and Wang
(2010) show the SED of one unlensed RIFSCz galaxy in this class (IRAS
F15307+3252, z = 0.926) with SFR = 8100 $M_{\odot} yr^{-1}$, and
Dowell et al report an object (FLS1, z = 4.29) with SFR = 9700
$M_{\odot} yr^{-1}$.

Our starting point is the {\it HerMES}-SWIRE
(Lock+XMM+ES1) galaxies with SFR $> 5000 M_{\odot} yr^{-1}$, according to our
automated infrared template fitting.  There are 70 candidates in all (details given in Tables A1-A3), 
but we have taken a robust approach to the reliability of the redshift estimates, rejecting sources
with lower-redshift aliases which give acceptable SED fits, and to the possibility of alternative
associations with lower redshift counterparts or blends (see below), resulting in a final list of 38 reliable extreme starbursts.  Details of the rejected sources and the reasons for rejection are given in Table A3.  Sources from Table A3 have been excluded from Figs 4,11,12.

\subsection{Reliability of redshift estimates}
For the 70 candidate objects, we refine the redshift estimates
adopting the approach of Rowan-Robinson et al (2016), who showed that
fitting our starburst templates to the 250-350-500\,$\mu$m data gives
an effective estimate of submillimetre redshift, $z_{subm}$. Combining
the $\chi^2$ distributions for the photometric and submillimetre
redshifts gives a best fit combined redshift $z_{comb}$. The values of
$z_{subm}$ and $z_{comb}$ are given in Tables A1 and A2.  
If $z_{comb}$ is significantly less than $z_{phot}$ and gives an acceptable 
SED fit, we have removed the object from the extreme starburst category and the
SED is not shown here (17 objects in all).  It is possible that in some cases the higher redshift is
correct, but we prefer to err on the side of caution. The source 9.17274-43.34398 ($z_{phot}$ = 3.06) has
a spectroscopic redshift of 1.748, which agrees well with $z_{subm}$, and so the source has been excluded.  

We have also examined the $\chi^2$
distribution for the photometric redshift fit to see if any lower redshift aliases are
present and the SEDs have also been examined for these aliases.  Again we have erred on the side of caution
and removed 7 objects with lower redshift aliases.    70 and 160 $\mu$m
fluxes have been included in the SED plots only if they have a signal-to-noise ratio
of at least 4. 
For three sources
(35.73369-5.62305, 7.98209-43.29812 and 161.89894+58.16401), 
$z_{comb}$ is significantly less than
$z_{phot}$, but the source remains in the extreme starburst category even with $z = z_{comb}$,
so we have shown the SED with $z_{comb}$ above the corresponding SED for $z_{phot}$. 

For 36.84426-5.31016 the photometric redshift (3.07) agreed well with the $z_{subm}$ (3.01) and
with $z_{comb}$ (3.07), but the $\chi^2$ for the photometric redshift fit was very poor, and the
template fit to the far infrared and submillimetre data was also poor, so we have preferred an
alternative association with a SWIRE $z_{phot}$ = 1.49 galaxy, which gives a good overall fit to
the SED, and so have excluded the object from the extreme starburst category.  There are four
other objects where detailed modelling of the SED gave solutions differing from the automated fit, which
did not confirm them as extreme starbursts.

As a further check on our photometric redshifts we have fitted our extreme starburst sample using the 
{\it CIGALE} code (Noll et al 2009). For 12 of our objects this yielded a lower preferred redshift.  For each
of these cases we have examined their overall optical-to-submillimetre SED to see if this alternative
redshift provides a plausible fit.  One of these we had already omitted due to a lower redshift alias
in the photometric redshift $\chi^2$ distribution and for one the lower redshift alias still yields a star-formation 
rate in the extreme category. For 2 objects the {\it CIGALE} redshift offered a plausible alternative
fit to the overall SED and these have been excluded.

We should also consider the reliability of the associations of {\it Spitzer} 3.6-24$\mu$m sources with 
optical and near infrared counterparts.  Confusion is not an issue at 24$\mu$m and the astrometric accuracy of 
the merged 3.6-24$\mu$m sources is $\pm$0.5 arcsec (Shupe et al 2007, Vaccari 2015).  The average number of galaxies to
i=25.5, the limit of associations considered here, is 0.014 per sq arcsec (Kashikawa et al 2004), so multiple
associations of optical-nir galaxies with Spitzer sources are extremely rare.

Generally the SED fits for the remaining 38 sources are reliable and plausible,
though only one is based on an optical spectroscopic
redshift. Spectroscopic confirmation of the remaining objects would
be highly desirable.  Almost all of our objects with $z_{phot} > 4$
have S350$>$S250, but it is worth noting that the range of $z_{subm}$
for our galaxies with $S250>S350>S500$ is 1.16-4.09, so the latter
condition does not imply low redshift.

To summarise the reliability of our redshift estimates for the 38
extreme starbursts: 1 has a spectroscopic redshift (indicated by 4
decimal places in Tables A1), a further 2 have photometric
redshift estimates $z\le$1.5 determined from at least 6 photometric
bands, so the rms uncertainty in (1+z) is $<4\%$ and the probability
of a catastrophic outlier is $<3\%$ (Rowan-Robinson et al 2013).  For
the 35 remaining objects with 1.5 $< z_{phot} <$ 5.2, 10 of which are
based on only 3 or 4 photometric bands, the photometric redshift
estimates are more uncertain, but are in most cases reinforced by the
estimates of $z_{subm}$.  Rowan-Robinson et al (2016) found that the
rms uncertainty in $(1+z_{comb})/(1+z_{spect})$ for 28 {\it Herschel}
galaxies with spectroscopic redshifts is $\sim21\%$.  
From the $\chi^2$ distributions for our photometric redshift
estimates we have estimated the corresponding redshift uncertainty, and hence estimated the
uncertainty in the star-formation estimate.  For 35/38 objects
the uncertainty is $\le$0.1 dex.
This uncertainty
could mean that a few of the objects could move out of the extreme
starburst category, probably balanced by others whose redshift has been
underestimated, but our overall conclusions are unlikely to be significantly affected.

For comparison we have listed the 19 {\it IRAS} RIFSCz objects with $SFR > 5000 M_{\odot} yr^{-1}$
in Table 6 (excluding the objects listed in Table 5).  We have modelled the SEDs 
of these objects individually (not shown here).  11 of the 19 objects have
spectroscopic redshifts, so for these objects the redshift uncertainty is not a 
major issue.  However the starburst component is usually fitted out to only 60 or
100 $\mu$m so the star-formation rates are uncertain by a factor of $\sim$2.  It
would be valuable to observe these galaxies at submillimetre wavelengths.

\subsection{Source blending, reliability of SWIRE associations}
Because we threshold at 5 $\sigma_{tot}$, where $\sigma_{tot}$ is the
total noise including confusion noise, problems of source blending and
confusion will be reduced (section 3).  As a check of this, for
each of our 500 $\mu$m sources we looked at any other possible
associations with 24 $\mu$m SWIRE sources within our 20 arc sec search
radius. 
For 18 of our 38 sources there was only one 24$\mu$m-detected SWIRE photometric
redshift catalogue counterpart within our 20 arcsec search radius. For the 
remainder we have summed the predicted 500$\mu$m fluxes (based on template fitting
to the SWIRE data) for all 24$\mu$m-detected counterparts within the search radius
and estimated the fraction of the predicted flux provided by our selected
association. For 34$/$38 sources this fraction is $\ge 90\%$ and for a further
2 sources it is in the range 80-90$\%$.  The overwhelming majority
of the alternative associations are unlikely to contribute significantly to the observed 500$\mu$m flux.

We looked at the SEDs of all the alternative associations to see whether any of these provided 
plausible SEDs when combined with our submillimetre sources.  
For 5 of our 38 sources
there was a plausible alternative lower redshift association and these
could be possible cases of misidentification or blending.  
These have been shown in Tables A1 and A2 in brackets.
One of
these sources (164.64154+58.09799) has a radio association (see below and Table 7)
which strongly supports the higher redshift association (the radio position is 0.7 arcsec from
high redshift position, 15 arcsec from lower redshift position).  
For two other sources (164.28366+58.43524 and 164.52054+58.30782) the
separation of the lower-z 24 $\mu$m association from the {\it Herschel}
position is much less than for the higher-z association (1.9 and 0.6 arcsec 
compared with 16.0 and 15.3 arcsec, respectively), so the
lower-z association may be correct. For 160.50839+58.67179 the {\it Herschel} position 
is 4.1 arcsec from our $z_{phot}$=3.81 association, 6.9 arcsec from an alternative 
$z_{phot}$=1.08 association, while for 35.28232-4.1400 the {\it Herschel} position 
is 2.4 arcsec from our $z_{phot}$=3.28 association, 5.3 arcsec from an alternative 
$z_{phot}$=2.55 association, so either association is plausible. 
We have shown the SEDs of these 5 alternative
associations in Figs 6-9. 

One option would be to split the
500 $\mu$m (and associated 350 and 250 $\mu$m fluxes) equally between
the two possible associations.  If this is done 2 of the 5 objects
move out of the extreme starburst category defined here.  Thus source
blending or misassociation is a relatively small problem in this
sample.

Confirmation of the correctness of our SWIRE associations with the
250-350-500 $\mu$m sources can be found through radio maps of some of
these sources.  Table 8 lists 7 of the 38 extreme starbursts for
which we have radio data. The positional offsets of the radio sources
from the SWIRE (3.6-24 $\mu$m) positions are all $<$ 1-2 arcsec.  We
can also calculate the q-values for these sources, where q =
$log_{10} (L_{FIR}/L(1.4GHz))$.  These lie in the range 2.0-2.6, with
a mean of 2.33, in good agreement with the values found for lower
redshift {\it Herschel} galaxies by Ivison et al (2010) and with the mean
value 2.34 found for {\it IRAS} galaxies (Yun et al 2001).  We have also
shown positional offsets for 1.2mm MAMBO observations of
161.75087+59.01883 by Lindner the al (2011) and for 850 $\mu$m
observations of the same source by Geach et al (2017). 
Hill et al (2018) have observed the same source with the SMA
interferometer and confirm that it is a single source.
These fluxes
are included in the SED of this source plotted in Fig 5.  This is our
best-case object, with a spectroscopic redshift (and $z_{subm}$
closely agreeing with this), radio confirmation of the SWIRE
association and the radio estimate of the star-formation rate agreeing
well with that from the submillimetre data (q=2.36).

It is worth commenting that the positional uncertainties of our 500 $\mu$m
sample are greatly improved by requiring also $5-\sigma$ detections at 350 $\mu$m.
In most cases ($39/44$) there is a detection at 250 $\mu$m as well and this is 
the position used, where available.

While we believe we have presented strong arguments for the reality of
these {\it Herschel} extreme starbursts, especially those confirmed by radio
observations both positionally and in the ratio of far-infrared to
radio luminosities, it will be important to confirm the correctness of
our SWIRE associations through ALMA and other submillimetre mapping,
and through further radio mapping (e.g. by LOFAR, GMRT, MeerKAT and SKA).
The correctness of our lensing candidates can be confirmed by HST and
JWST mapping.  For the IRAS extreme starbursts (Table 7) confusion and source blending
are not an issue.

In the 2.9 deg$^{2}$ of the SWIRE-CDFS area (not used in this
study), we find 8 extreme starbursts, consistent with the
surface-density of 1.9 per sq deg found in Lockman+XMM+ES1.
Unfortunately none of these lie in the 0.25 sq deg area surveyed at
870 $\mu$m with LABOCA by Weiss et al (2009), and followed up with
ALMA by Hodge et al (2013).

\subsection{Role of lensing}
Although we believe we have removed all the lensed systems from our
sample (section 4) we need to consider whether any of these 38 extreme
starbursts could be lensed.  From the analyses of Negrello et al
(2010) and Wardlow et al (2013) it is the brightest 500 $\mu$m sources
that are most likely to be lensed.  None of our extreme starbursts
have S500 $>$ 80 mJy but 6 have 60 $<$ S500 $<$ 80 mJy.  Because there
is reasonable agreement between $z_{phot}$ and $z_{subm}$ for these
sources (exact agreement for 3 of the objects), they could only be
lensed if the optical and submillimetre emission was also from the
lensed galaxy.  This would make them distinct from the known
submillimetre lenses.  Also 3 of these bright sources are amongst
those confirmed by radio surveys (Table 7) and not reported as lensed.
The lensing galaxy candidates found by Rowan-Robinson et al (2014)
typically have i-magnitudes in the range 19-22, significantly brighter than
the optical counterparts of our extreme starburst sample.  We have also checked 
whether known clusters lie close to any of our 38 objects, in case cluster lensing was
an issue, but have found none within one arcmin of our objects.
Our expectation is that few or none of our 38 objects will turn out to be
lensed systems.

\begin{table*}
\caption{Radio positional offsets and q-values for extreme starbursts}
\begin{tabular}{llllllll}
\hline\hline
\smallskip
 RA(J2000) & Dec(J2000) & frequency & flux-density & offset & reference & q & $log_{10}$ SFR\\
  & & & & arcsec & & & ($M_{\odot} yr^{-1}$)\\
\hline
36.10986 & -4.45889 & 1.4GHz & 0.219$\pm$0.026mJy & 0.5 & (2) & 2.46 & 4.20\\
161.75087 & 59.01883 & 324.5MHz & 687$\pm$72$\mu$Jy & 0.7 & (3) & 2.36 & 3.75\\
  & & 1.4GHz & 278.8$\pm$15.2$\mu$Jy & 1.2 & (4) & & \\
  & & 1.2mm & 3.5$\pm$0.6mJy & 1.5 & (5) & & \\
  & & 850$\mu$m & 9.29$\pm$1.05mJy & 1.4 & (6) & &\\
161.98271 & 58.07477 & 1.4GHz & 0.125mJy & 1.1 & (7) & 2.12 & 3.76\\
162.33324 & 58.10657 & 1.4GHz & 1.36mJy & 1.1 & (7) & 2.45 & 3.73\\
162.46065 & 58.11701 & 1.4GHz & 0.064mJy & 1.7 & (7) & 2.50 & 3.88\\
162.91730 & 58.80596 & 324.5MHz & 1013$\pm$96$\mu$Jy & 1.6 & (3) & 2.47 & 3.75\\
  & & 1.4GHz & 0.504mJy & 1.0 & (7) & & \\
164.64154 & 58.09799 & 1.4GHz & 0.182mJy & 0.7 & (7) & 2.37 & 3.72\\
\hline
\end{tabular}
\tablefoot{
(1): Zinn et al (2012),
(2): Bondi et al (2003),
(3): Owen et al (2009),
(4): Owen et al (2008),
(5): Lindner et al (2009),
(6): Geach et al (2017), Hill et al (2018) (7): Prandoni I. 2017 (personal communication)}
\end{table*}

\begin{figure*}
\includegraphics[width=14cm]{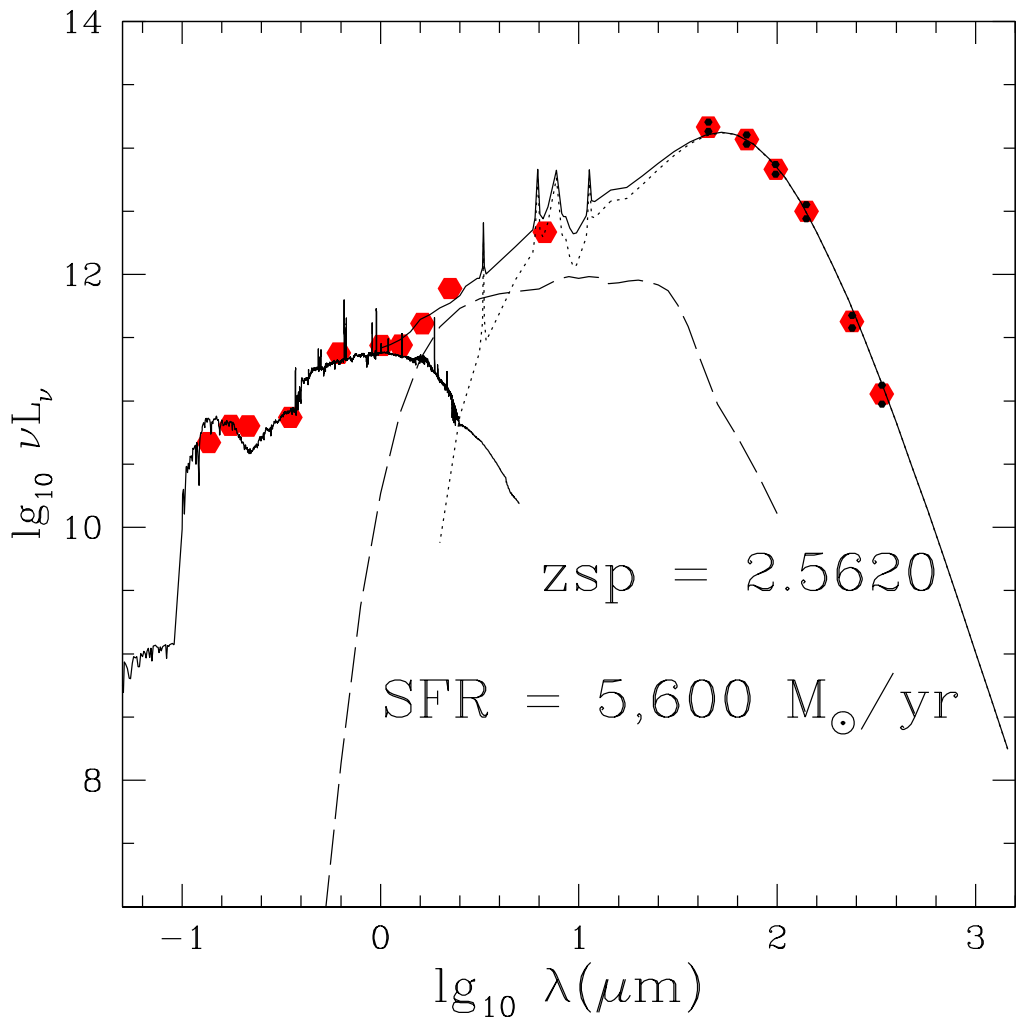}
\caption{
Rest-frame SED of 161.75087+59.01883, {\it Herschel}-SWIRE 500 $\mu$m source with spectroscopic redshift and extreme starburst luminosity. SWIRE association is confirmed by radio, 1.2mm, and 850 $\mu$m positions (latter two fluxes shown in SED).
Errors for submillimetre fluxes indicated by black dots.  
Dotted loci: M82 starburst, long-dashed loci: AGN dust torus.
}
\end{figure*}

\begin{figure*}
\includegraphics[width=4.0cm]{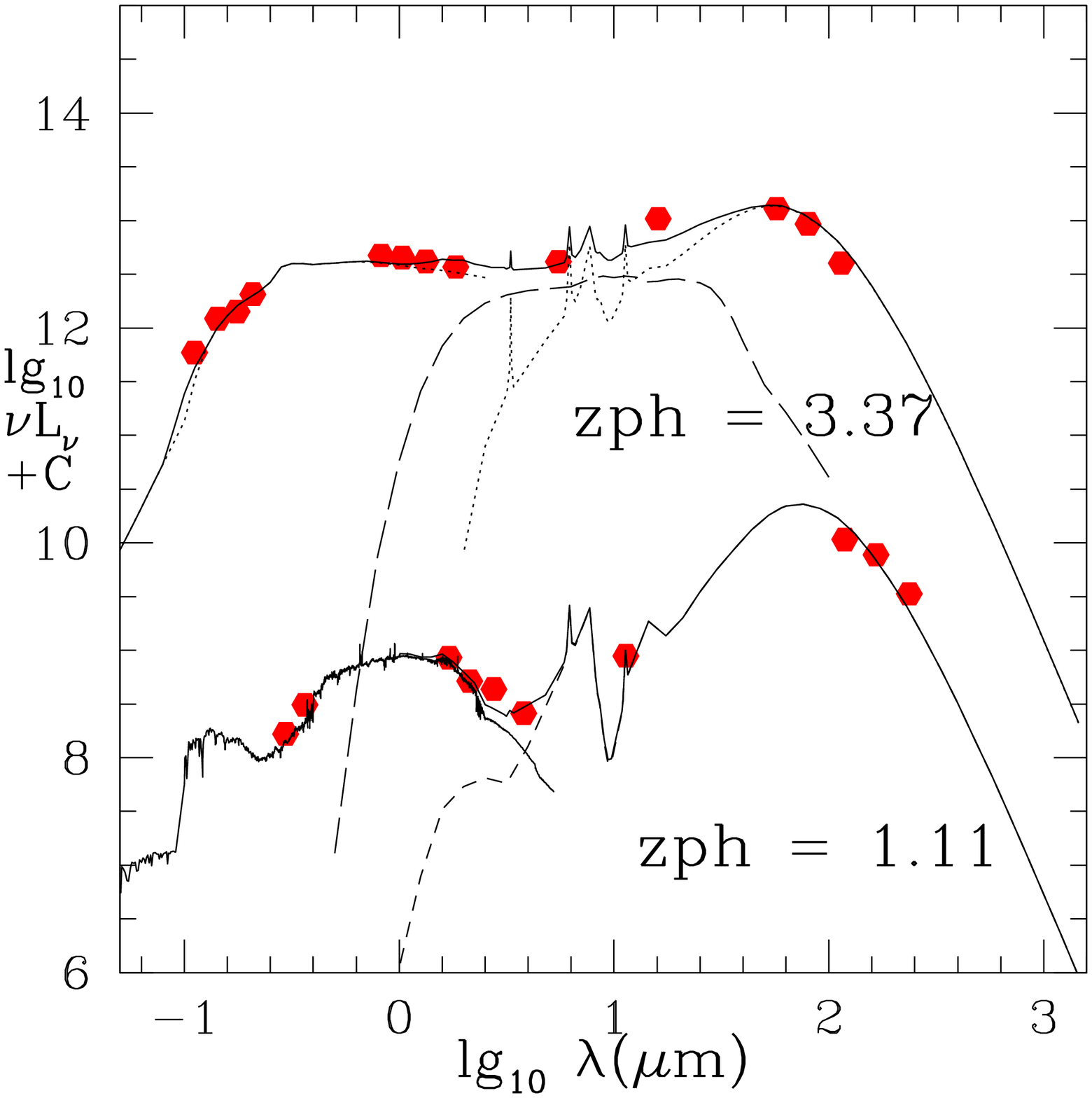}
\includegraphics[width=4.0cm]{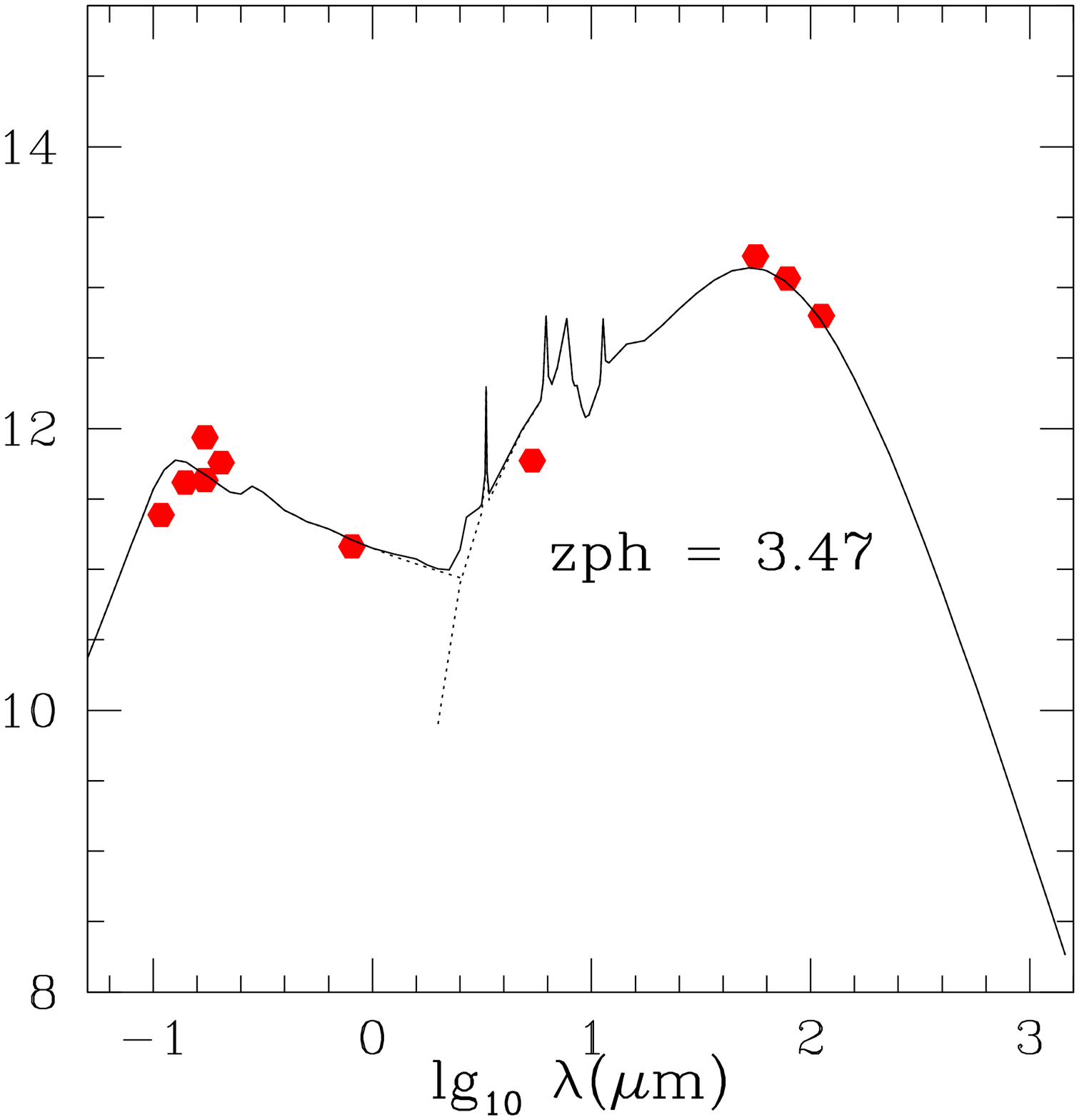}
\includegraphics[width=4.0cm]{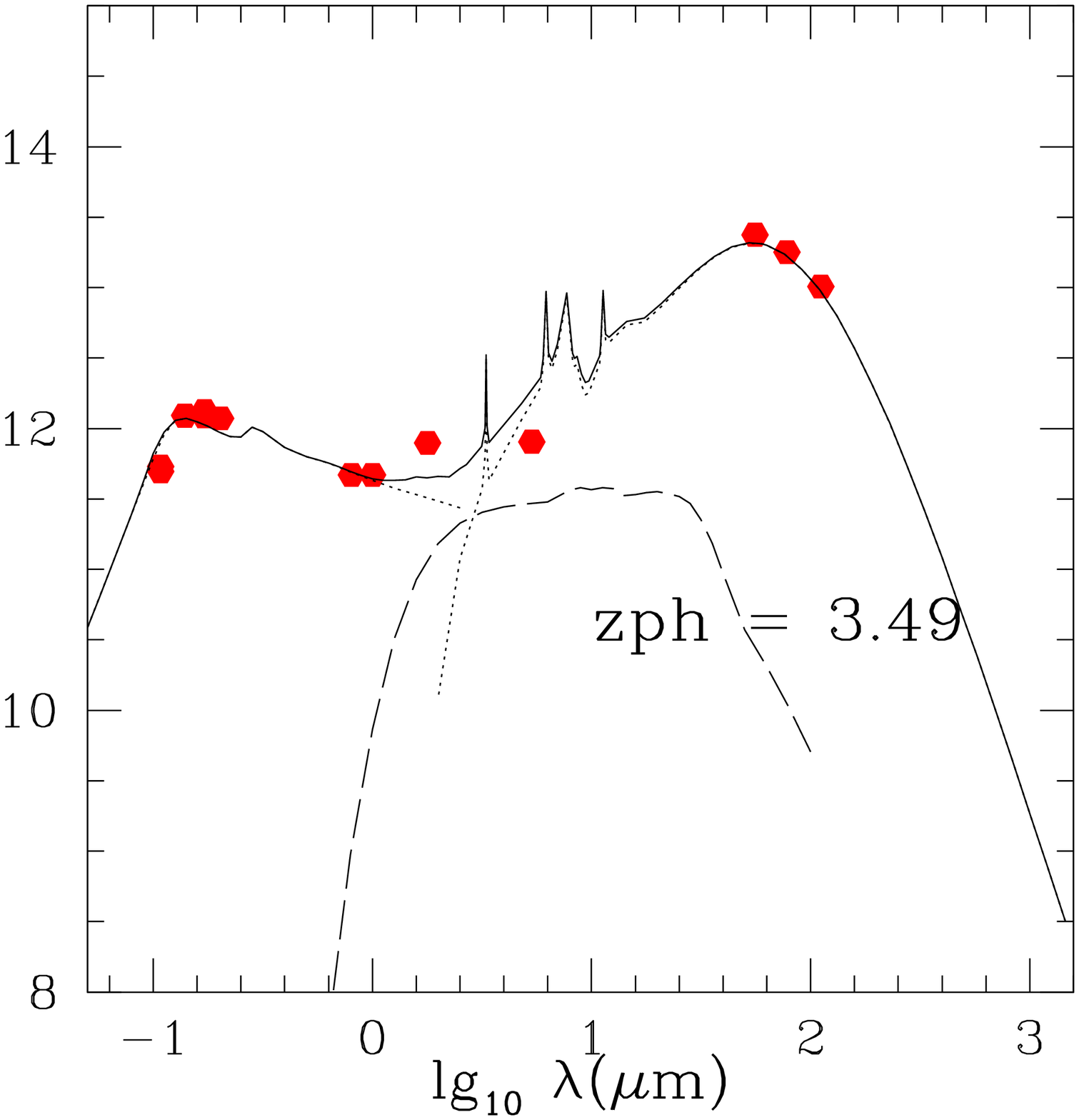}
\includegraphics[width=4.0cm]{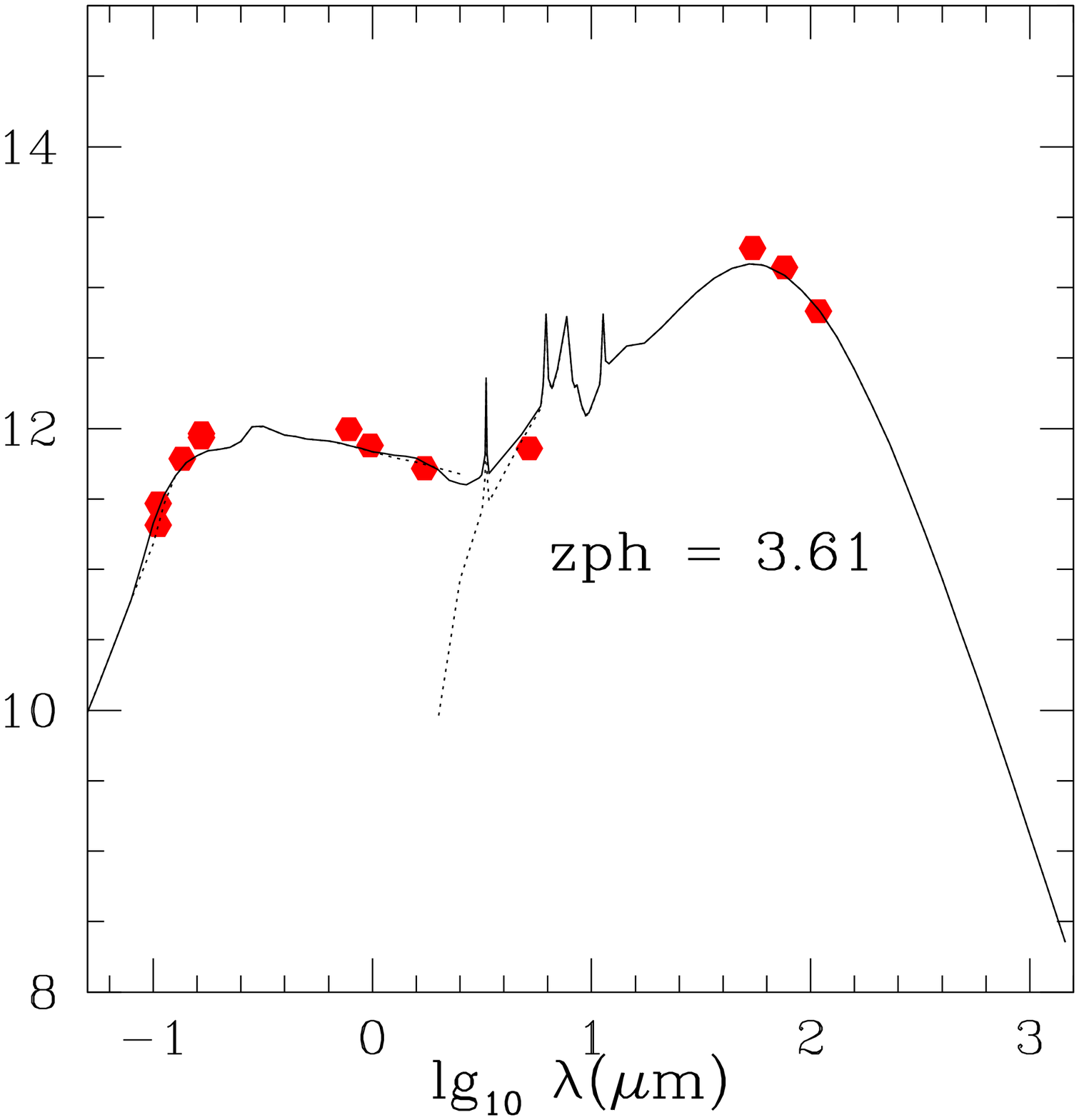}
\includegraphics[width=4.0cm]{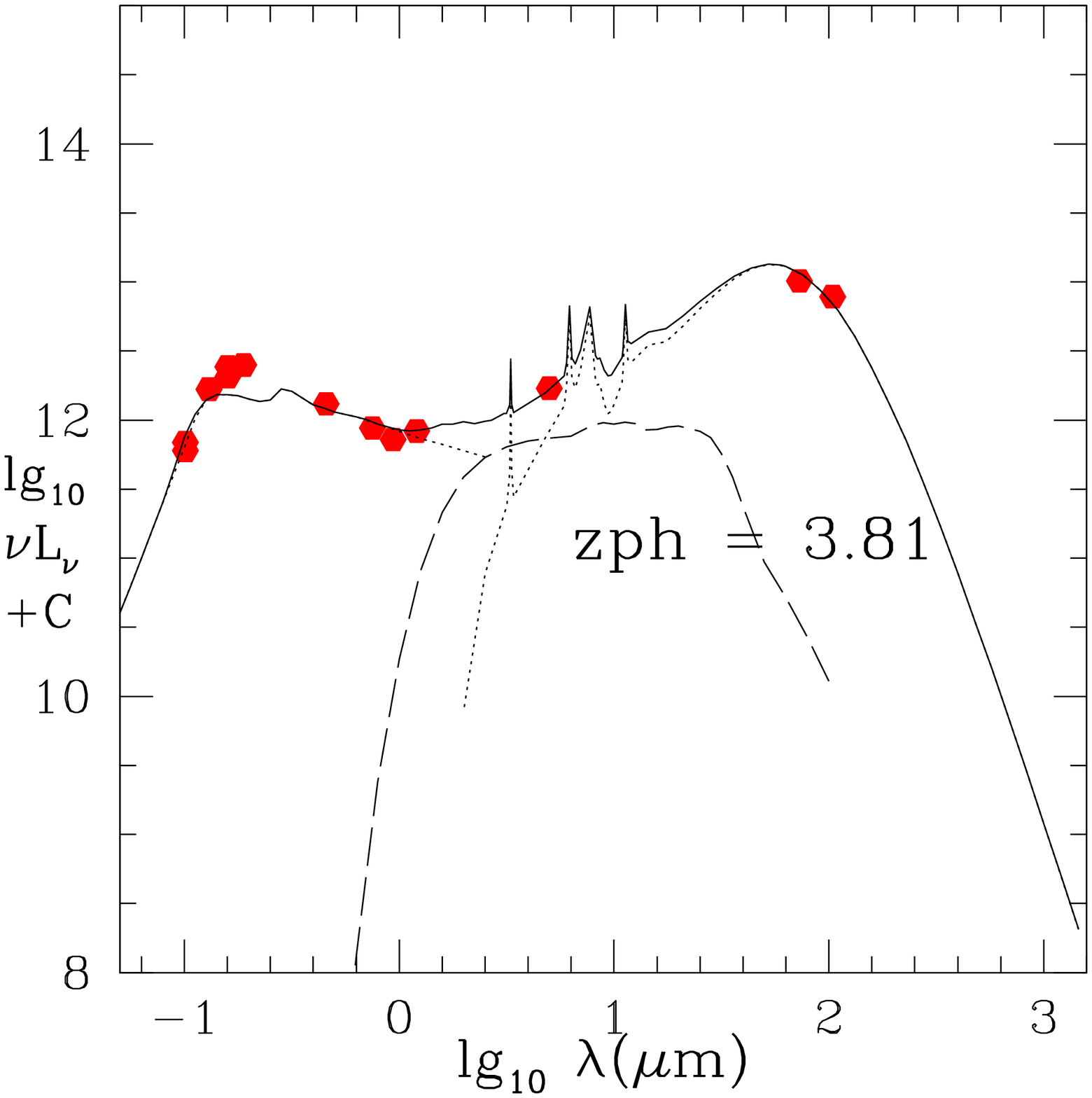}
\includegraphics[width=4.0cm]{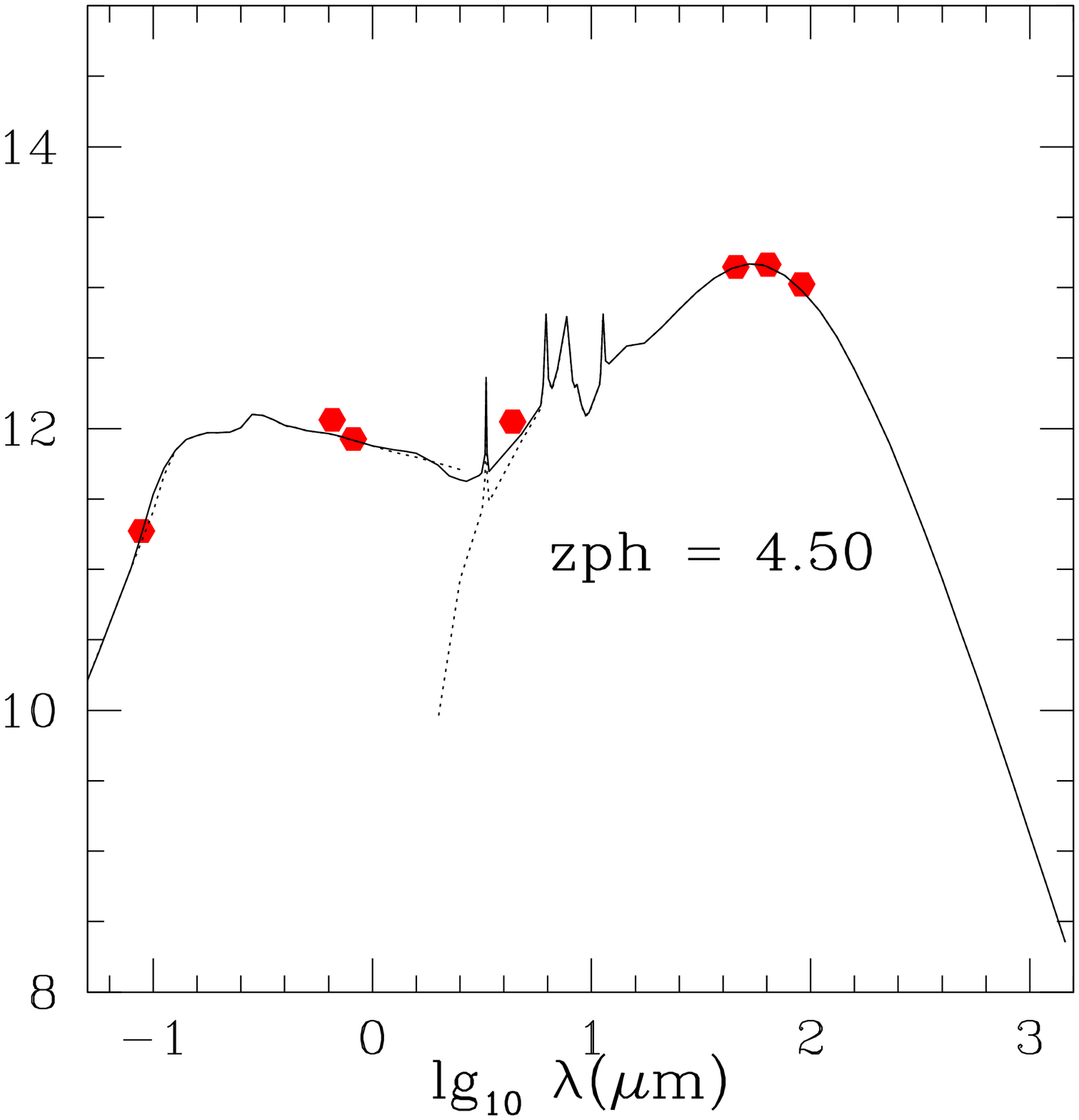}
\caption{
Rest-frame SEDs of {\it Herschel}-SWIRE 500 $\mu$m sources with extreme starburst luminosities (SFR$> 5000\,M_{\odot} yr^{-1}$), labelled with the redshift, whose optical through near-infrared SEDs are best-fitted by a QSO template.  
Alternative SWIRE associations are shown plotted below the extreme starburst solution.  Possible redshift aliases
($z_{comb}$)are shown plotted above the preferred $z_{phot}$ solution.
Dotted loci: M82 starburst, dashed loci: Arp 220 starburst, long-dashed loci: AGN dust torus.
The constant C = 0 except in cases where SEDs are shown for both the photometric redshift and for a lower redshift alternative SWIRE association.  In these cases C = -2 for the lower SED and +1 for the upper one.
}
\end{figure*}

\subsection{SEDs of extreme starbursts}
We present SEDs for these extreme starbursts in the following figures.
Figure 6 shows SEDs of extreme starbursts in the {\it Herschel}-SWIRE fields
whose optical and near infrared data is fitted with a QSO template.
Pitchford et al (2016) have studied a sample of 513 Type 1 QSOs detected by {\it
  Herschel} at 250 $\mu$m, some in the {\it HerMES}-SWIRE areas, and found
star-formation rates ranging up to 5000 $M_{\odot}/yr$.  In Figure 7 we show SEDs of
objects whose optical-nir data is fitted with a galaxy template, but
whose mid ir data show the presence of an AGN dust torus.  These
sources are plausibly Type 2 AGN whose host galaxies exhibit extreme
rates of star formation. Of the 38 objects in our sample, 19 have
optical through mid-infrared SEDs consistent with Type 1 or Type 2
AGN. In no case however does the luminosity of the AGN exceed that of
the starburst.

In contrast, there exist many examples of `pure' extreme starbursts in
our sample.  Figure 8 shows SEDs of objects whose optical and near
infrared data are fitted with galaxy templates and whose mid ir, far
ir and submillimetre data are fitted with M82 or Arp220 starburst
templates.  Figure 9 shows an especially interesting set of examples
of pure extreme starbursts, whose infrared SEDs are best fit with
young starburst templates. None of these objects show any evidence for
significant AGN activity.  Altogether 19$/$38 objects are pure
starbursts.

As a check on our star-formation rate estimates we have also fitted the overall
SEDs with the {\it CIGALE} code, using our preferred redshifts.  We find
that the {\it CIGALE} SFR estimates are in broad agreement with ours.

The star formation rates in these extreme starbursts all lie in the
range $5,000-30,000\,M_{\odot} yr^{-1}$. As noted in the introduction,
such high star formation rates are not predicted by any current
semi-analytic model for galaxy formation, so these objects pose a
serious challenge to theoretical models. Our 38 {\it Herschel}-SWIRE
objects correspond to a surface density of 1.9 extreme starbursts per
sq deg.  The $500\,\mu$m sources which are not associated with SWIRE
galaxies could add up to a further $\sim9$ extreme starbursts per sq
deg.

In Tables A1,A2 we have also shown our stellar mass estimates.  They lie in the
range $lg_{10} M_*$ = 11.28-12.50, so these are exceptionally massive galaxies.
For 3.5$<z<$4.5, 11.5$<lg_{10} M_*<$12.5, there are 16 objects, yielding a space-density
of $10^{-7.12} Mpc^{-3} dex^{-1}$, which fits nicely on an extrapolation of the
mass-function given by Davidzon et al (2017, their Figs 8,11).

\begin{figure*}
\includegraphics[width=4.0cm]{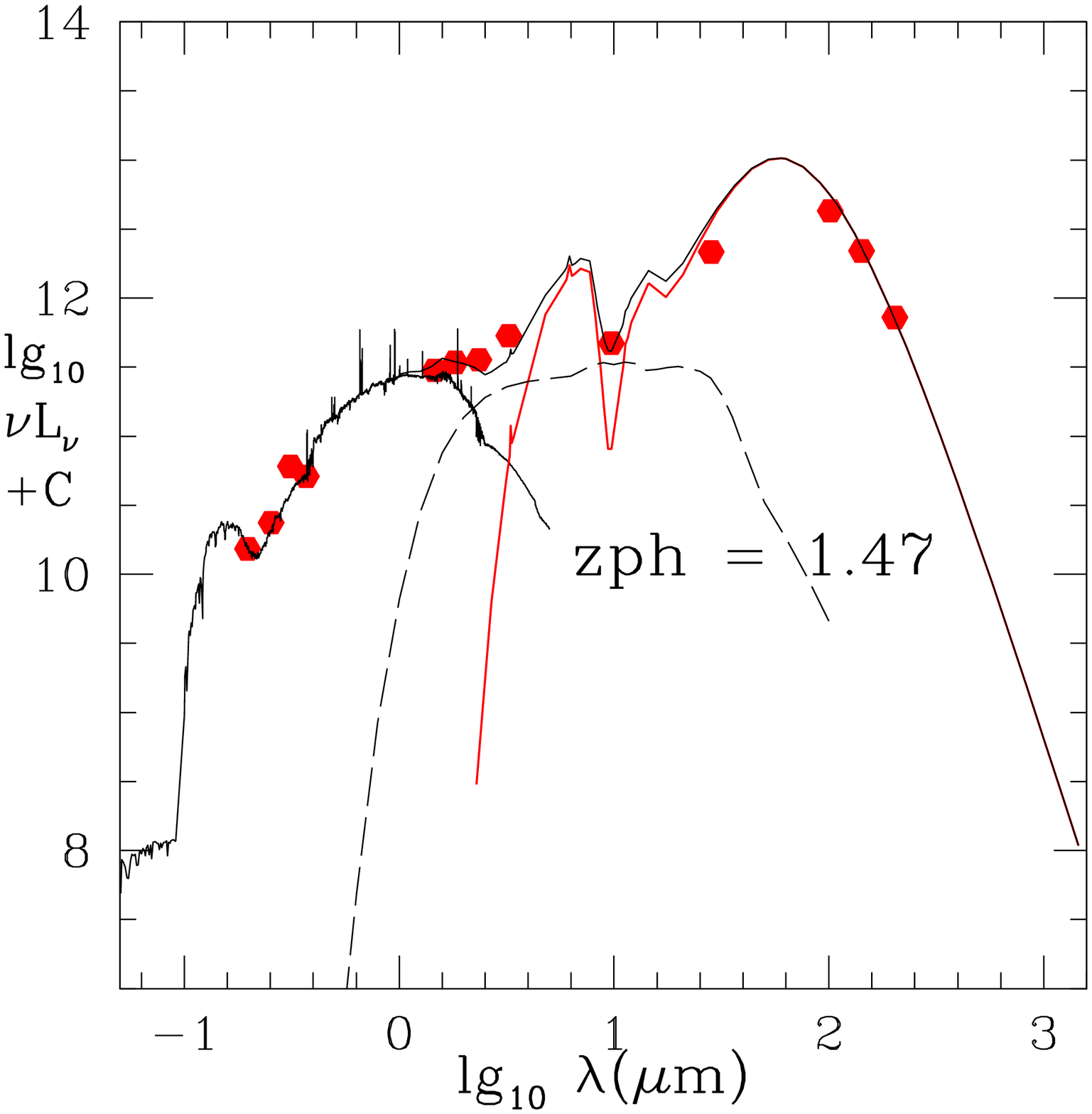}
\includegraphics[width=4.0cm]{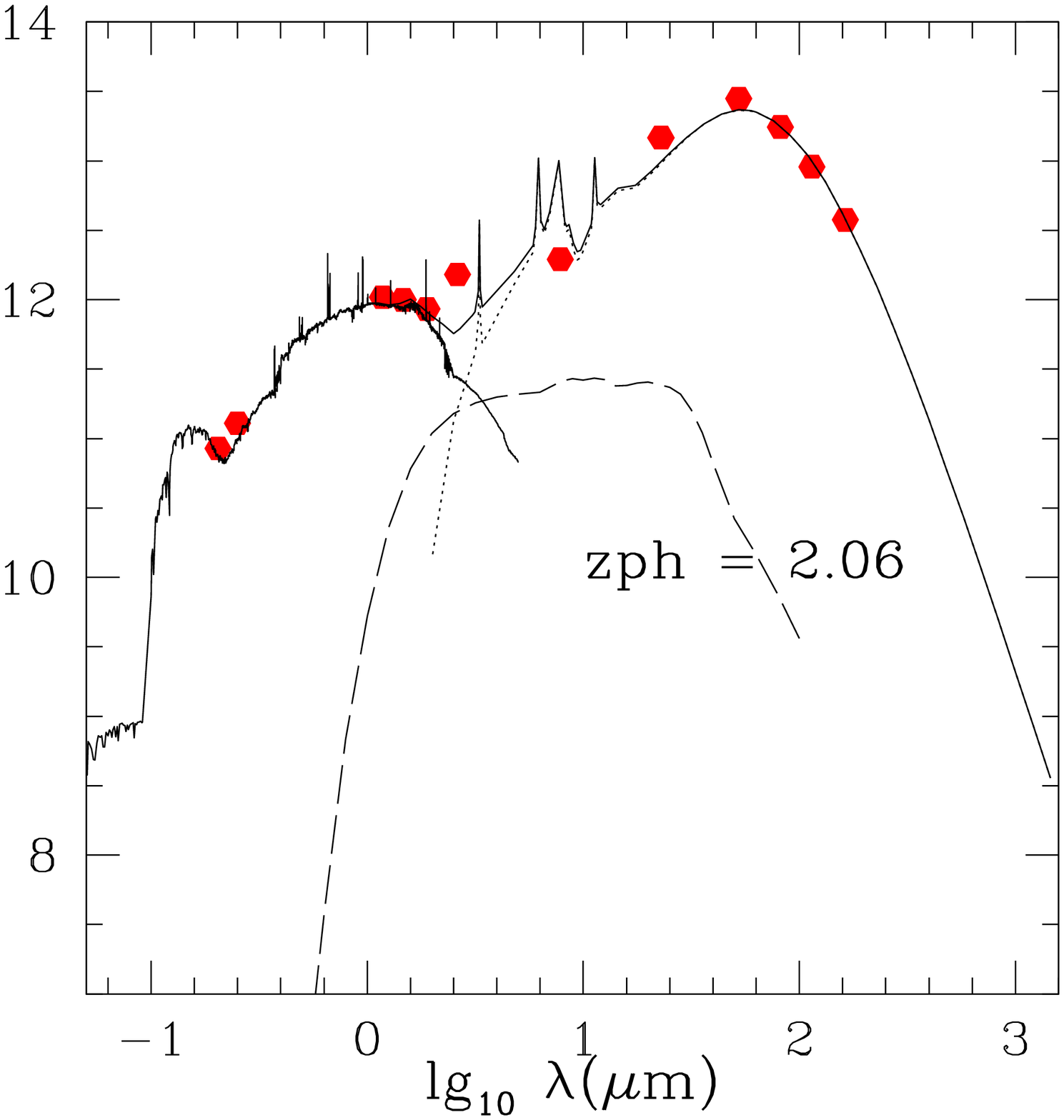}
\includegraphics[width=4.0cm]{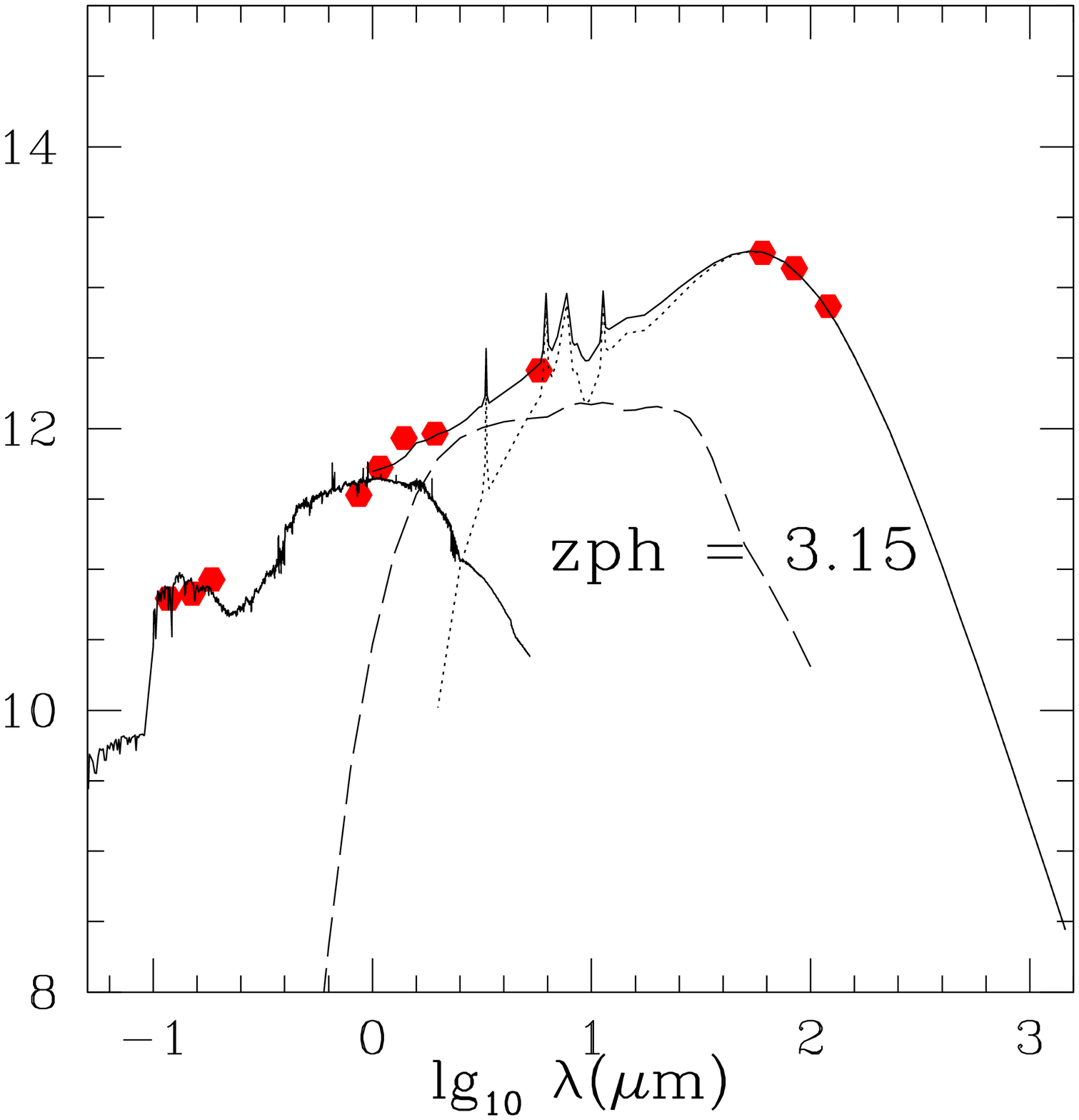}
\includegraphics[width=4.0cm]{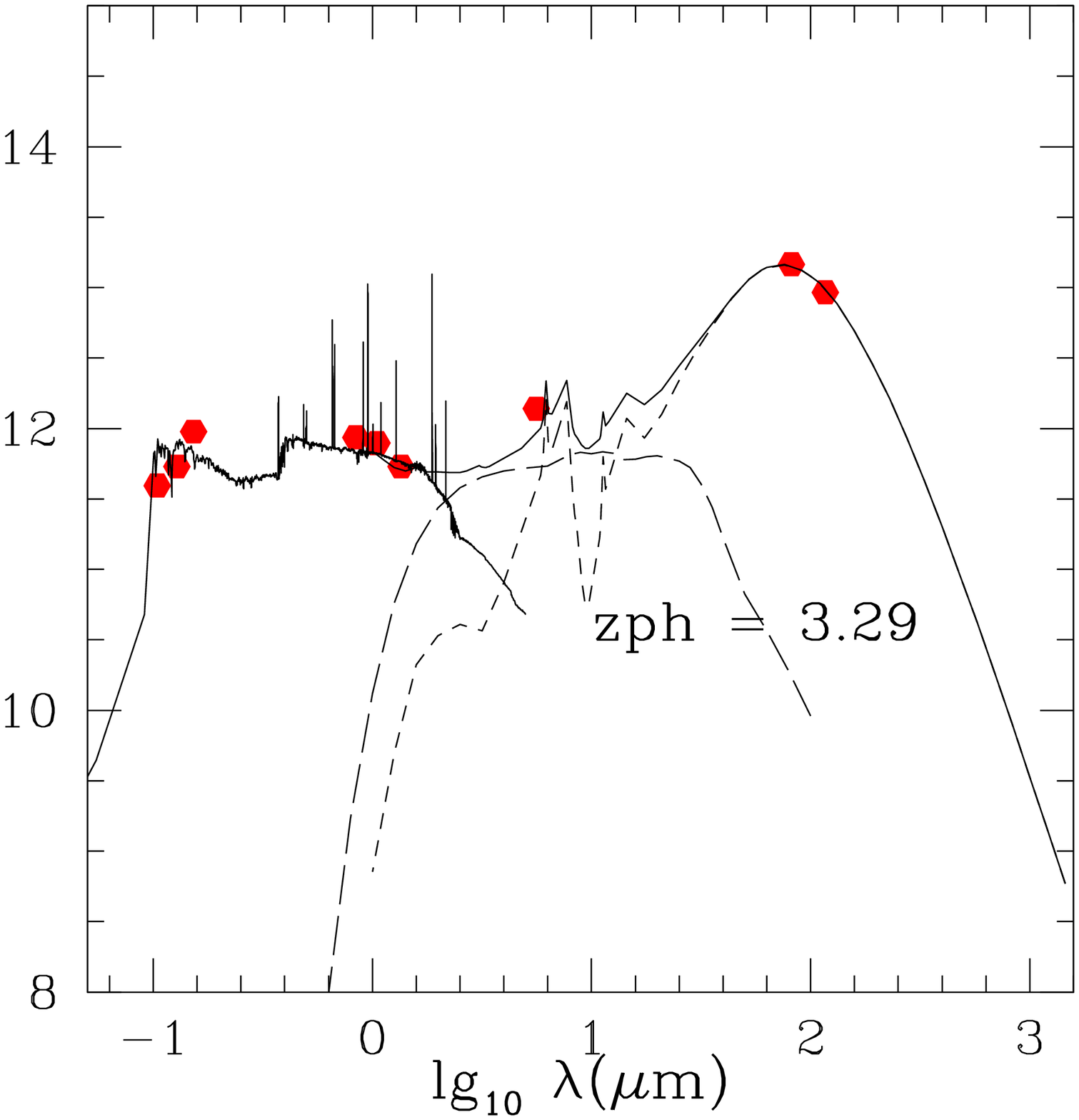}
\includegraphics[width=4.0cm]{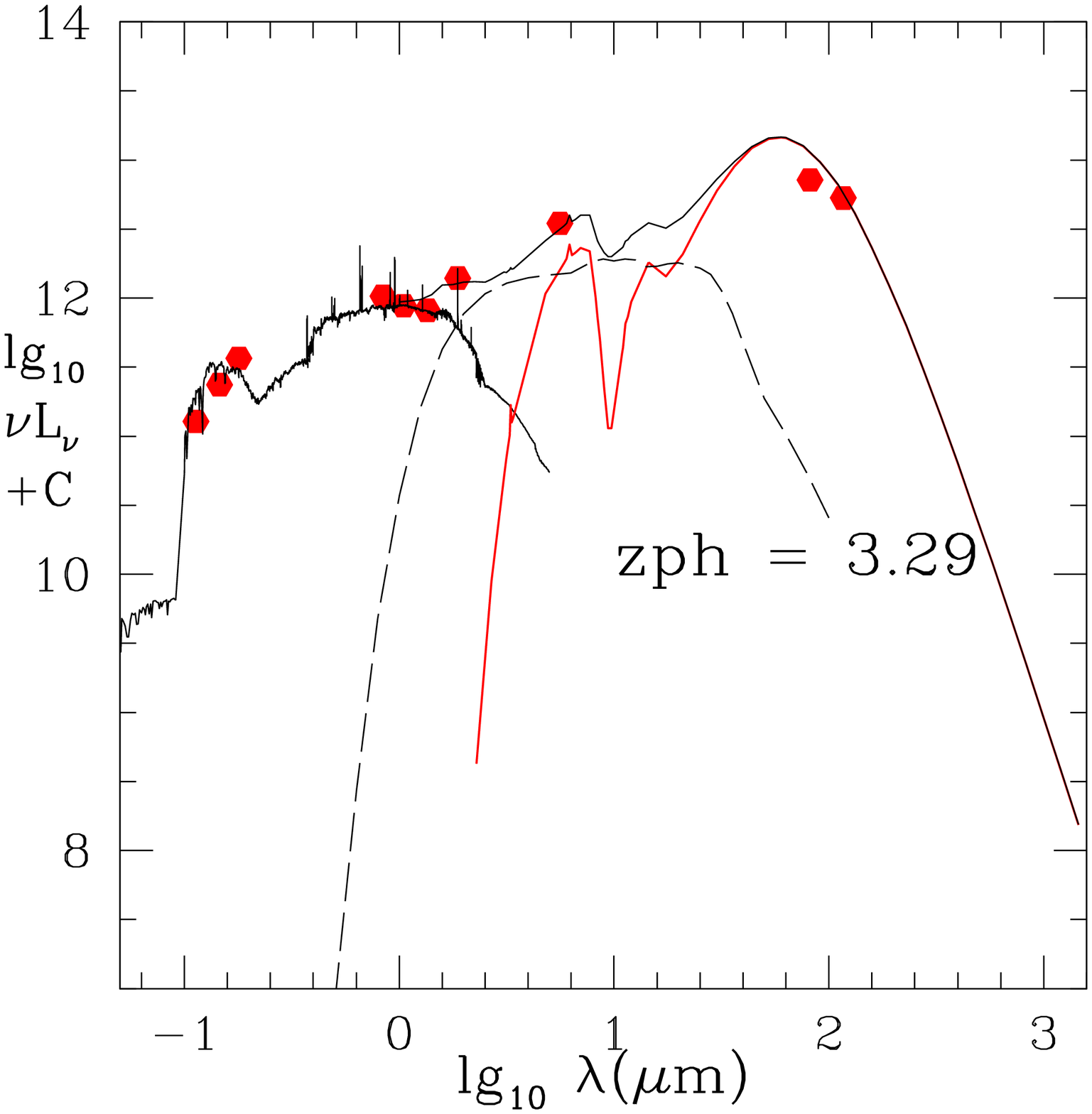}
\includegraphics[width=4.0cm]{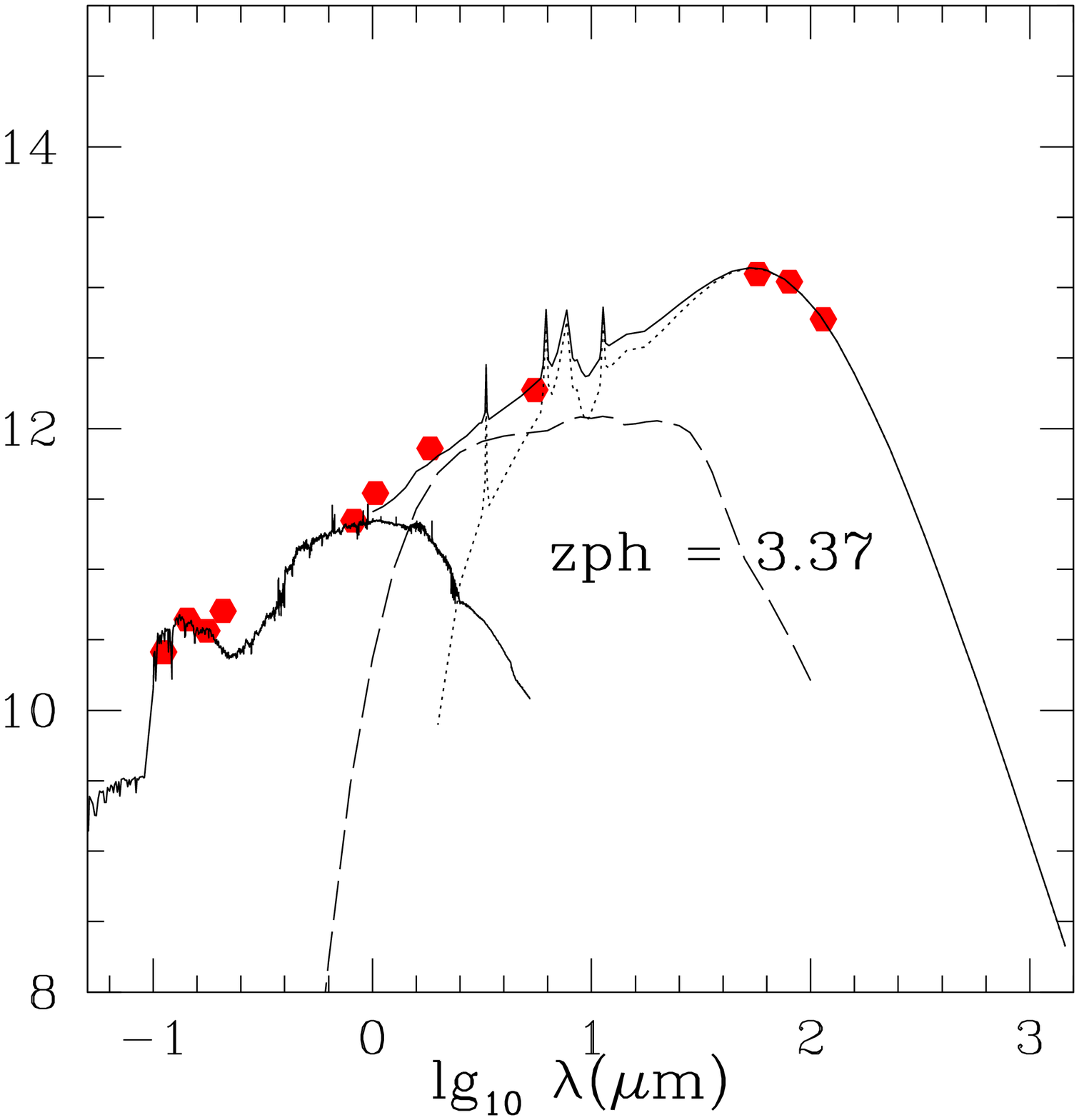}
\includegraphics[width=4.0cm]{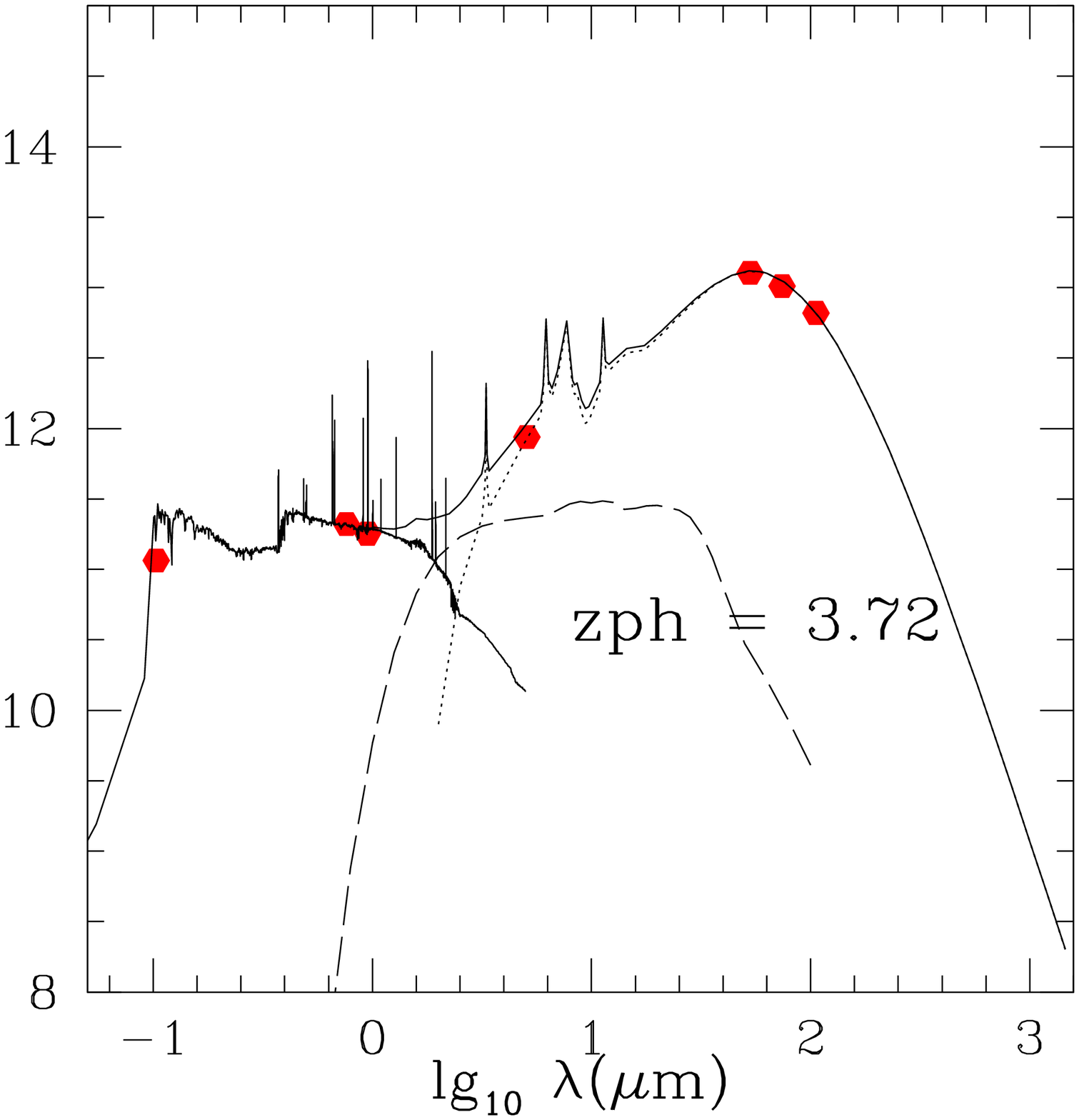}
\includegraphics[width=4.0cm]{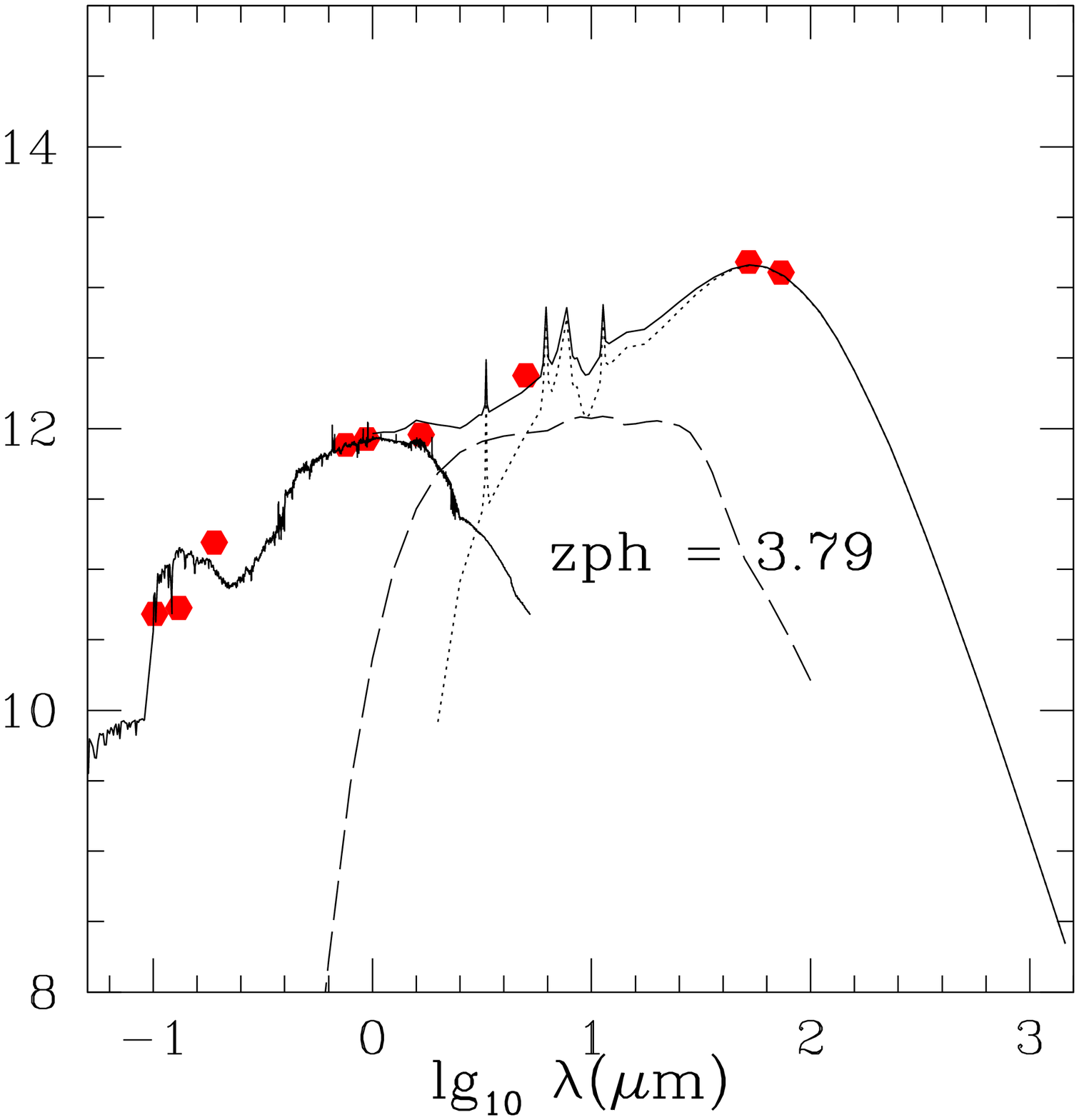}
\includegraphics[width=4.0cm]{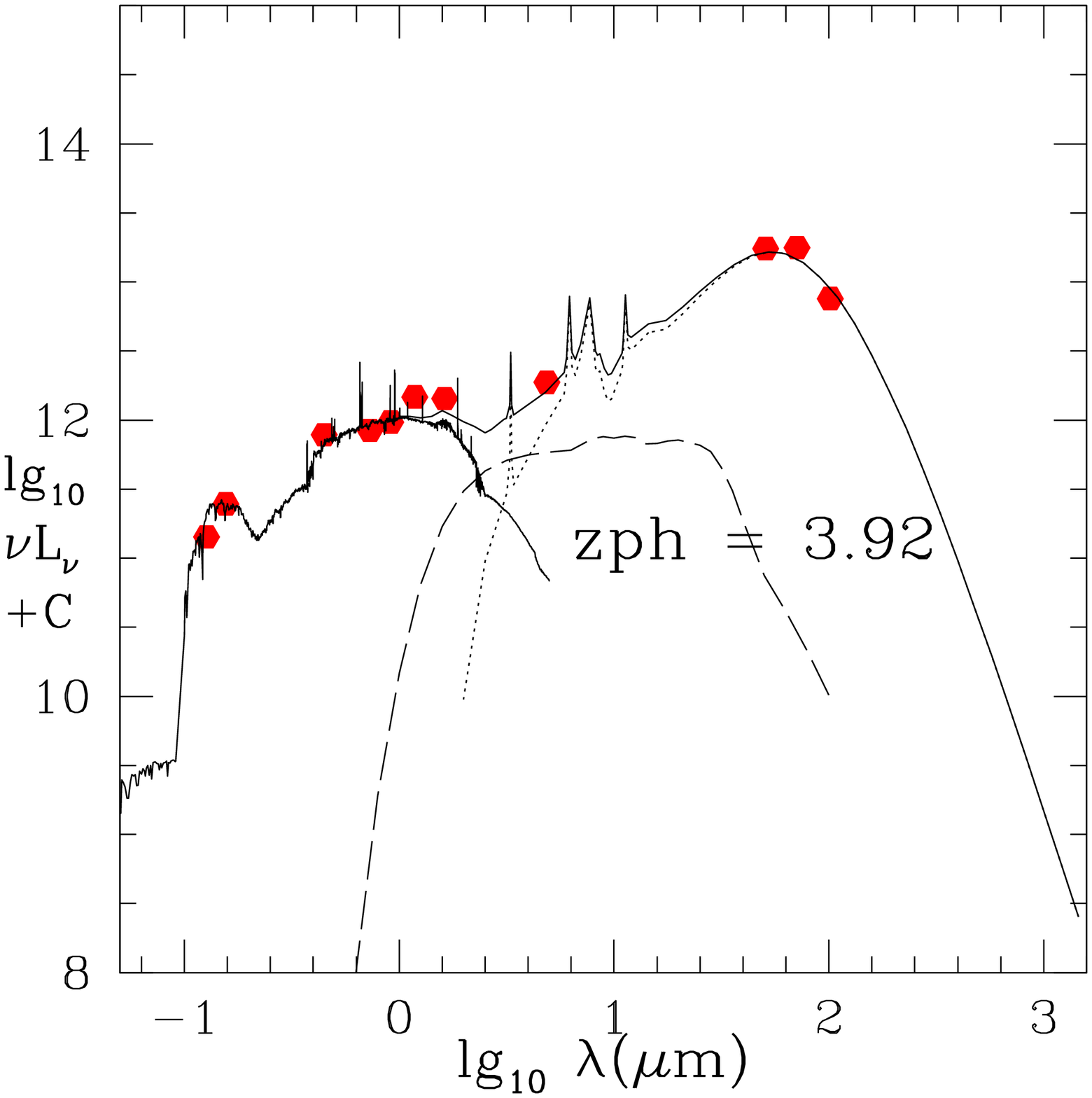}
\includegraphics[width=4.0cm]{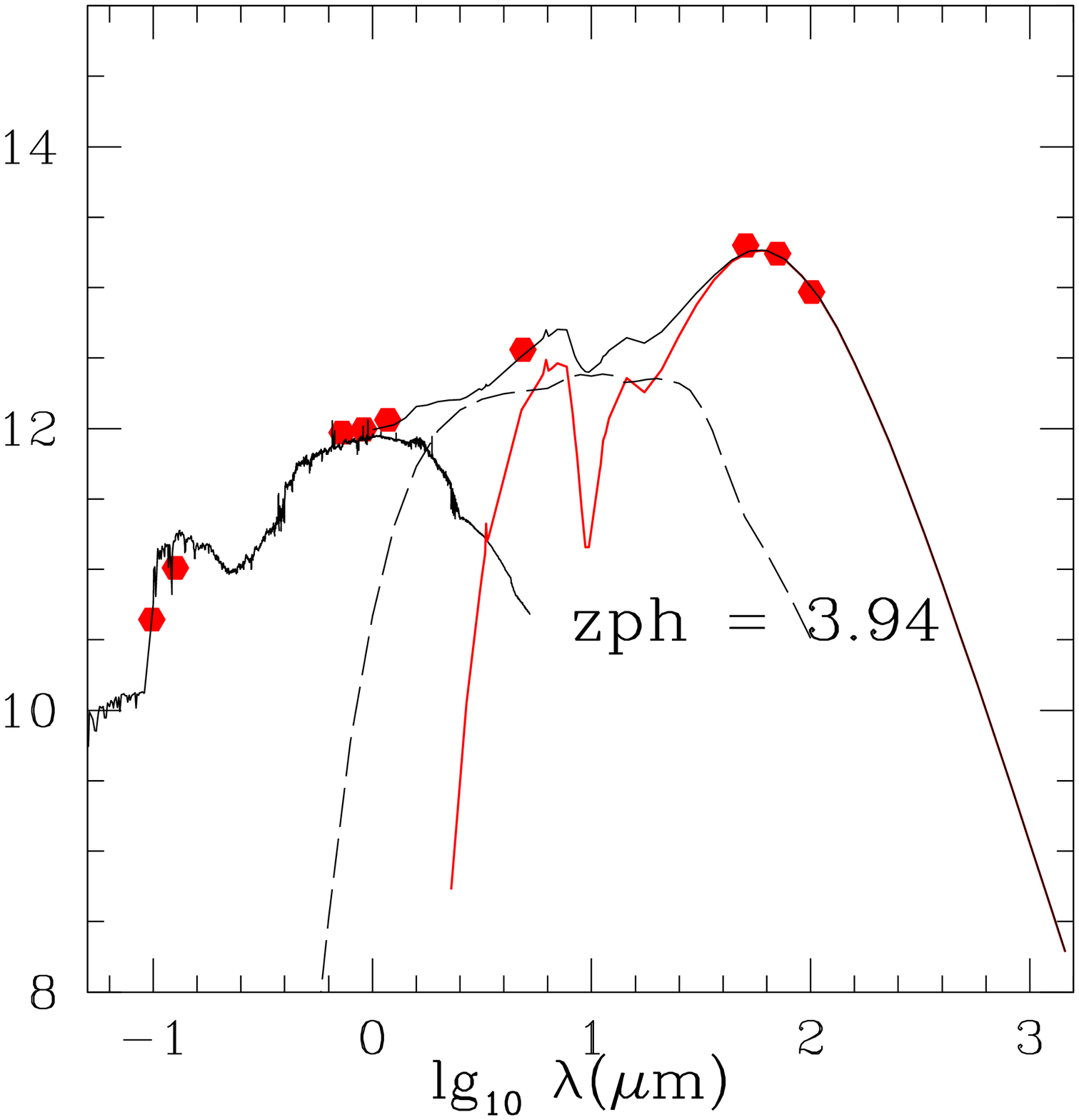}
\includegraphics[width=4.0cm]{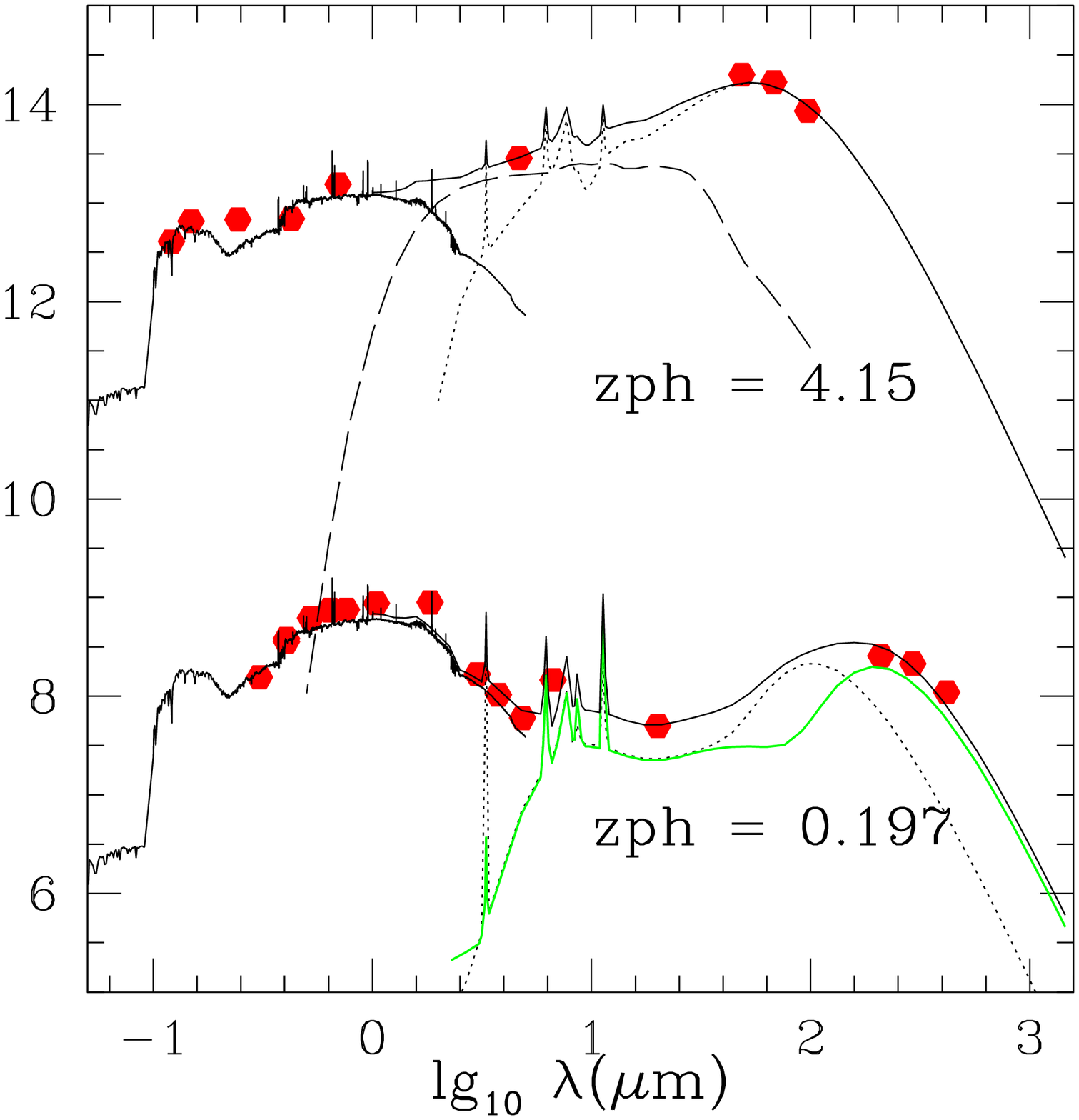}
\includegraphics[width=4.0cm]{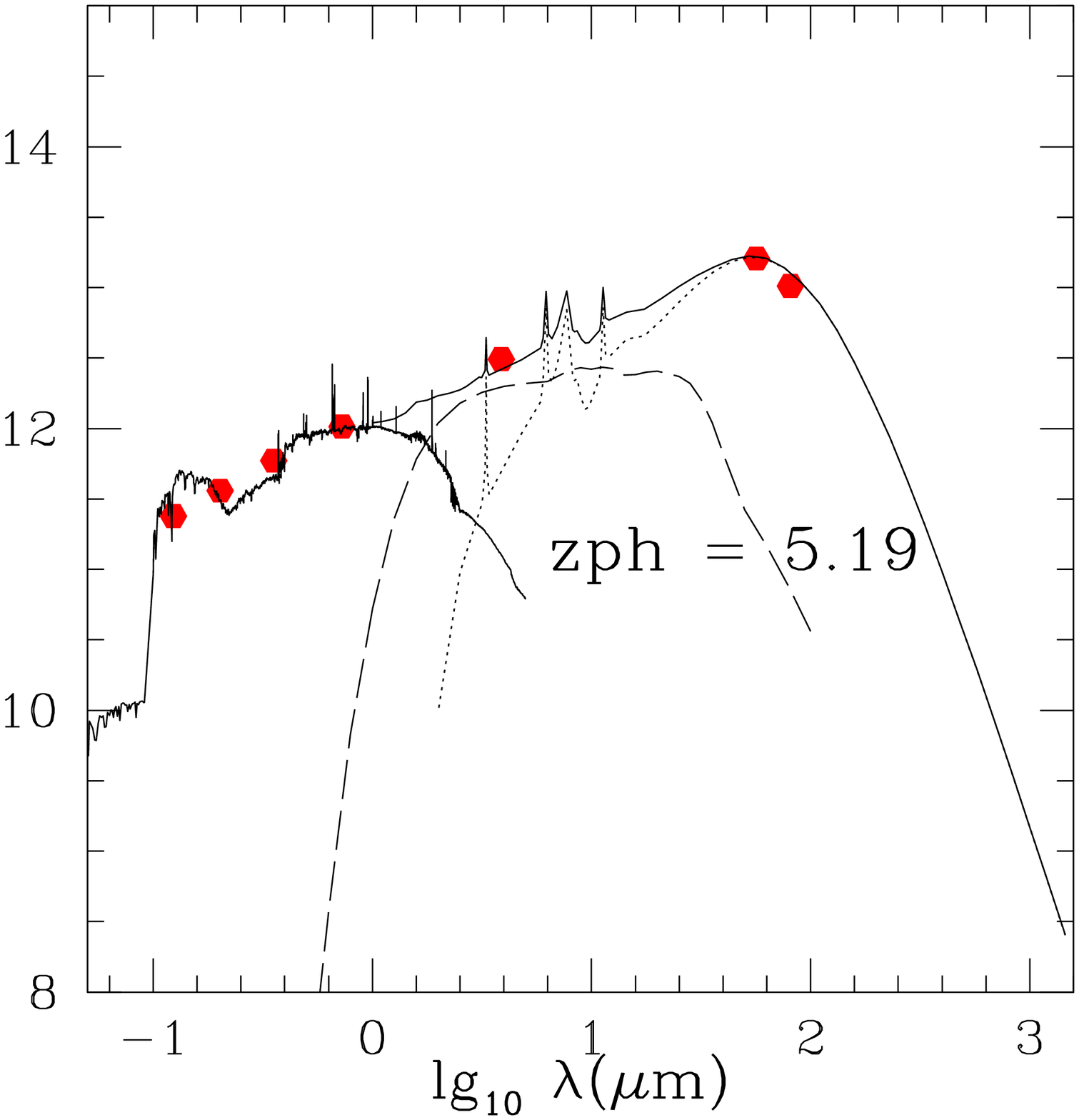}
\caption{
Rest-frame SEDs of {it Herschel}-SWIRE 500 $\mu$m sources with extreme starburst luminosities, labelled with the redshift, whose optical through near-infrared SEDs are best-fitted by a galaxy template, but whose mid-infrared SEDs require an AGN dust torus template.
Red loci: young starburst template, other details as in Figure 6.
}
\end{figure*}

\begin{figure*}
\includegraphics[width=4.0cm]{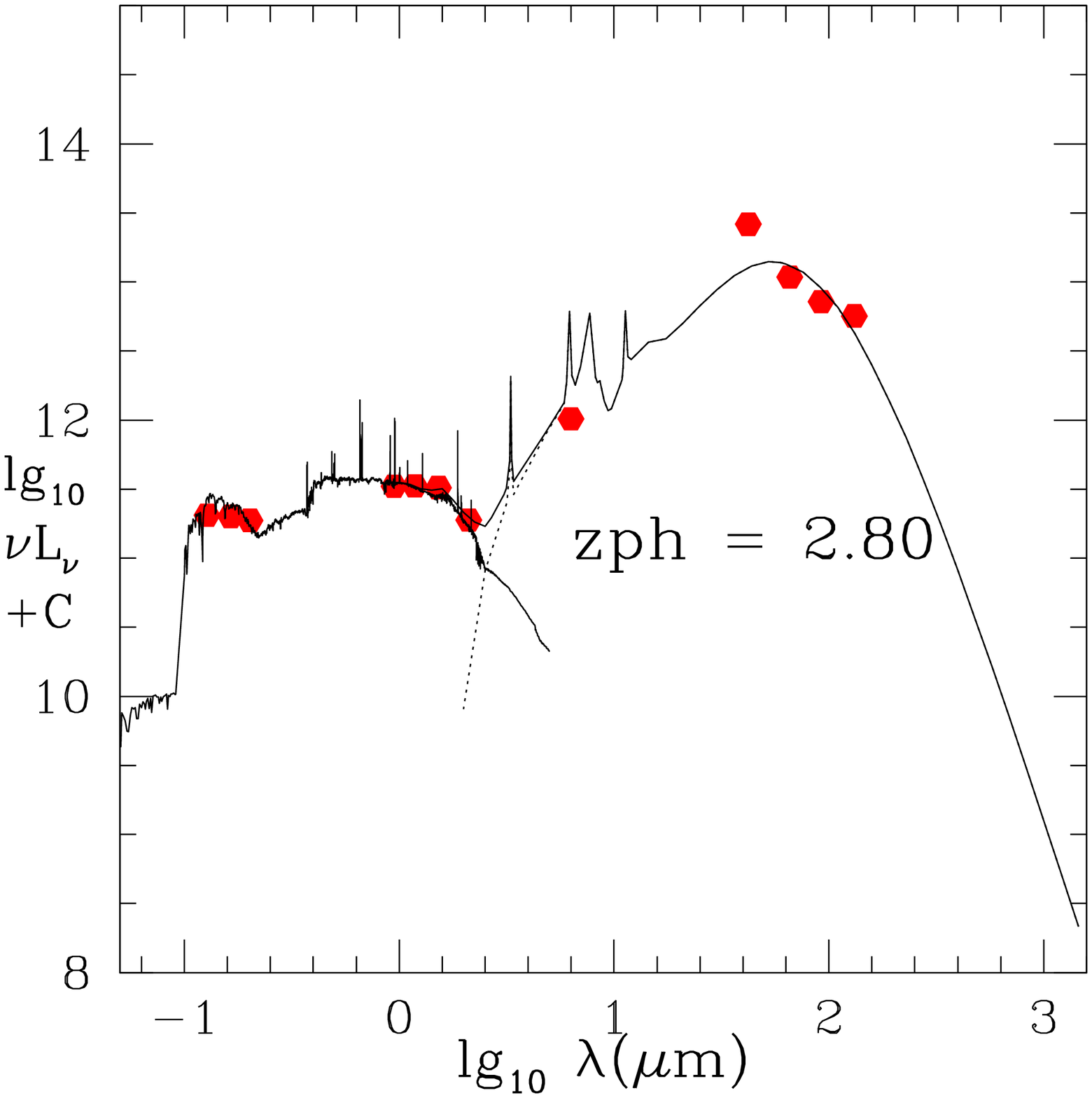}
\includegraphics[width=4.0cm]{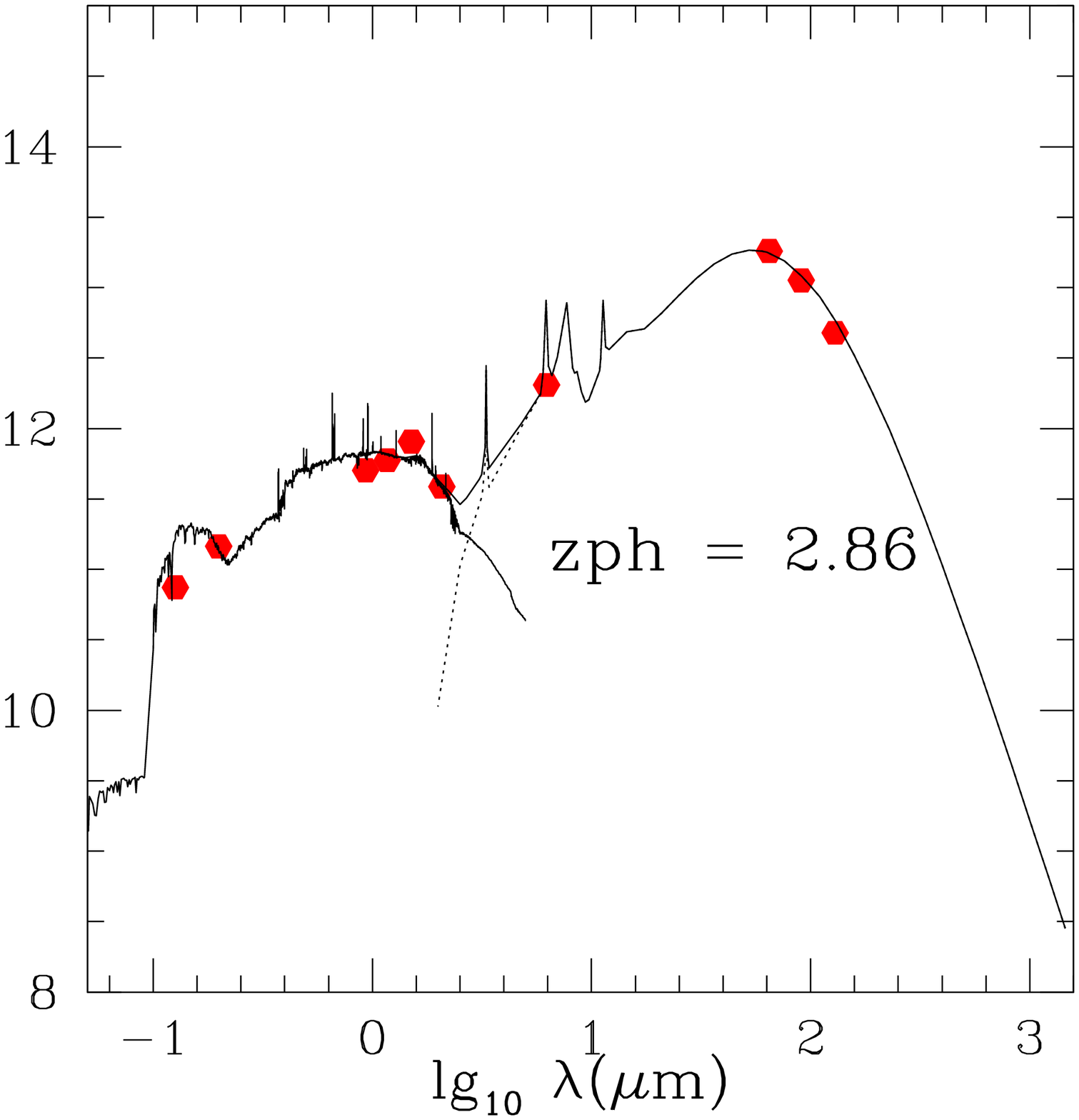}
\includegraphics[width=4.0cm]{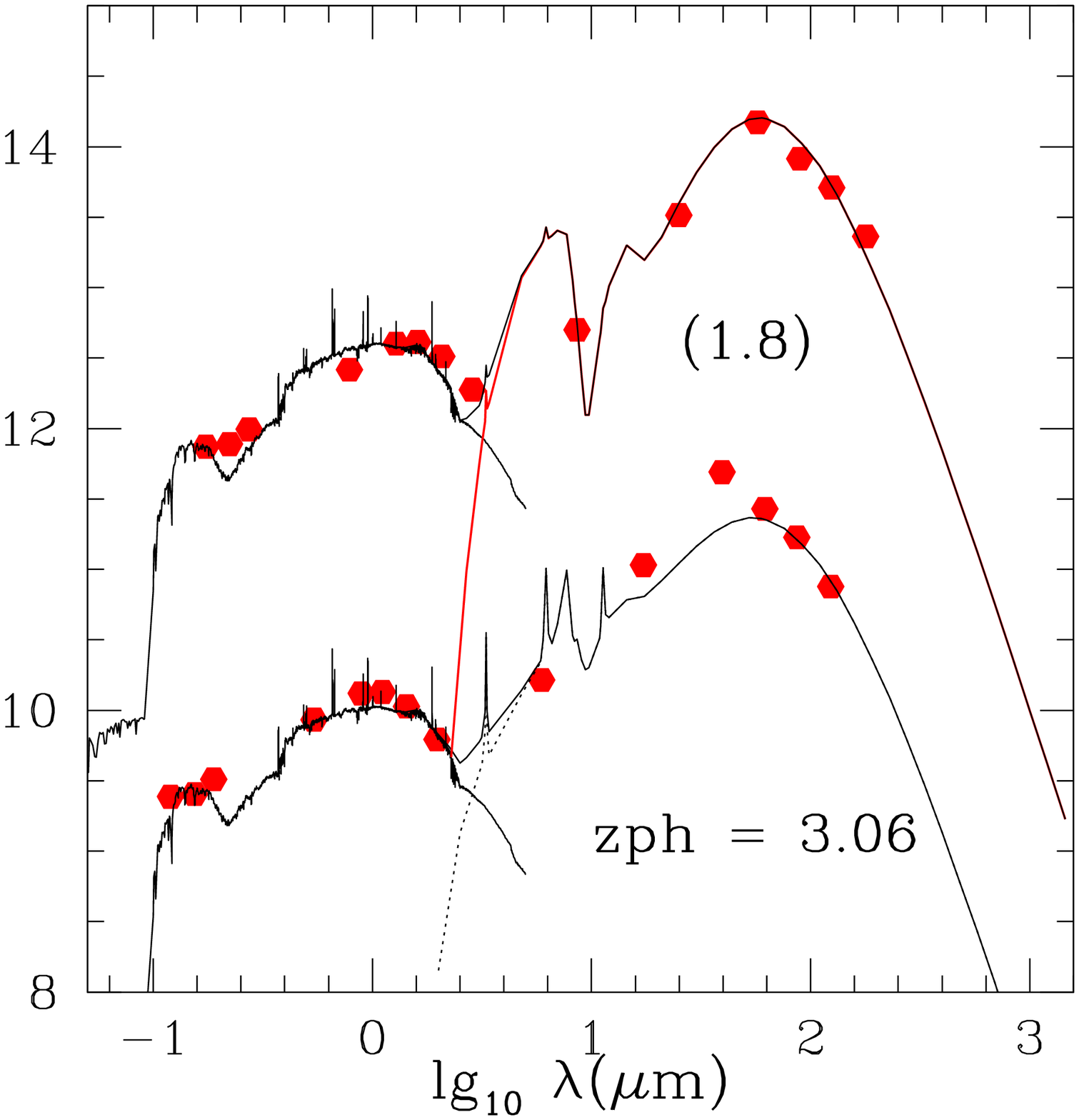}
\includegraphics[width=4.0cm]{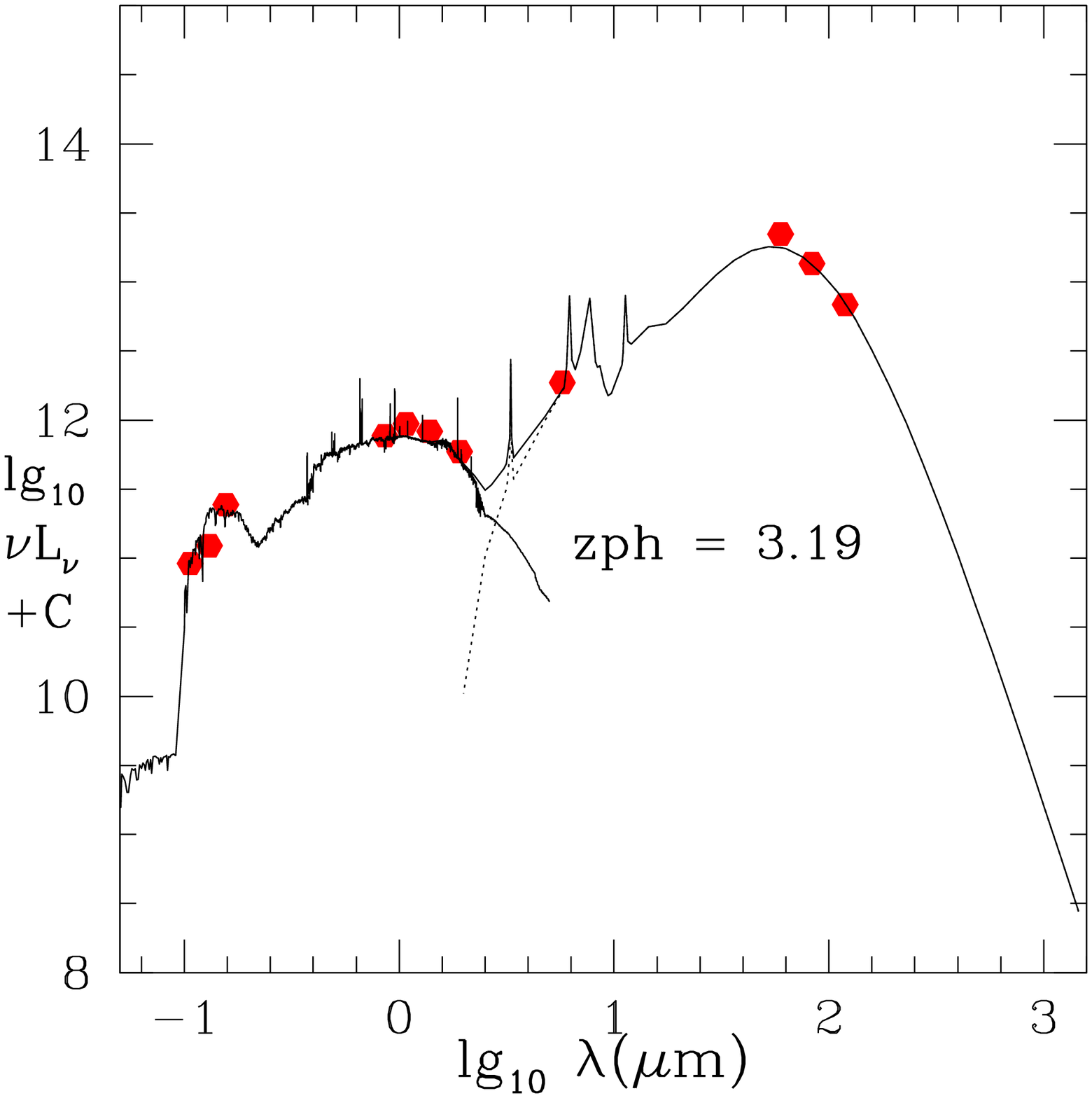}
\includegraphics[width=4.0cm]{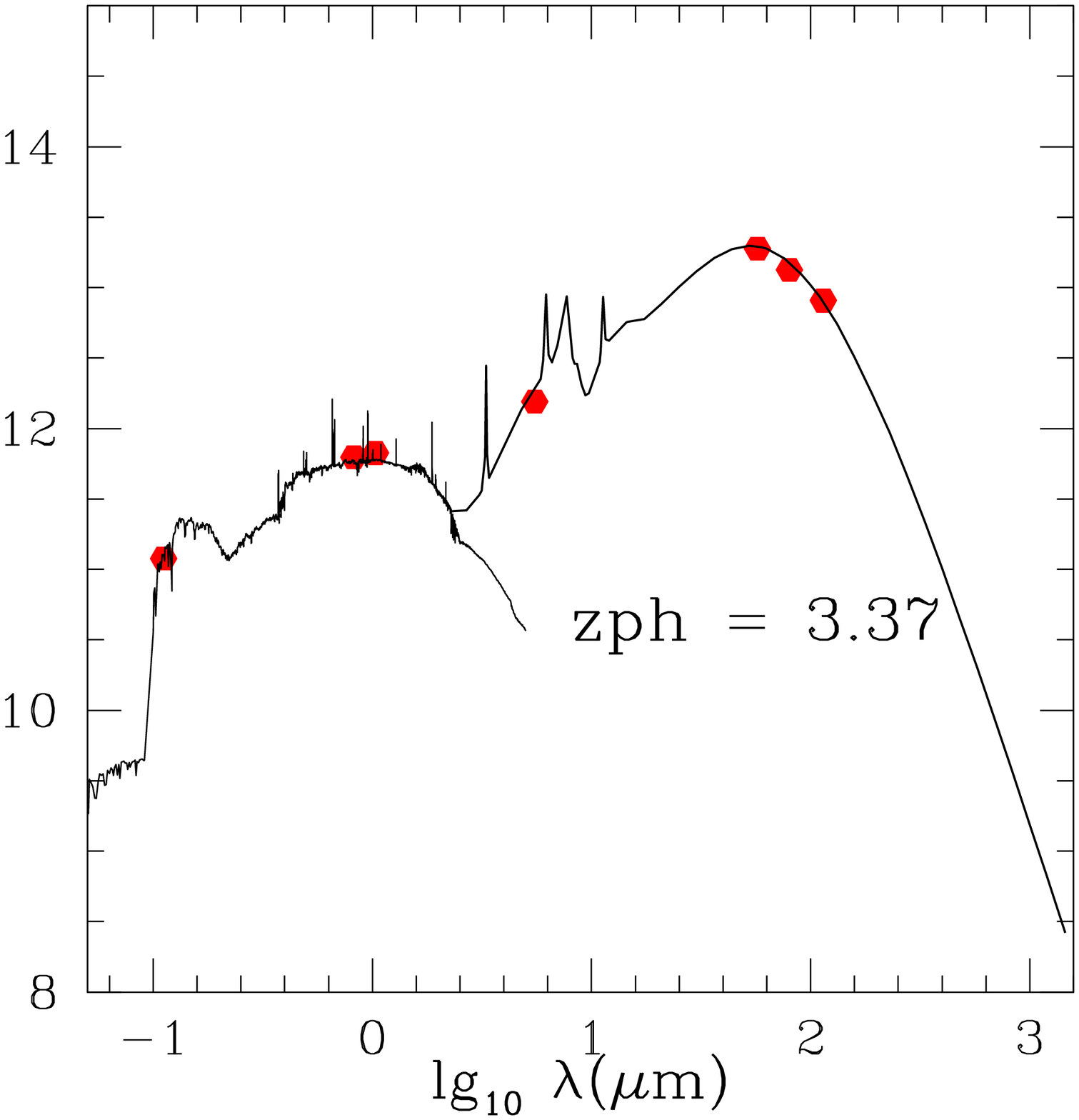}
\includegraphics[width=4.0cm]{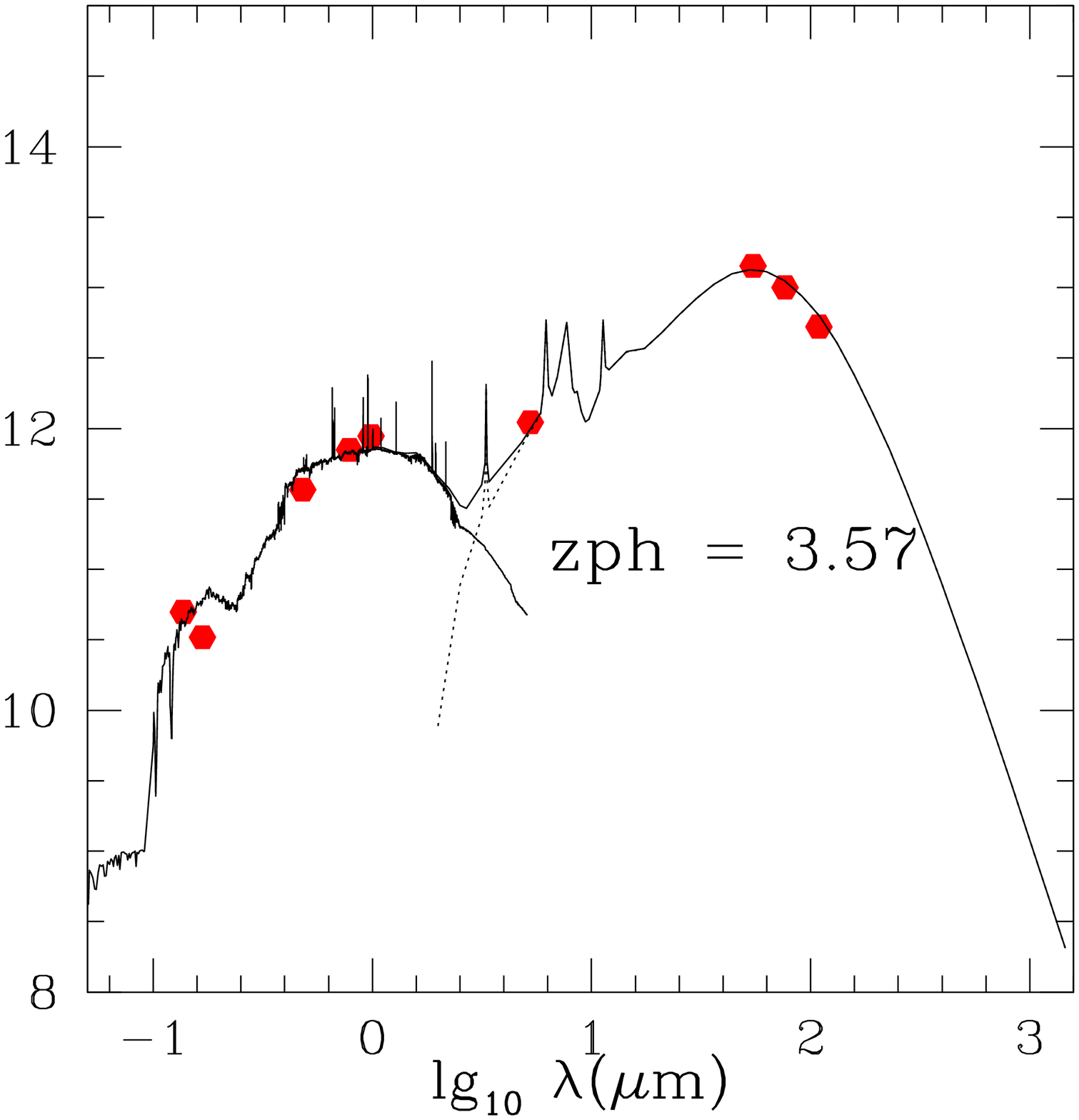}
\includegraphics[width=4.0cm]{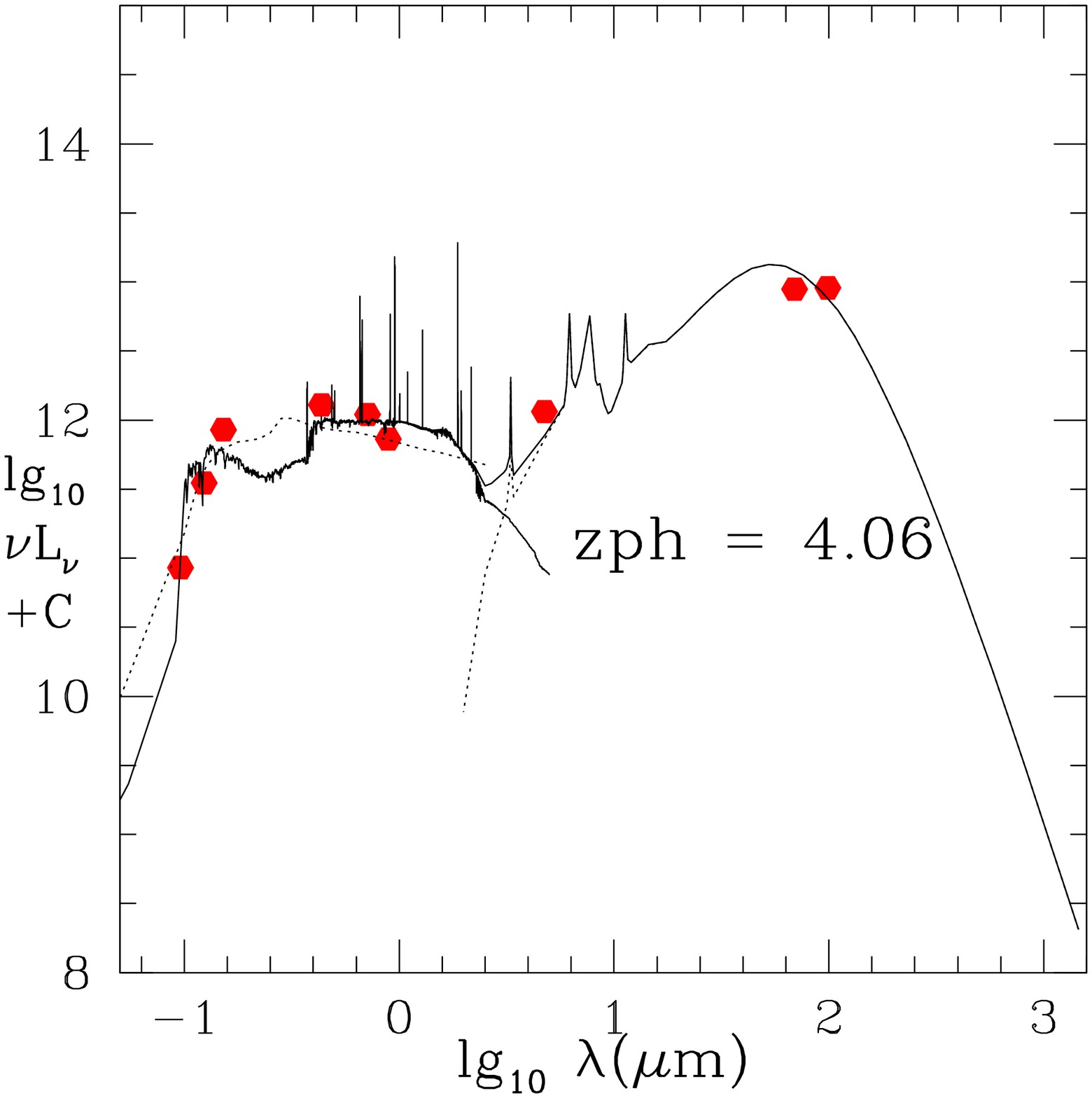}
\includegraphics[width=4.0cm]{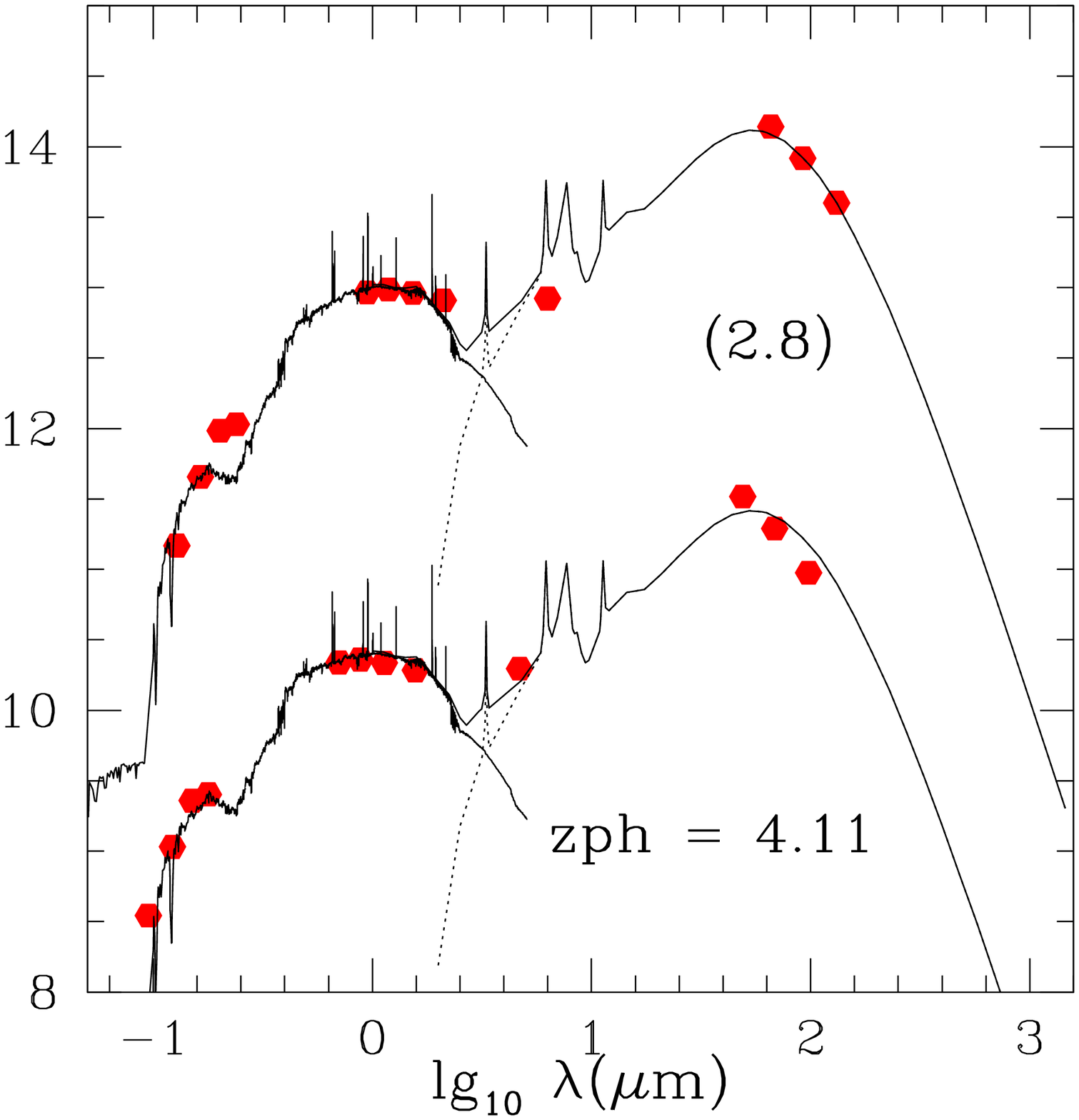}
\includegraphics[width=4.0cm]{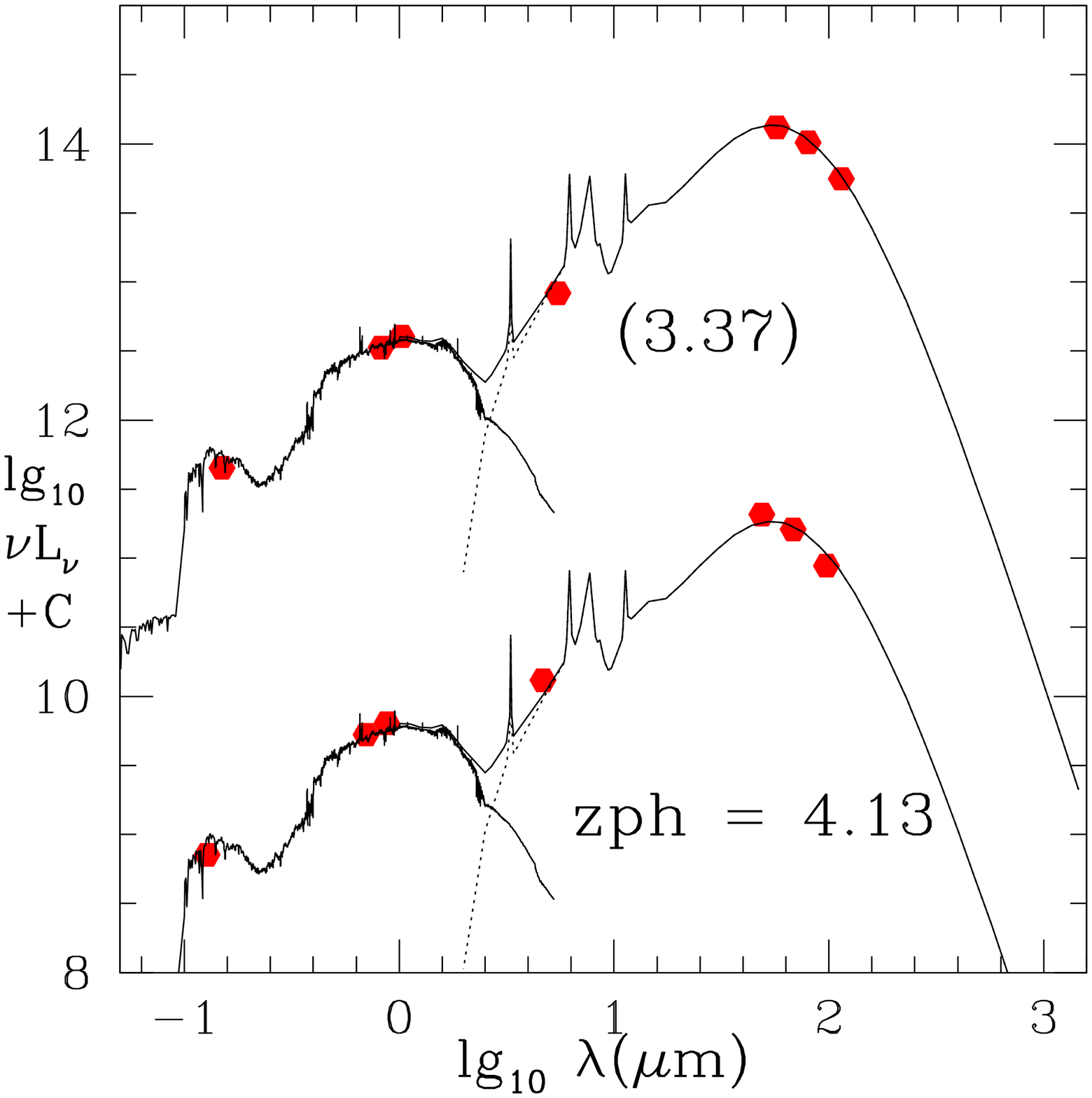}
\includegraphics[width=4.0cm]{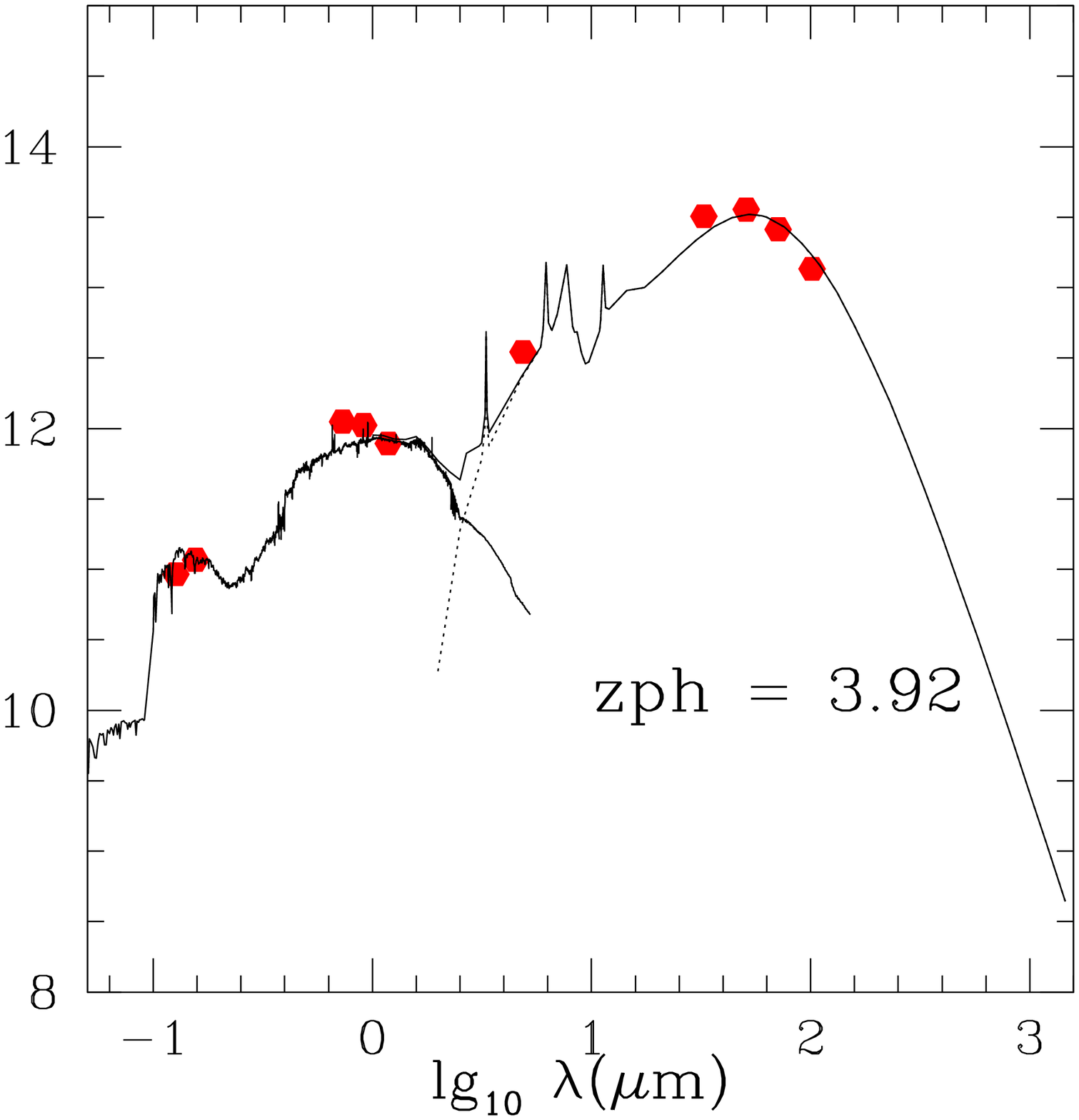}
\includegraphics[width=4.0cm]{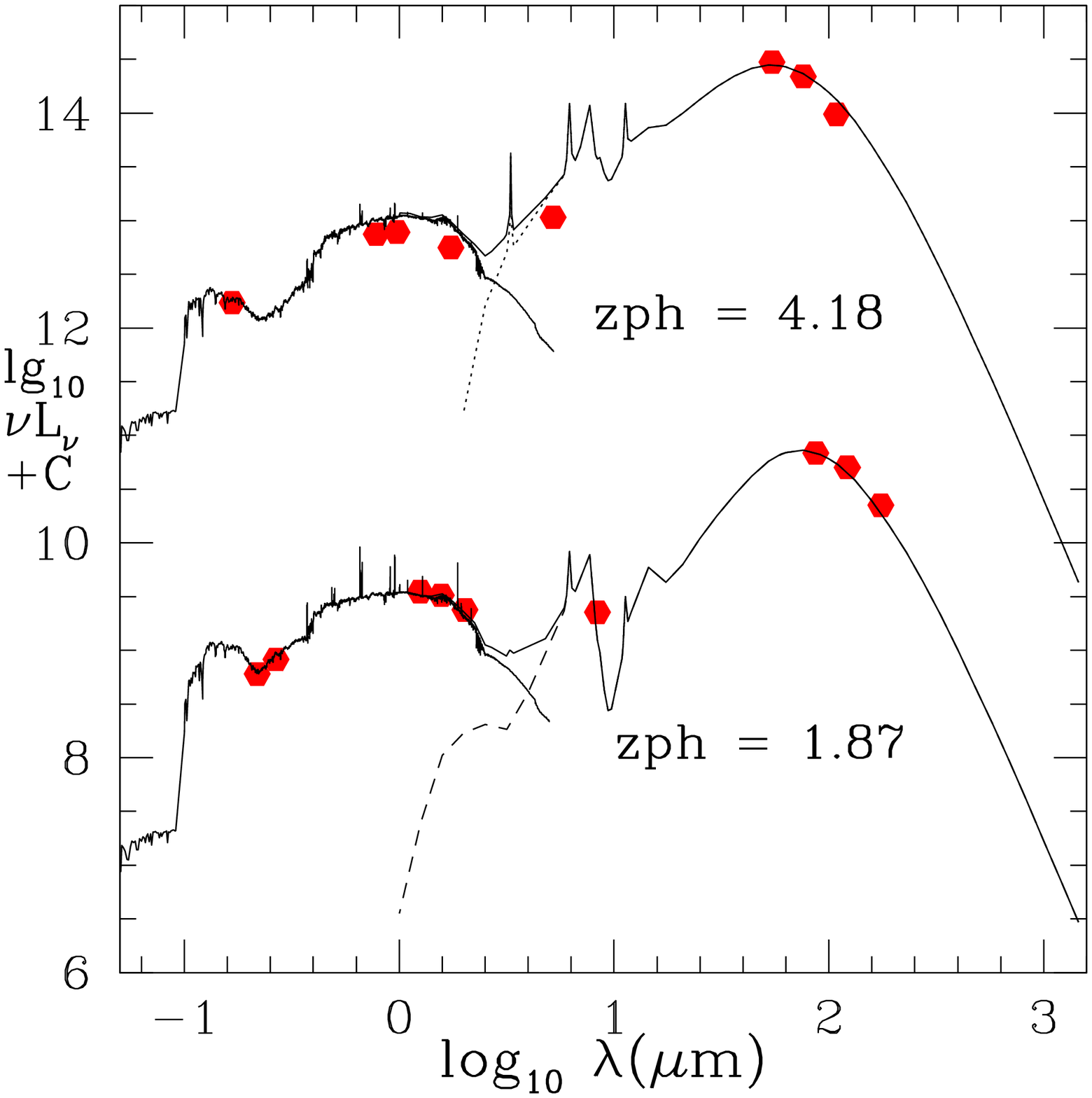}
\includegraphics[width=4.0cm]{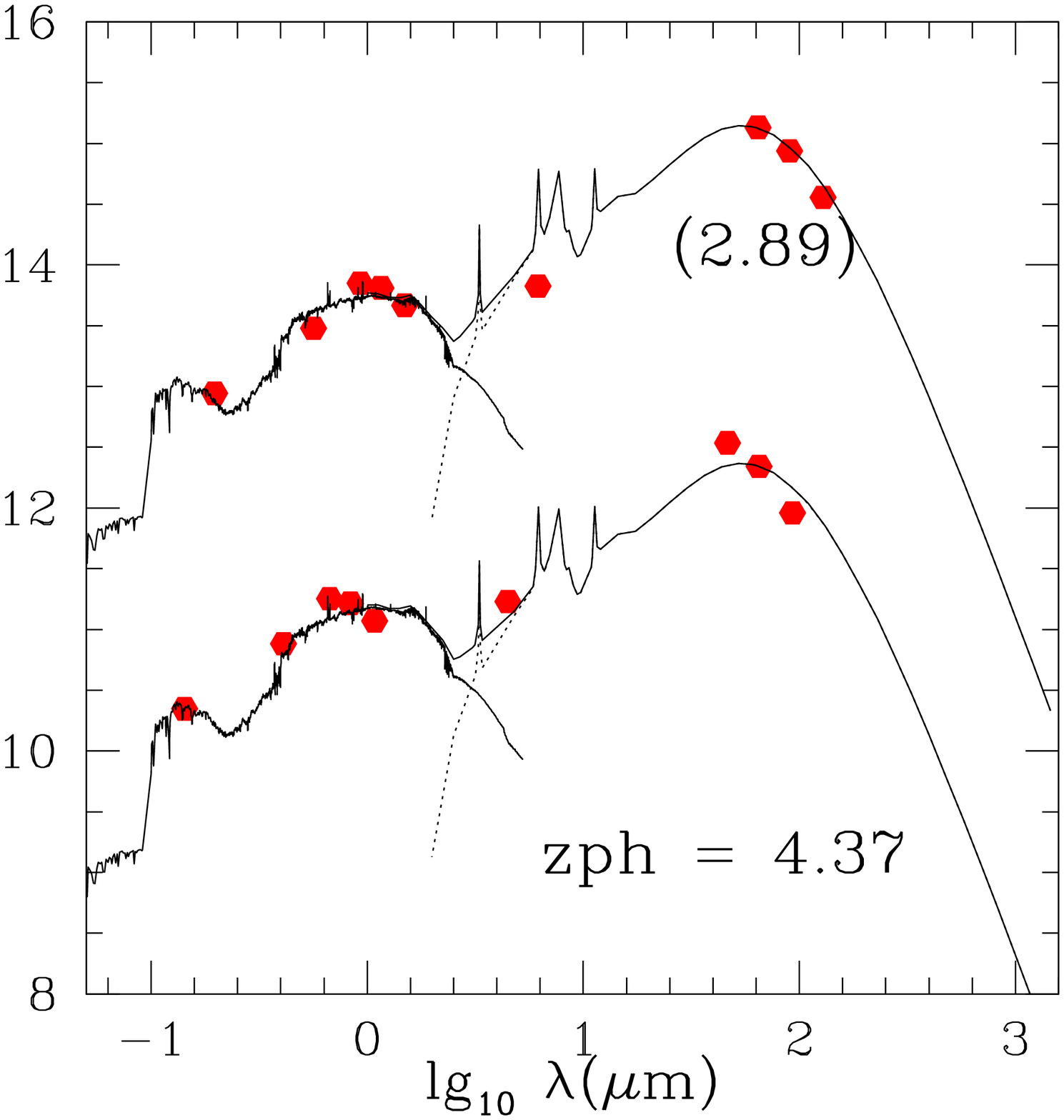}
\includegraphics[width=4.0cm]{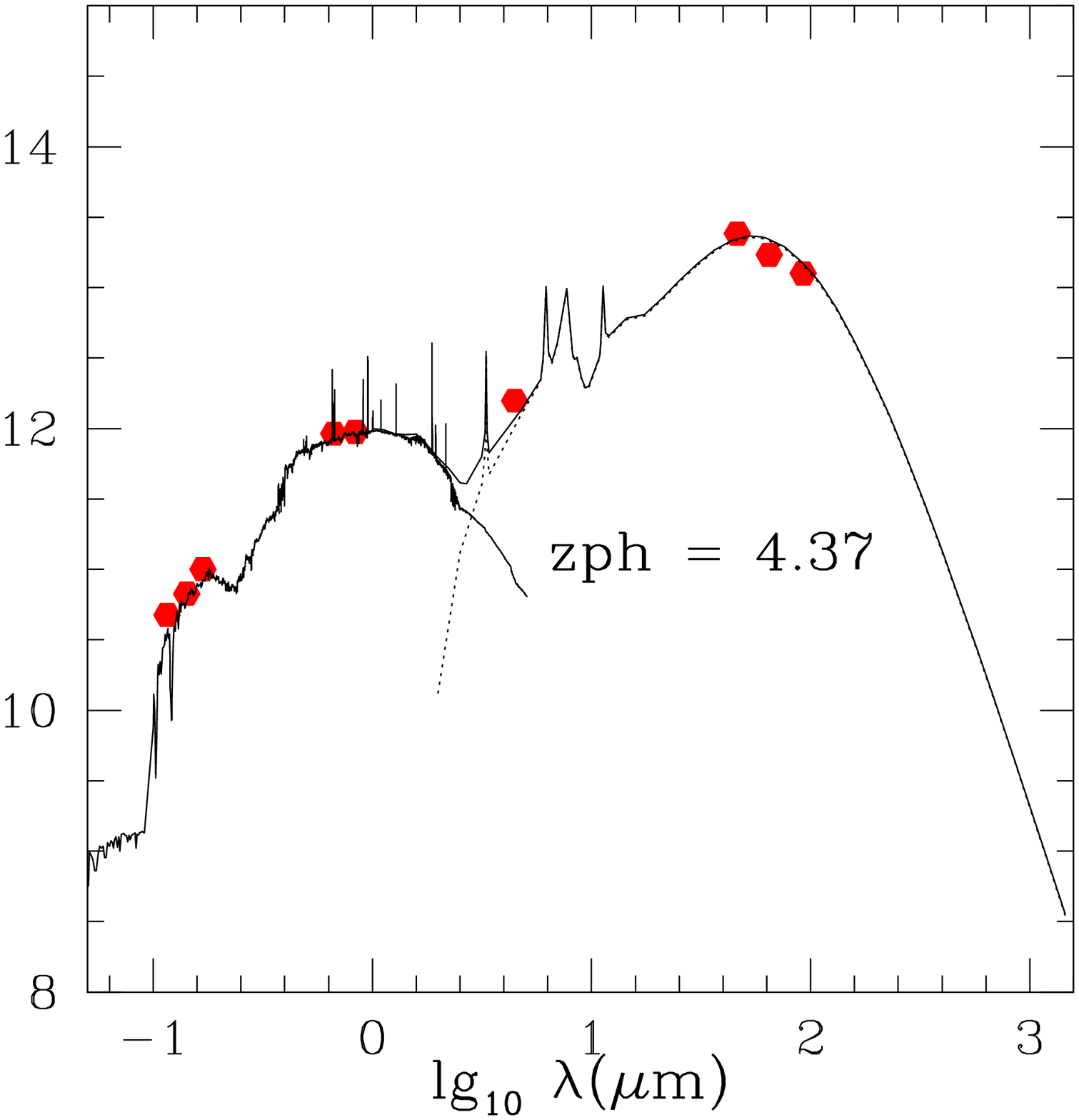}
\includegraphics[width=4.0cm]{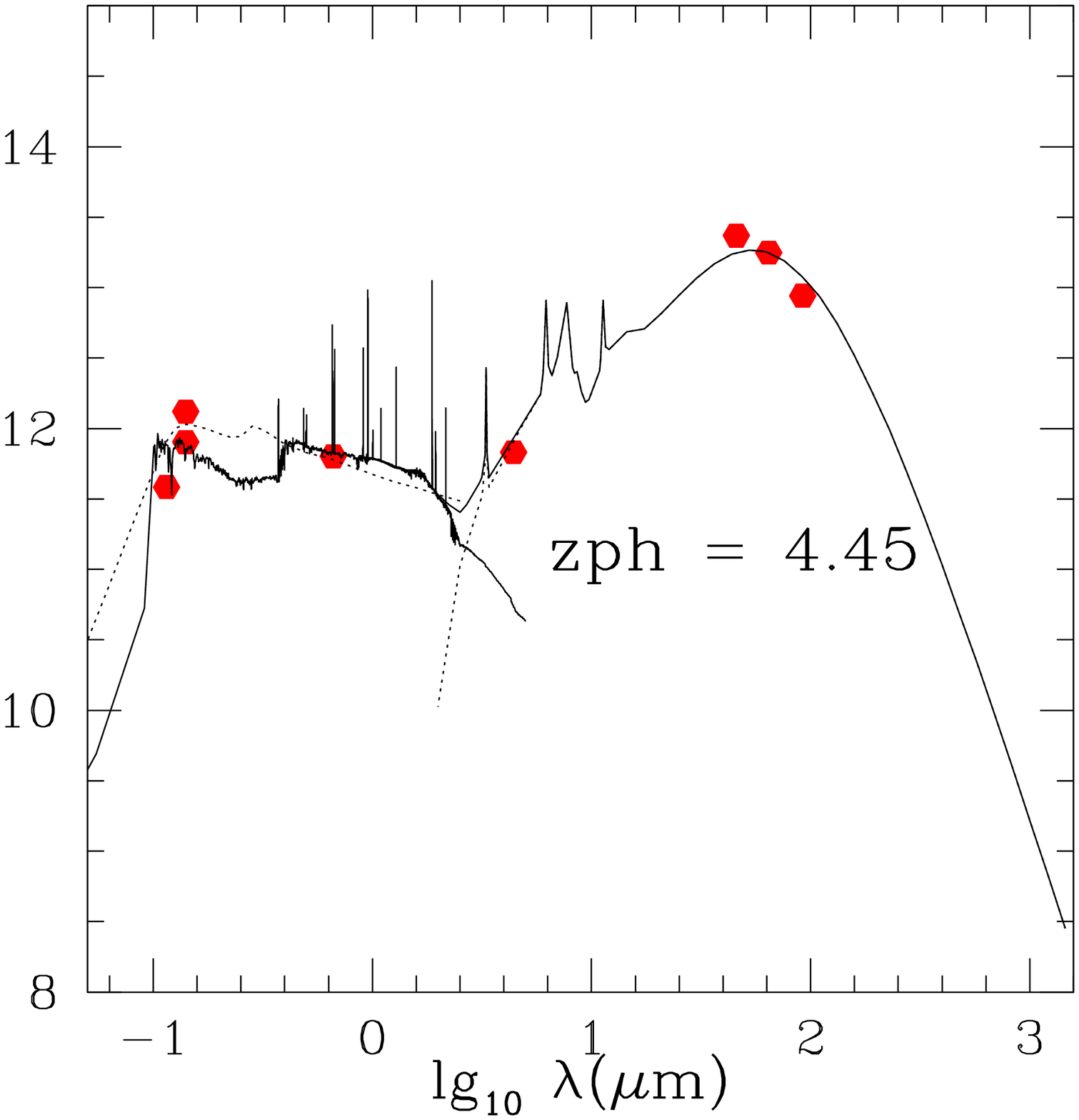}
\caption{
The rest-frame SEDs of Herschel-SWIRE 500\,$\mu$m sources with extreme starburst luminosities, labelled by redshift, whose optical through near-infrared SEDs are best-fitted by a galaxy template, and whose mid- through far-infrared SEDs are fitted with M82 or Arp220 starburst templates. 
Other details are the same as in Figure 6.
}
\end{figure*}

\begin{figure*}
\includegraphics[width=4.0cm]{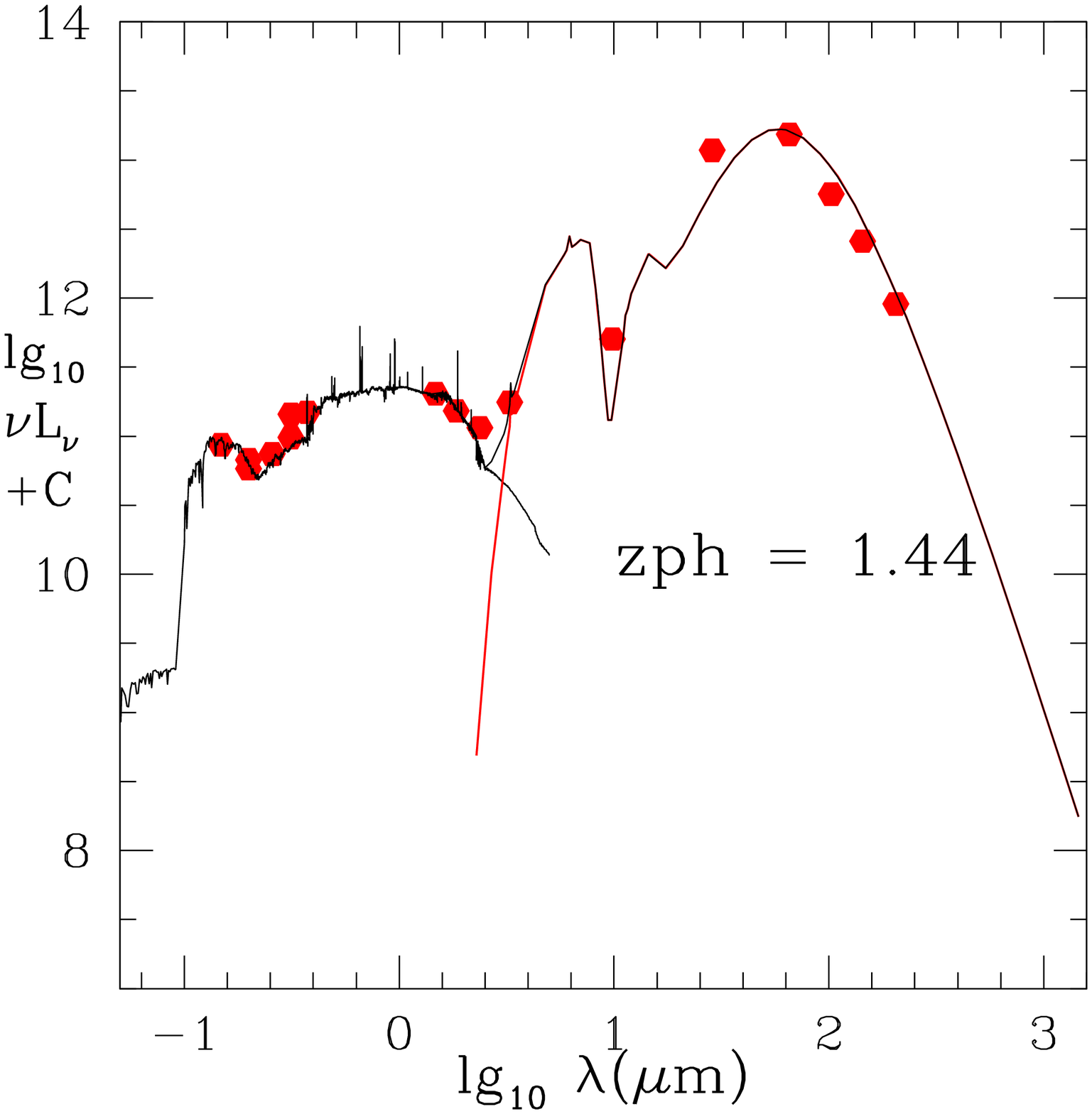}
\includegraphics[width=4.0cm]{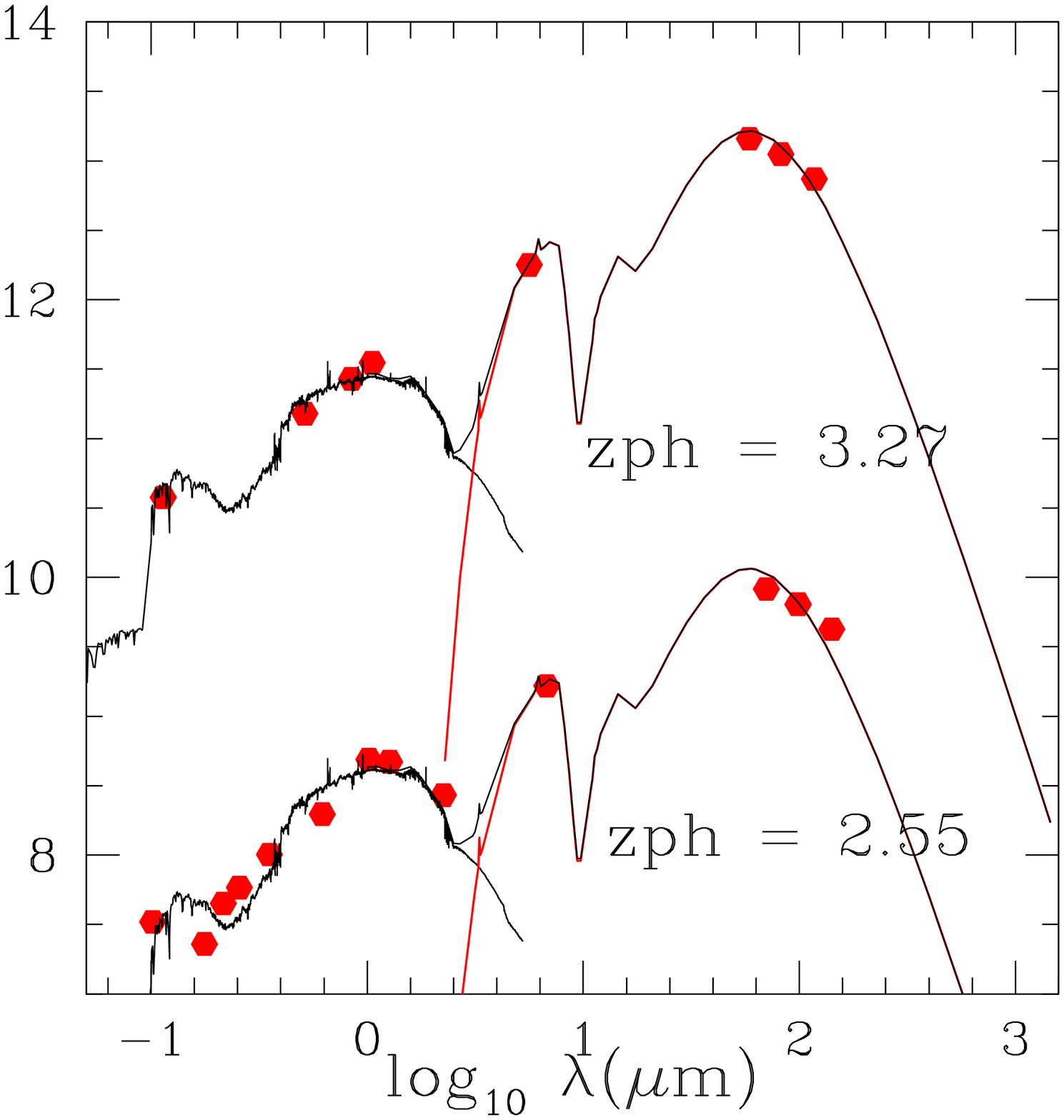}
\includegraphics[width=4.0cm]{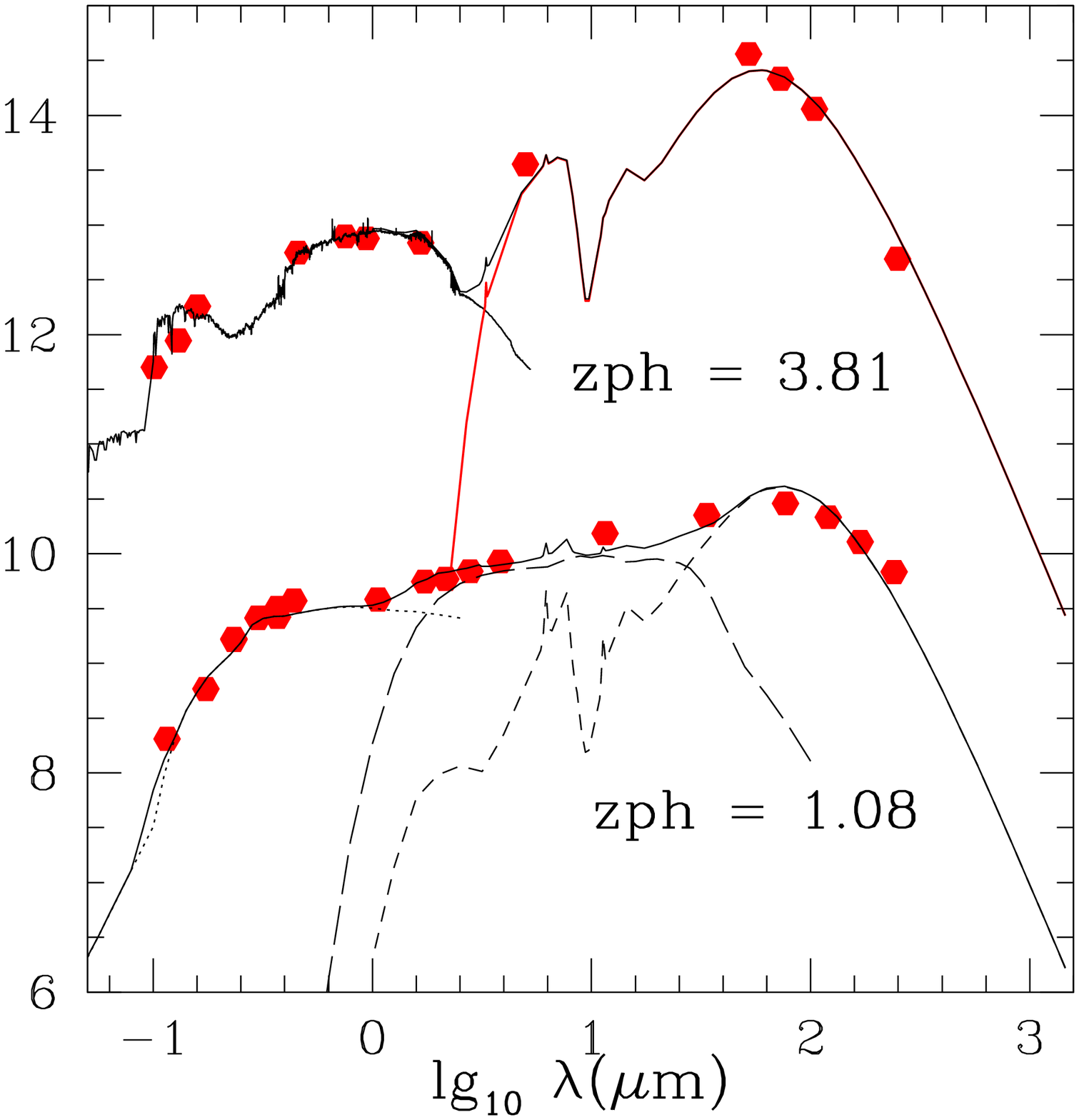}
\includegraphics[width=4.0cm]{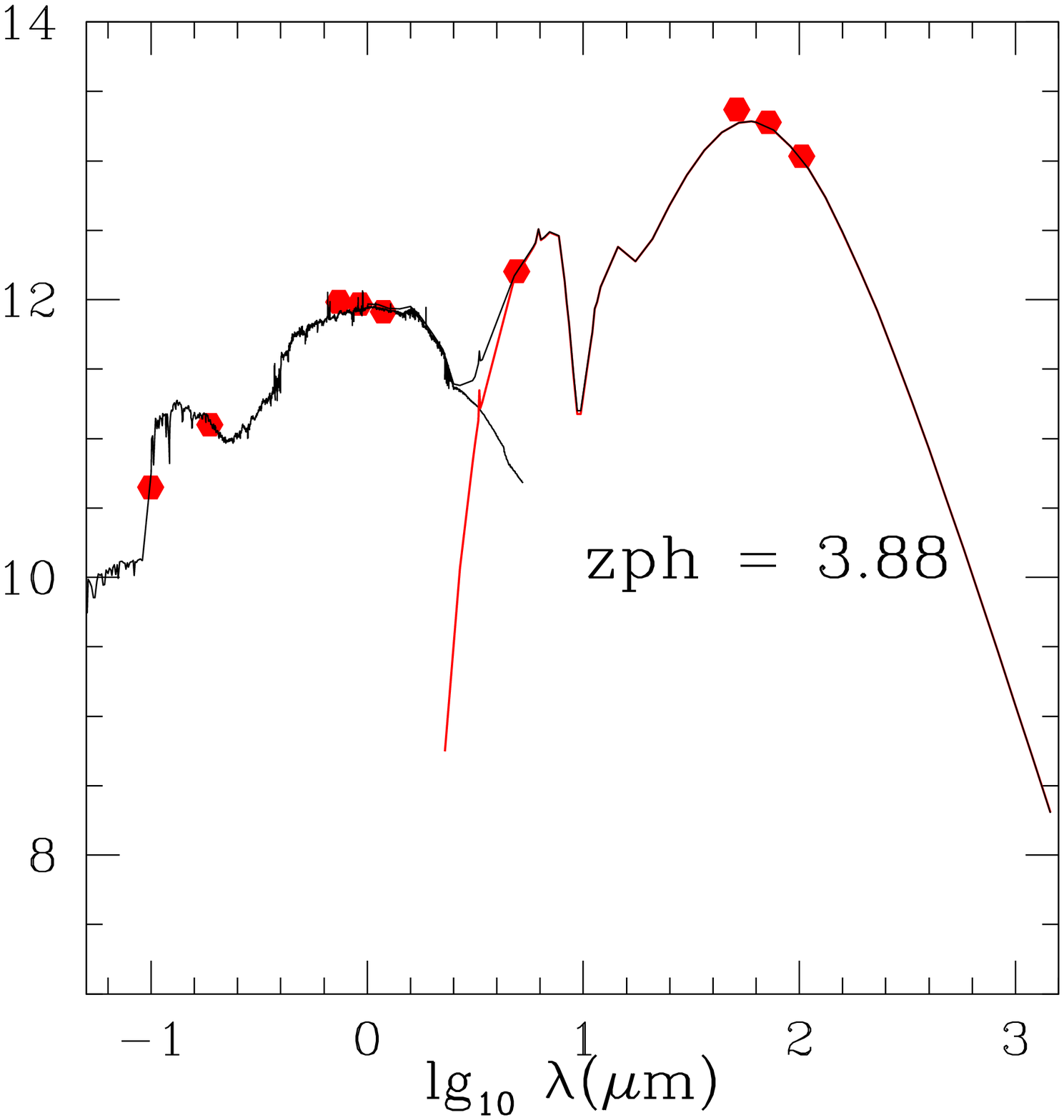}
\includegraphics[width=4.0cm]{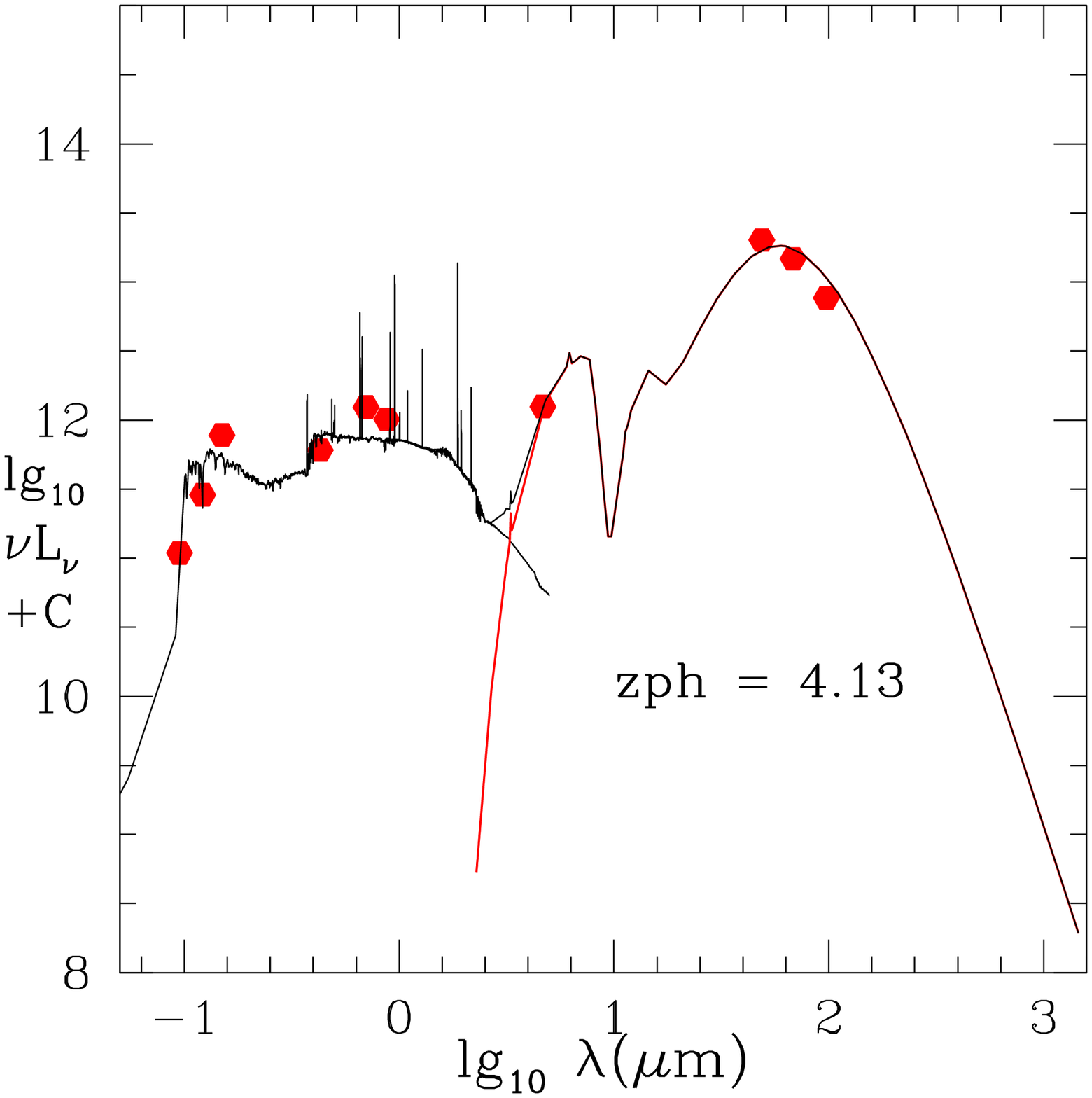}
\caption{
The SEDs of {it Herschel}-SWIRE 500 $\mu$m sources with extreme starburst luminosities, labelled by redshift, whose optical through near-infrared SEDs are best-fitted by a galaxy template, and whose mid- through far-infrared SEDs are fitted with young starburst templates.
Red loci: young starburst template.  Other details are the same as in Figure 6.
}
\end{figure*}

\section{Role of AGN}
A surprisingly high proportion of {\it Herschel} extreme starbursts
have an inferred AGN dust torus component (50$\%$).  The dust tori
are, however, quite weak and in no case does $L_{tor}$ exceed
$L_{sb}$, nor does the dust torus contribute significantly to the
submillimetre emission.  We can use the ratio of luminosity in the
dust torus to that in the bolometric uv-optical-nir luminosity of the
QSO, $L_{tor}/L_{bh}$, as a measure of the covering factor by dust, {\it f},
which is independent of the geometry of the dust, assuming the thermal
uv-optical-nir emission from the accretion disk is radiated
isotropically.  In the case of a toroidal dust distribution, {\it f} would
be a measure of the opening angle of the torus.  Figure 10 shows
$L_{tor}/L_{opt}$ versus redshift for SWIRE QSOs, where $L_{opt}$ is
the 0.1-2$\mu$m luminosity of the QSO.  Assuming the bolometric output
of the black hole, $L_{bh} = 2.0 L_{opt}$ (Rowan-Robinson et al 2009),
the average covering factor, {\it f}, is $\sim$0.4 for z$>$2, declining to
$\sim$0.16 at z = 0.  This trend can also be interpreted as a decline
in dust torus covering factor with declining optical (and bolometric)
luminosity (see Rowan-Robinson et al 2009 and references quoted
therein).

Using this relation, Fig 11L shows black-hole mass,
$M_{bh} \beta^{-1}$, versus total stellar mass, $M_*$, for {\it
  Herschel} galaxies and for {\it IRAS}-FSS galaxies with z $<$ 0.3, where
$M_{bh}$ is estimated from $L_{bh}$ assuming that the AGN is radiating
at a fraction $\beta$ of the Eddington luminosity:
\begin{equation}
L_{bh} = \beta L_{Edd} 
= 4 \pi \beta G M_{bh} m_p c/\sigma_T$ = 3.2 x $10^4 \beta (M_{bh}/M_{\odot})  (L_{\odot})  
\end{equation}
$L_{bh}$ is estimated as 2.0 $L_{opt}$ for QSOs, and from $L_{tor}/f$
for galaxies with AGN dust tori.  A wide range of values of the
Eddington ratio $\beta$ is found in the literature (Babic et al 2007,
Fabian et al 2008, Steinhardt and Elvis 2009, Schutze and Wistotzki
2010, Suh et al 2015, Pitchford et al 2016, Harris et al 2016), with a
typical range of 0.01-1 for z $>$ 1 (Kelly et al 2010, Lusso et al
2012).  Since QSOs are excluded from Fig 11L by the requirement for a
measurement of stellar mass, these are all Type 2 AGN. The mean value
of $lg_{10} L_{bh}/(\beta M_*)$ for 500 {\it HerMES}-SWIRE AGN is -4.11,
with an rms dispersion of 0.56. There will be a contribution to this
rms from the dispersion in values of the covering factor f.  The
distribution of AGN in Fig 11L is broadly similar to the equivalent
plot by Reines and Volonteri (2015) for broad-line AGN, though we have
a higher proportion of high mass galaxies and we do not have objects
corresponding to their elliptical and S0 galaxies.

Figure 11R shows $M_{bh} \beta^{-1}/M_*$ versus redshift for the same
galaxies.  If we take $\beta \sim$ 0.1 as a characteristic value, then
$M_{bh}/M_* \sim 0.001$ at all redshifts, with a range of $\sim\pm$ 1
dex.  This is reminiscent of the Magorrian et al (1998) relation
between black-hole mass and bulge mass (see also review by Kormendy
and Ho 2013). This ratio is set by the very high star-formation (and
black-hole build-up) at redshift 2-5.  The Milky Way, with
$M_* = 6\times10^{10} M_{\odot}$ and $M_{bh} = 4\times10^6 M_{\odot}$
lies on the lower end of this range.

\begin{figure}
\includegraphics[width=9cm]{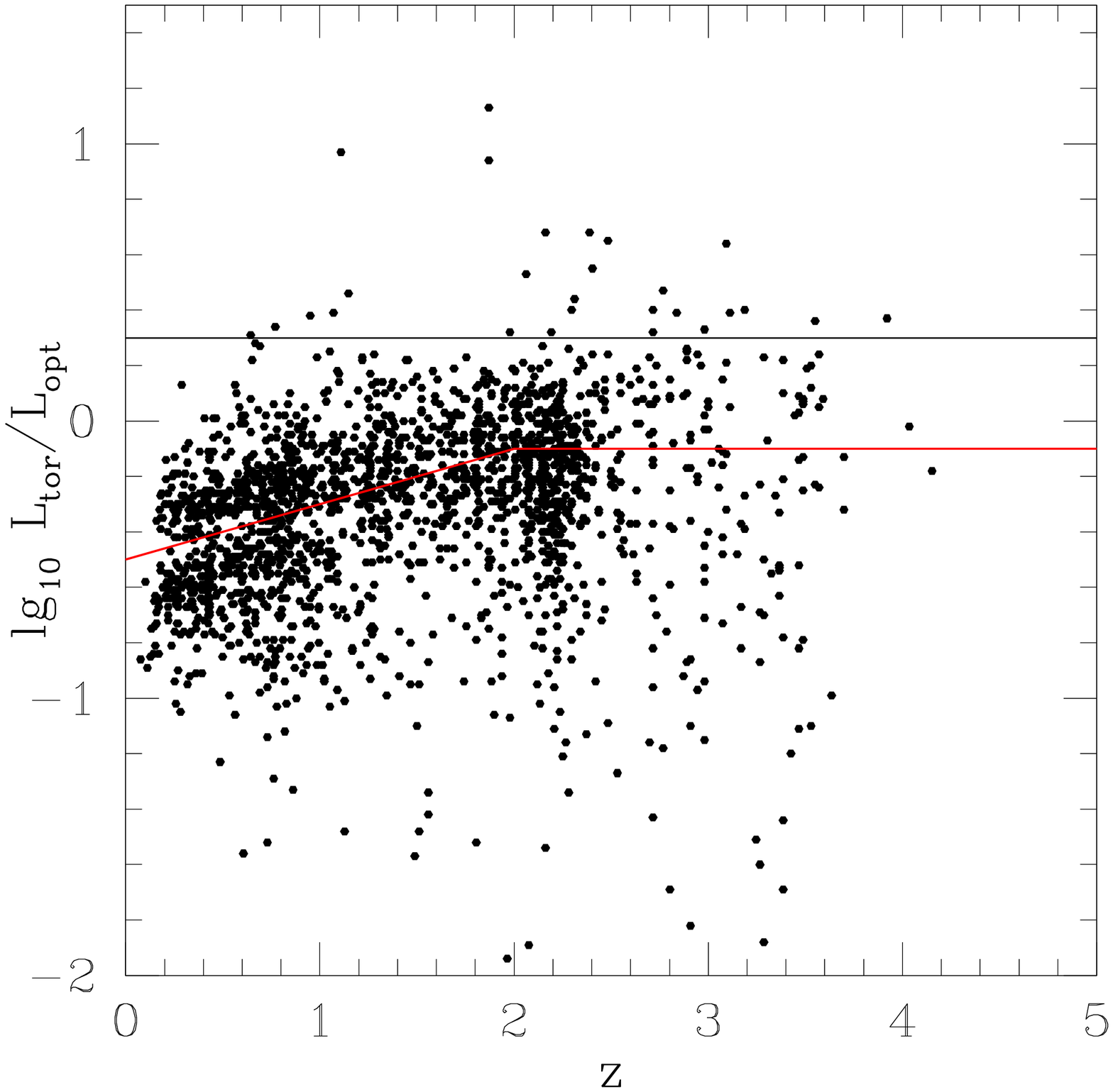}
\caption{
The behaviour of the torus covering factor ($L_{tor}/L_{opt}$) as a
function of redshift for the {\it HerMES} extreme starbursts with QSO-like
optical SEDs. The solid black line corresponds to $L_{tor} = L_{bh}$
with an assumed optical bolometric correction of 2.0 (Rowan-Robinson
et al 2009). The red line shows a relation that approximately
reproduces the trend seen among the plotted population.  }
\end{figure}

\begin{figure*}
\includegraphics[width=8cm]{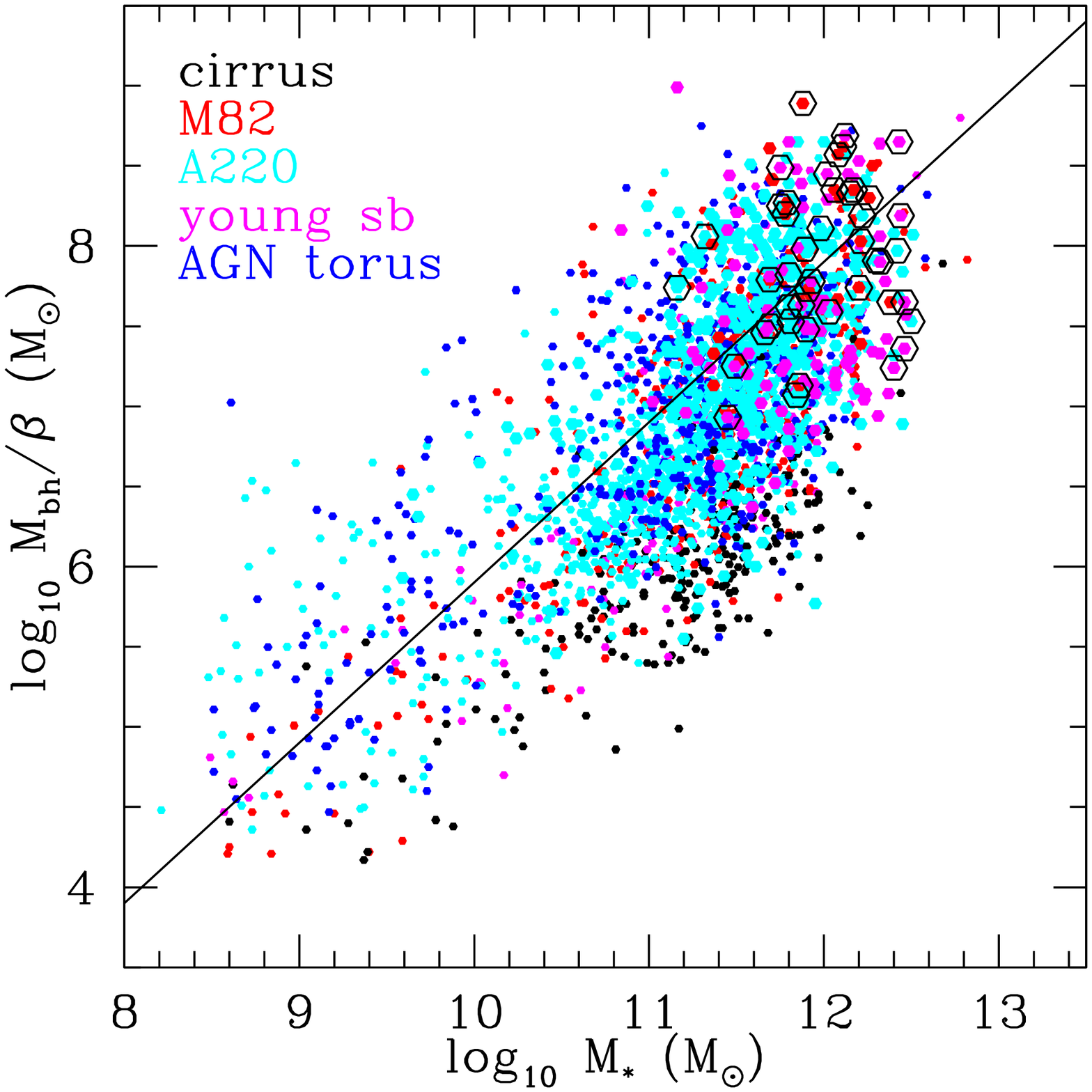}
\includegraphics[width=8cm]{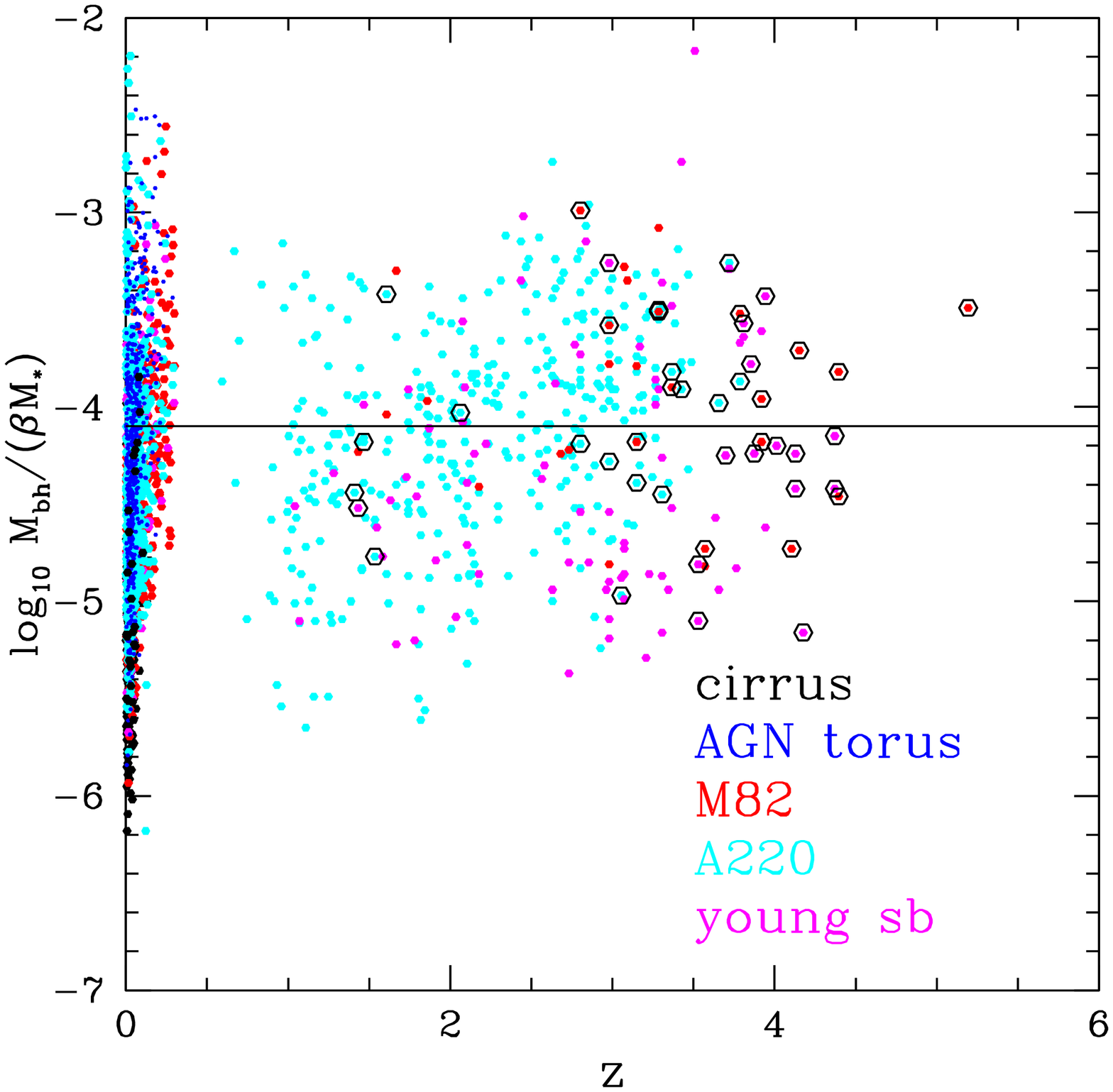}
\caption{L: Black-hole mass, $M_{bh}\beta^{-1} (M_{\odot})$, versus total stellar mass, $M_* (M_{\odot})$, for {\it HerMES}-SWIRE sources and for RIFSCz sources with z$<$0.3 (smaller points), with AGN dust tori. 
QSOs are excluded by the requirement that there be a stellar mass estimate, so these are all Type 2 AGN.  Circled points are {\it Herschel} extreme starbursts.
R: $M_{bh}/(\beta M_*)$ versus z for {\it HerMES}-SWIRE sources and for RIFSCz sources with z$<$0.3, with AGN dust tori.
}
\end{figure*}

\begin{figure}
\includegraphics[width=9cm]{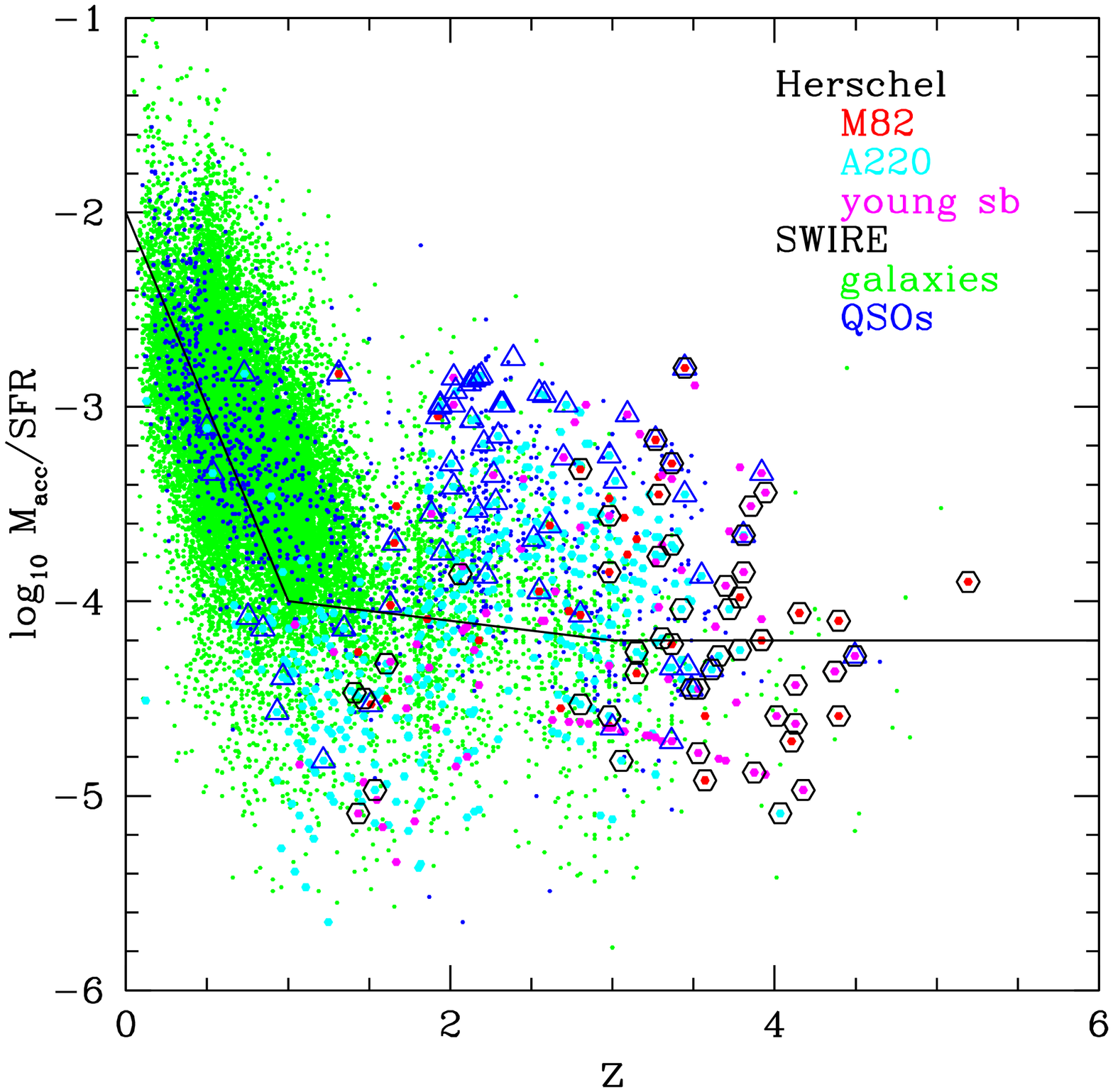}
\caption{
$log_{10} M_{acc}/SFR$ versus z for SWIRE (small green and blue dots) and {\it HerMES} sources (larger red cyan and magenta dots). 
 Circled points are extreme starbursts and blue open triangles denote {\it HerMES} QSOs.
}
\end{figure}

Figure 12 shows $M_{acc}/SFR$ versus redshift for {\it Herschel}
galaxies and for (non-{\it Herschel}) SWIRE galaxies (smaller
symbols), where the black-hole accretion rate $M_{acc}$ is calculated
assuming conversion efficiency of accreting mass to radiation is
$\epsilon$=0.1:
\begin{equation}
L_{bh} = \epsilon {M_{acc}} c^2
\end{equation}

Note that the combination of equations (6) and (7) gives the Salpeter
time-scale for black hole growth $t_S = 4.10^8 \epsilon \beta^{-1}$
yrs (Salpeter 1964). QSOs have been indicated in Fig 11 by open blue
triangles.

Figure 12 shows that $M_{acc}/SFR$ is $\sim 10^{-4}$ at z = 2-5, but
that this ratio has increased by a factor of 30 by z $<$0.5.  The
star-formation rates in $z<0.5$ galaxies are 1000 times lower than
those seen in the extreme starbursts, but the black hole accretion
rates are only 30 times lower. This is consistent with source count
models that find shallower evolution for AGN compared to that for
starbursts (e.g. Rowan-Robinson 2009).
A recent apparent exception to this has been presented by Barnett et
al (2015), who quote a much higher value of $M_{acc}/SFR \sim$ 0.2 for
a redshift 7.1 QSO, based on a SFR derived from the CII 158 $\mu$m
line.  However they also quote a bolometric luminosity of
$6.7\times10^{13} L_{\odot}$, which could yield a SFR of $\sim$ 13,000
$M_{\odot}/yr$, about 100 times their estimate from CII.  This would
move $M_{acc}/SFR$ into the range seen in Fig 12.

Figure 12 shows that there is an intimate and evolving connection
between black hole accretion and star formation.  A plausible
interpretation of this result is as follows. In these high redshift,
high luminosity submillimetre galaxies we are seeing major mergers
(Chakrabarti et al 2008, Hopkins et al 2010, Haywards et al 2011,
Ivison et al 2012, Aguirre et al 2013, Wiklind et al 2014, Chen et al
2015), in which the star formation is taking place close to ($<\,$ 1
kpc) the galactic nucleus, so it is not surprising that there is a
strong connection between star-formation and black-hole growth.
However at recent epochs ($z < 1$) star-formation is mainly fed by accretion
from the cosmic web, by minor mergers and interactions, and by spiral
density waves, so is taking place further from the galactic nucleus.
This uncouples the direct connection between star-formation and
black-hole growth.  The gas feeding the black hole is fed to the
galactic nucleus more gradually and may include gas fed by mass-loss
from stars.  It is however still surprising that it is so much easier
to feed a black hole at the present epoch than it is to form stars.
Another possible interpretation of Fig 12 is that the
emission from the AGN provides a limit to star-formation, forcing
$SFR < 10^5 M_{acc}$.

It is possible that the high proportion of AGN amongst these extreme
starbursts is pointing to the influence of AGN jet-induced star
formation in these extreme objects (Klamer et al. 2004; Clements et
al. 2009). While we currently have no information on the prevalence of
jets in the sample discussed here, there are individual extreme
starbursts such as J160705.16+533558.5 (Clements et al., 2009, and in
prep) and 4C41.17 (de Breuck et al., 2005; Steinbring, 2014) where
there are strong indications that jets are triggering star
formation. Furthermore, Klamer et al. (2004) presents a sample of 12
z$>$3 star forming AGN where star formation appears to be triggered by
relativistic jets. More information on the AGN and gas distribution in
the sources in the current paper is clearly needed, but we note that
the time-scales for these starbursts and the time-scale for black hole
growth are remarkably well matched at $\sim$ 10$^{7}$ years (Rigopoulou et
al. 2009). However, the greatly enhanced gas supply to the nucleus
associated with violent mergers may be a sufficient explanation.

If there is a connection between black hole accretion and
star-formation, why do only half of our extreme starbursts harbour AGN
?  Firstly the non-detection of an AGN dust torus sets only a modest
upper limit on $M_{bh} \beta^{-1}/M_*$ of $\sim 10^{-5}$, i.e. at the
lower end of the observed distribution.  It is possible that there is
a phase-lag between star-formation and black-hole growth and this is
supported by the fact that of the 11 galaxies fitted with a young
starburst template in the infrared only two also have an AGN dust
torus.  To grow a black hole there has to be a black hole present in
the first place and perhaps some galaxies have not yet formed a
massive nuclear black hole.

Finally, we note an interesting disconnection between X-ray detected
AGN, and the {\itshape Herschel} sources. The SWIRE-Lockman area
includes the CLASX X-ray survey. Rowan-Robinson et al (2009) gave a
detailed discussion of the associations of CLASX and SWIRE sources.
Only two of the 400 CLASX-SWIRE sources are detected by {\it
  Herschel}-SPIRE.  This is consistent with the idea that, while AGN
are present in the {\it Herschel} submillimetre galaxy population,
they make a negligible contribution to the submillimetre flux.

\section{Conclusions}
After careful exclusion of lensed galaxies and blazers, we have
identified samples of extreme starbursts, with star-formation rates in
the range 5000-30,000 $M_{\odot}/yr$, from the {\it IRAS}-FSS 60 $\mu$m
galaxy catalogue (RIFSCz) and from the {\it Herschel}-SWIRE ({\it HerMES})
500 $\mu$m survey.  The correctness of our SWIRE associations is
confirmed for 8 objects by radio maps.  ALMA submillimetre mapping
and deeper radio mapping by LOFAR, GMRT, MeerKAT and SKA will help confirm
the reality of the remaining sources.

There do not seem to be any genuine cases with SFR$>$30,000
$M_{\odot}/yr$ and this may be essentially an Eddington-type
limit. The SEDs of 38 {\it HerMES} extreme starbursts have been modelled in
detail. The photometric redshifts are, in almost all cases, supported
by redshift estimates from the 250-500 $\mu$m colours.  The proportion
of 500 $\mu$m sources which may be subject to blending or association
with the wrong 24 $\mu$m source is $<12\%$.  Using dust mass as a
proxy for gas mass, extreme starbursts are found to be very gas rich
systems, which will double their stellar mass in 0.3-3 x $10^8$ yrs.

About half of the {\it Herschel} extreme starburst systems also
contain an AGN, but in no case do these dominate the bolometric
output. With assumptions about the Eddington ratio and accretion
efficiency, we find a universal relation between black-hole mass and
total stellar mass, with $M_{bh} \sim 0.001 M_*$.  This is driven by
the episode of extreme star-formation and black hole growth at
z=2-5. However while the star formation rate has fallen by a factor of
1000 between redshift 5 and the present epoch, the black hole
accretion rate has fallen by a factor of only 30, suggesting a
decoupling between star formation and the feeding of the nuclear black
hole.  

\begin{acknowledgements}
{\it Herschel} is an ESA space observatory with science instruments
provided by European-led Principal Investigator consortia and with
important participation from NASA. SPIRE has been developed by a
consortium of institutes led by Cardiff University (UK) and including
Univ. Lethbridge (Canada); NAOC (China); CEA, LAM (France); IFSI,
Univ.  Padua (Italy); IAC (Spain); Stockholm Observatory (Sweden);
Imperial College London, RAL, UCL-MSSL, UKATC, Univ. Sussex (UK); and
Caltech, JPL, NHSC, Univ. Colorado (USA).  This development has been
supported by national funding agencies: CSA (Canada); NAOC (China);
CEA, CNES, CNRS (France); ASI (Italy); MCINN (Spain); SNSB (Sweden);
STFC, UKSA (UK); and NASA (USA).

We thank an anonymous referee for comments that allowed us to substantially improve the paper.
\end{acknowledgements}

\begin{appendix} 
\section{Tabulated data for Extreme Starbursts}
Tables from section 6.
\begin{sidewaystable*}
\caption{Extreme starbursts: QSOs and Type 2 AGN}
\begin{tabular}{lllllllllllllll}
\hline\hline
\smallskip
RA & Dec & $m_i$ & S24 & S250 & S350 & S500 & $z_{phot}$ & type & $\chi^2$ & $n_{b}$ & $z_{subm}$ & $z_{comb}$ & $M_*$ & SFR\\
&&& ($\mu$Jy) & (mJy) & (mJy) & (mJy) &&&&  &&& $\log(M_{\odot})$ & $\log(M_{\odot} yr^{-1})$\\
\hline
QSOs, Fig 6 \\
\hline
164.64154 & 58.09799 & 20.95 & 1386.3 & 44.9 & 45.4 & 28.0 & 3.37 & QSO & 2.1 & 11 & 2.01 & 3.07 & & 3.71 \\
(164.64879 & 58.09513 & & & & & & 1.11 & & & & & & & 2.74)\\
162.77605 & 58.52327 & 22.32 & 184.6 & 54.5 & 53.0 & 41.1 & 3.47 & QSO & 3.2 & 8 & 3.19 & 3.37 & & 3.71 \\
161.85410 & 57.91928 & 21.11 & 245.8 & 76.1 & 80.3 & 65.3 & 3.49 & QSO & 1.3 & 10 & 3.42 & 3.47 & & 3.86 \\
162.20728 & 58.28162 & 21.59 & 205.5 & 56.6 & 57.7 & 40.3 & 3.61 & QSO & 8.5 & 10 & 3.14 & 3.37 & & 3.83 \\
161.36092 & 58.03157 & 20.67 & 428.6 & - & 37.5 & 41.2 & 3.81 & QSO & 5.2 & 11 & 5.05 & 3.90 & & 3.83 \\
162.68120 & 57.55606 & g=24.35 & 194.9 & 25.5 & 37.0 & 38.3 & 4.50 & QSO & 1.3 & 3 & 4.79 & 4.50 & & 3.79 \\
\hline
Type 2 AGN, Fig 7\\
\hline
33.71100  & -4.17344 & 22.53 & 1016.3 & 96.2 & 69.0 & 32.5 & 1.47 & Scd & 9.3 & 6 & 1.92 & 1.57 & 11.80 & 3.81 \\
162.91730 & 58.80596 & 22.03 & 1964.8 & 183.8 & 133.1 & 78.7 & 2.06 & Sab & 45.6 & 6 & 2.11 & 2.09 & 12.22 & 3.81 \\
161.75087 & 59.01883 & 23.66 & 1329.6 & 75.0 & 61.0 & 40.6 & 2.5620 & Scd & 1.7 & 7 & 2.53 &  & 11.78 & 3.70 \\
35.95578  & -5.08144 & 23.85 & 1002.1 & 71.2 & 77.6 & 59.5 & 3.15 & Sab & 7.3 & 5 & 3.32 & 3.27 & 11.91 & 3.80 \\
10.23271 & -44.07592 & R=21.50 & 487.0 & 75.3 & 68.1 & 45.7 & 3.29 & sb & 13.3 & 5 & 2.67 & 3.07 & 11.75 & 3.77 \\
160.33716 & 59.40493 & 22.36 & 1225.7 & - & 36.8 & 39.0 & 3.29 & Scd & 6.7 & 6 & 4.93 & 3.27 & 12.08 & 3.71 \\
36.15258  & -5.10250 & 24.92 & 627.2 & 43.5 & 53.7 & 41.5 & 3.37 & Sab & 2.0 & 5 & 3.60 & 3.47 & 11.69 & 3.71 \\
159.78395 & 58.55888 & g=24.41 & 231.6 & 35.3 & 39.9 & 36.7 & 3.72 & Sdm & 0.0 & 3 & 3.91 & 3.68 & 11.32 & 3.79 \\
36.96840  & -5.02193 & 24.92 & 610.3 & 40.4 & 47.8 & 39.3 & 3.79 & Sab & 7.5 & 6 & 3.66 & 3.79 & 11.79 & 3.94 \\
160.85139 & 58.02007 & 23.22 & 443.7 & 43.1 & 61.2 & 37.4 & 3.92 & Sbc & 2.6 & 5 & 2.24 & 3.37 & 12.21 & 3.94 \\
35.92307  & -4.73225 & 25.55 & 849.5 & 48.5 & 59.5 & 45.3 & 3.94 & Sab & 2.9 & 5 & 3.56 & 3.57 & 12.12 & 3.83 \\
164.28366 & 58.43524 & 22.30 & 596.0 & 43.5 & 51.0 & 37.4 & 4.15 & Scd & 60.0 & 5 & 2.46 & 3.79 & 12.06 & 4.14 \\
(164.28227 & 58.43064 & & & & & & 0.20 & & & & & & & 0.40)\\
161.63013 & 59.17688 & 23.94 & 391.4 & -    & 29.6 & 27.0 & 5.19 & Scd & 4.8 & 4 & 3.05 & 4.75 & 12.11 & 4.22 \\
\hline
\end{tabular}
\tablefoot{Spectroscopic redshifts shown with four decimal places.}
\end{sidewaystable*}

\begin{sidewaystable*}
\caption{Extreme starbursts: young starbursts, M82 and A220 type starbursts}
\begin{tabular}{lllllllllllllll}
\hline\hline
\smallskip
RA & dec & i & S24 & S250 & S350 & S500 & $z_{phot}$ & type & $\chi^2$ & $n_{bands}$ & $z_{subm}$ & $z_{comb}$ & $M_*$ & SFR \\
&&& ($\mu$Jy) & (mJy) & (mJy) & (mJy) &&&&  &&& $\log(M_{\odot})$ & $\log(M_{\odot} yr^{-1})$\\
\hline
young sbs,& Fig 9 \\
\hline
159.03456 & 58.44533 & 21.37 & 1127.8 & 131.6 & 84.2 & 42.2 & 1.44 & Scd & 4.8 & 4 & 1.16 & 1.19 & 11.62 & 3.89 \\
35.28232 & -4.14900 & g=25.31 & 636.1 & 53.7 & 58.3 & 55.1 & 3.27 & Sab & 2.4 & 4 & 3.84 & 2.80 & 11.28 & 3.70 \\
(35.28037 & -4.14839 & & & & & & 2.55 & & & & & & & 3.59)\\
160.50839 & 58.67179 & 23.49 & 904.6 & 95.7 & 79.5 & 60.5 & 3.81 & Sab & 4.1 & 6 & 2.86 & 3.07 & 12.02 & 3.99 \\
(160.51041 & 58.67371 & & & & & & 1.08 & & & & & & & 3.05)\\
36.59817  & -4.56164 & 25.11 & 389.5 & 59.3 & 67.2 & 54.7 & 3.88 & Sab & 2.1 & 5 & 3.53 & 3.68 & 11.49 & 3.85 \\
161.98271 & 58.07477 & 22.10 & 264.4 & 44.2 & 45.3 & 33.6 & 4.13 & sb & 25.4 & 6 & 3.15 & 3.68 & 11.87 & 3.76 \\
\hline
M82, A220 sbs, & Fig 8\\
\hline
162.33324 & 58.10657 & 22.61 & 516.4 & 56.9 & 52.9 & 59.6 & 2.80 & Scd & 6.1 & 8 & 4.09 & 2.89 & 11.69 & 3.72 \\
9.64405 & -44.36636 & 23.02 & 987.2 & 91.7 & 79.4 & 48.1 & 2.85 & Scd & 1.8 & 4 & 2.43 & 2.72 & 11.89 & 3.70 \\
160.91940 & 57.91475 & 22.32 & 682.3 & 116.5 & 102.0 & 65.4 & 3.06 & Sab & 20.5 & 6 & 2.56 & 2.80 & 12.50 & 4.12 \\
8.81979  & -42.69724 & R=22.90 & 706.2 & 87.4 & 74.6 & 53.9 & 3.19 & Scd & 6.1 & 5 & 2.72 & 2.98 & 12.06 & 3.79 \\
162.38754 & 57.70547 & g=24.13 & 519.3 & 65.7 & 65.0 & 56.3 & 3.37 & Scd & 1.6 & 3 & 3.47 & 3.37 & 12.15 & 3.74 \\
36.71948 & -3.96377 & 25.18 & 322.8 & 43.4 & 42.8 & 32.1 & 3.57 & Sbc & 5.2 & 5 & 3.10 & 3.27 & 12.21 & 3.70 \\
162.46065 & 58.11701 & 21.96 & 252.5 & -    & 28.4 & 41.4 & 4.06 & sb (QSO?)& 7.9 & 6 & 4.81 & 4.13 & 11.97 & 3.98 \\
35.73369 & -5.62305 & 23.42 & 422.0 & 73.1 & 60.9 & 42.1 & 4.11 & Sbc & 4.1 & 6 & 2.61 & 2.80 & 12.38 & 4.10 \\
&&&&&&& 2.80 & Sbc &&&&& & 3.70 \\
7.98209 & -43.29812 & R=24.87 & 275.8 & 45.6 & 49.7 & 38.8 & 4.13 & Sab & 0.02 & 3 & 3.37 & 3.37 & 11.90 & 3.82 \\
&&&&&&& 3.37 & Sab &&&&& & 3.70 \\
36.10986  & -4.45889 & 24.59 & 828.6 & 88.4 & 89.5 & 67.0 & 3.92 & Sab & 2.0 & 5 & 3.14 & 3.27 & 12.26 & 4.21 \\
164.52054 & 58.30782 & 23.40 & 306.9 & 81.9 & 92.1 & 58.2 & 4.18 & Sab & 0.03 & 3 & 2.16 & 3.07 & 12.40 & 3.99\\
(164.52888 & 58.30814 & & & & & & 1.87 & & & & & & & 3.35)\\
161.89894 & 58.16401 & 23.60 & 315.0 & 66.4 & 59.7 & 35.3 & 4.37 & Sab & 22.7 & 4 & 1.82 & 2.89 & 12.32 & 3.97 \\
&&&&&&& 2.89 & Sab &&&&& & 3.70 \\
36.65871 & -4.14628 & 24.91 & 288.8 & 46.1 & 45.7 & 48.1 & 4.40 & Sbc & 2.1 & 6 & 3.92 & 4.01 & 12.20 & 4.03\\
162.42290 & 57.18750 & 22.25 & 121.0 & 43.5 & 46.1 & 32.6 & 4.45 & sb (QSO?) & 5.6 & 4 & 3.22 & 4.37 & 11.79 & 3.78 \\
\hline
\end{tabular}
\end{sidewaystable*}

\begin{sidewaystable*}
\caption{Rejected extreme starburst candidates}
\begin{tabular}{lllllllll}
\hline\hline
\smallskip
RA & dec & $z_{phot}$ & type & $\chi^2$ & $n_{bands}$ & $z_{subm}$ & $z_{comb}$ & reason for rejection\\
&&& &&  &&& \\
\hline
159.24428 & 57.85775 & 3.06 & Scd & 3.1 & 5 & 1.80 & 1.51 &  $z_{comb}$ = 1.51 gives acceptable fit\\
35.49809 & -5.92264  & 3.13 & Sab & 1.9 & 6 & 2.49 & 1.69 &  $z_{comb}$ = 1.69 gives acceptable fit\\
161.21138 & 58.11261 & 3.33 & Sbc & 26.8 & 5 & 2.02 & 1.75 &  $z_{comb}$ = 1.75 gives acceptable fit\\
162.55945 & 57.19608 & 3.70 & sb & 3.6 & 5 & 2.01 & 1.19 &  $z_{comb}$ = 1.19 gives acceptable fit\\
159.67438 & 58.55686 & 3.72 & Scd & 0.0 & 3 & 2.68 & 2.02 &  $z_{comb}$ = 2.02 gives acceptable fit\\
162.52769 & 57.28142 & 3.92 & QSO & 8.4 & 8 & 1.93 & 1.95 & $z_{comb}$ = 1.95 gives acceptable fit\\
35.11967  & -5.73062 & 3.94 & Sbc & 14.8 & 6 & 2.90 & 1.69 &  $z_{comb}$ = 1.69 gives acceptable fit \\
34.26031 & -4.95556 & 4.15 & Sbc & 3.9 & 4 & 2.03 & 2.16 &  $z_{comb}$ = 2.16 gives acceptable fit\\
164.02647 & 52.07153 & 4.18 & sb & 2.5 & 4 & 2.74 & 1.82 &  $z_{comb}$ = 1.82 gives acceptable fit\\
9.28433 & -44.23750 & 4.18 & Scd & 0.03 & 3 & 2.40 & 2.39 &  $z_{comb}$ = 2.39 gives acceptable fit \\
9.08571 & -42.59628 & 4.25 & Scd & 0.85 & 3 & 2.33 & 2.09 &  $z_{comb}$ = 2.09 gives acceptable fit\\
36.25277 & -5.59534 & 4.27 & Scd & 1.2 & 5 & 2.65 & 1.95 &  $z_{comb}$ = 1.95 gives acceptable fit\\ 
9.11142 & -42.84052 & 4.60 & Scd & 0.03 & 3 & 2.36 & 2.31 &  $z_{comb}$ = 2.31 gives acceptable fit\\
9.30474 & -43.03506 & 4.86 & Scd & 0.09 & 3 & 2.87 & 2.80 &  $z_{comb}$ = 2.80 gives acceptable fit\\
161.58835 & 59.65826 & 5.05 & Scd & 4.7 & 5 & 3.03 & 2.98 &  $z_{comb}$ = 2.98 gives acceptable fit\\
34.53469 & -5.00769 & 4.40 & sb  & 4.0 & 5 & 3.18 & 3.17 &  $z_{comb}$ = 3.17 gives acceptable fit\\
8.70199 & -44.48560 & 4.73 & Scd & 0.01 & 3 & 3.45 & 3.47 & $z_{comb}$ = 3.47 gives acceptable fit \\
9.17274  & -43.34398 & 3.06 & Sab & 15.7 & 6 & 1.84 & 2.47 &  $z_{spect}$ = 1.748 (Bongiorno et al 2014)\\
35.73605 & -4.88950 & 4.37 & Scd & 11.4 & 6 & 3.69 & 2.55 &  alias at $z_{phot}$ = 1.9 gives acceptable fit\\ 
35.33492  & -5.74307 & 3.81 & Sbc & 2.7 & 4 & 2.06 & 2.09 &  alias at $z_{phot}$ = 1.7 gives acceptable fit \\
164.55716 & 58.65286 & 3.92 & sb & 2.3 & 6 & 2.96 & 3.57 &  alias at $z_{phot}$ = 1.2 fits 24-500$\mu$m\\
9.19681 & -44.42382 & 5.00 & sb (QSO?) & 0.04 & 3 & 3.89 & 4.13 &  alias at $z_{phot}$ = 1.7 gives acceptable fit\\
159.95905 & 57.18814  & 3.37 & Scd & 0.1 & 3 & 4.88 & 3.37 & alias at $z_{phot}$ = 1.95 gives acceptable fit \\
163.98088 & 57.81277 & 3.74 & Scd & 4.4 & 5 & 5.03 & 3.68 & alias at $z_{phot}$ = 2.3 gives acceptable fit \\
162.84616 & 58.00514 & 4.01 & sb & 13.0 & 5 & 3.56 & 4.01 & alias at $z_{phot}$ = 2.1 gives acceptable fit \\
160.16505 & 57.27072 & 3.70 & Sab & 31.2 & 4 & 3.45 & 3.68 & CIGALE alias at $z_{phot}$ = 1.9 gives acceptable fit \\
162.26817 & 58.46461 & 3.15 & Sbc & 5.0 & 6 & 3.16 & 3.17 & CIGALE alias at $z_{phot}$ = 1.7 gives acceptable fit \\
36.84426 & -5.31016  & 3.07 & QSO & 42.1 & 14 & 3.01 & 3.07 &  alternative SWIRE association with $z{phot}$ = 1.49 fits better\\
9.62462  & -43.74844 & 1.5670 & QSO & 2.9 & 9 & 2.40 &  &  SED fit implies SFR $<$ 3.70\\
164.96553 & 58.30081 & 2.3350 & QSO & 13.6 & 11 & 2.33 &  & SED fit implies SFR $<$ 3.70\\
8.04056  & -43.71930 & 1.51 & Sab  & 26.4 & 7 & 2.96 & 2.09 &  SED fit implies SFR $<$ 3.70\\
35.11776 & -5.50099  & 1.55 & Scd & 9.0 & 7 & 2.74 & 2.02 &  SED fit implies SFR $<$ 3.70\\
\hline
\end{tabular}
\end{sidewaystable*}

\end{appendix}

\end{document}